\documentclass[12pt, a4paper,twoside,american]{article}

\newcommand{\ttl}{Decorrelation of neural-network activity by inhibitory feedback}
\newcommand{\sttl}{Decorrelation by inhibitory feedback}

\usepackage{geometry}
\usepackage[intlimits]{amsmath}
\usepackage{amssymb}
\usepackage{graphicx}
\usepackage{calc}
\usepackage{float}
\usepackage{color}
\usepackage{prettyref}
\usepackage{setspace}
\usepackage{babel}
\usepackage[labelfont=bf,labelsep=period]{caption}
\usepackage{fancyhdr}
\usepackage{multirow}
\usepackage{colortbl}
\usepackage{tabularx}
\usepackage{twoopt}
\usepackage{ifthen}
\usepackage{bm}

\usepackage[pdftex,breaklinks=true,colorlinks=true,linkcolor=blue,citecolor=blue,urlcolor=blue,
filecolor=blue,pdffitwindow,backref=true,pagebackref=false,bookmarks=true,bookmarksopen=true,
bookmarksnumbered=true]{hyperref}
\usepackage[pdftex,displaymath,floats,graphics,textmath,sections,footnotes,showlabels,counters]{preview}

\PreviewEnvironment{itemize}

\makeatletter

\usepackage{fancyhdr}
\newrefformat{fig}{\hyperref[#1]{Fig.\,\ref*{#1}}}
\newrefformat{sec}{\hyperref[#1]{Sec.\,\ref*{#1}}}
\newrefformat{tab}{\hyperref[#1]{Tab.\,\ref*{#1}}}
\newrefformat{app}{\hyperref[#1]{App.\,\ref*{#1}}}
\newrefformat{subsec}{\hyperref[#1]{Sec.\,\ref*{#1}}}
\newrefformat{eq}{\hyperref[#1]{(\ref*{#1})}}

\renewenvironment{abstract}
{\noindent{\normalfont\large\textbf{Abstract}}%
\par\vspace{0.5\baselineskip}\noindent}%
{\par}

\makeatother

\newcommand{\modelhdr}[3]{%
  \multicolumn{#1}{|l|}{%
    \color{white}\cellcolor[gray]{0.0}%
    \textbf{\makebox[0pt]{#2}\hspace{0.5\linewidth}\makebox[0pt][c]{#3}}%
  }%
}
\newcommand{\parameterhdr}[3]{%
  \multicolumn{#1}{|l|}{%
    \color{black}\cellcolor[gray]{0.8}%
    \textbf{\makebox[0pt]{#2}\hspace{0.5\linewidth}\makebox[0pt][c]{#3}}%
  }%
}
\newcommand{\nettypehdr}[3]{%
  \multicolumn{#1}{|l|}{%
    \color{black}\cellcolor[gray]{0.9}%
    \textit{\makebox[0pt]{#2}\hspace{0.5\linewidth}\makebox[0pt][c]{#3}}%
  }%
}

\newcommandtwoopt{\citep}[3][\empty][\empty]{%
  \ifthenelse{\equal{#1}{\empty} \AND \equal{#2}{\empty}}
  {\cite{#3}}%
  {(#1~\cite{#3}#2)}%
}
\newcommandtwoopt{\citealp}[3][\empty][\empty]{%
  \ifthenelse{\equal{#1}{\empty} \AND \equal{#2}{\empty}}%
  {\cite{#3}}%
  {#1~\cite{#3}#2}%
}
\newcommand{\citet}[1]{\cite{#1}}

\newcommand{\adj}{^\ast}

\newcommand{\bz}{\mathbf{z}}

\newcommand{\cc}{^\ast}

\newcommand{\Cov}[2][]{\text{Cov}_{#1}\left[#2\right]}
\newcommand{\defeq}{\stackrel{\text{def}}{=}}

\newcommand{\dt}{\text{d}t}
\newcommand{\e}{\mathbf{e}}

\newcommand{\EW}[2][]{\text{E}_{#1}\left[#2\right]}

\newcommand{\Ex}{\text{E}}
\newcommand{\Exc}{\mathcal{E}}
\newcommand{\X}{\mathcal{X}}
\newcommand{\Y}{\mathcal{Y}}
\newcommand{\Z}{\mathcal{Z}}
\newcommand{\W}{\mathcal{W}}
\newcommand{\ext}{\text{ext}}
\newcommand{\eye}{\mat{1}}
\newcommand{\FF}{\text{FF}}

\newcommand{\Fourier}[2]{\mathfrak{F}\left[#1\right]\left(#2\right)}
\newcommand{\InvFourier}[2]{\mathfrak{F}^{-1}\left[#1\right]\left(#2\right)}
\newcommand{\Hz}{\text{Hz}}
\newcommand{\In}{\text{I}}
\newcommand{\Inh}{\mathcal{I}}
\newcommand{\inp}{\text{in}}
\newcommand{\Int}[4]{\int\limits_{#3}^{#4} \text{d}#2\ #1}
\newcommand{\la}{\leftarrow}

\newcommand{\mat}[1]{{\bm{#1}}}
\newcommand{\MOhm}{\text{M}\Omega}
\newcommand{\ms}{\text{ms}}
\newcommand{\mV}{\text{mV}}
\newcommand{\mynote}[1]{\noindent\fcolorbox{black}{yellow}{\parbox{\linewidth}{#1}}}

\PreviewMacro[!]{\mynote}
\newcommand{\nlra}{\not\rightleftarrows}
\newcommand{\pA}{\text{pA}}

\newcommand{\psc}{\text{psc}}
\newcommand{\ra}{\rightarrow}
\newcommand{\resetpot}{V_\text{reset}}
\newcommand{\RM}{R_{\text{m}}}
\newcommand{\s}{{\bm{s}}}
\newcommand{\seconds}{\text{s}}
\newcommand{\sfc}{\text{sfc}}
\newcommand{\SN}{\text{SN}}
\newcommand{\sps}{\text{s}^{-1}}
\newcommand{\syn}{\text{s}}
\newcommand{\tauM}{\tau_{\text{m}}}
\newcommand{\tauR}{\tau_{\text{ref}}{}}

\newcommand{\threshold}{\theta}
\newcommand{\transp}{^\text{T}}
\newcommand{\tpr}{t^\prime}
\newcommand{\tppr}{t^{\prime \prime}}

\newcommand{\Var}[2][]{\text{Var}_{#1}\left[#2\right]}
\newcommand{\w}{\bar{w}}
\newcommand{\g}{\bar{g}}
\renewcommand{\vec}[1]{\bm{#1}}

\newcommand{\oneandahalfcolumnfigure}[1]{\includegraphics[width=3\textwidth/4]{#1}}
\newcommand{\doublecolumnfigure}[1]{\includegraphics[width=\textwidth]{#1}}

\renewcommand{\titlepage}{
  \setcounter{page}{0}
  \vfill
  \begin{flushleft}
    {\bf Correspondence to:}\\[0.2cm]
    Tom Tetzlaff\\
    Inst.~of Neuroscience and Medicine (INM-6)\\
    Computational and Systems Neuroscience\\
    Research Center J\"ulich\\
    D-52425 J\"ulich\\
    Germany\\
    \href{mailto:t.tetzlaff@fz-juelich.de}%
    {t.tetzlaff@fz-juelich.de}\\
  \end{flushleft}
  
  \vfill
  
  \vfill
  \begin{center}
    \date{\today}
  \end{center}
  
  \thispagestyle{empty}
  
  \clearpage
}
\def\figpath{./figures}

\begin{document}
\renewcommand{\restylefloat}{figure}
\setcounter{page}{1}
\pdfbookmark[1]{Title}{TitlePage}
\title{\bfseries%
  \ttl
}
\author{%
  Tom Tetzlaff\;$^{1,2,\#,\ast}$,
  Moritz Helias\;$^{1,3,\#}$,
  Gaute T.~Einevoll\;$^{2}$,
  Markus Diesmann\;$^{1,3}$\\
  \parbox{\textwidth}{
    \begin{itemize}\small
    \item[$^{1}$]
      Inst.~of Neuroscience and Medicine (INM-6),\\ 
      Computational and Systems Neuroscience,\\
      Research Center J\"ulich, Germany
    \item[$^{2}$]
      Department of Mathematical Sciences and Technology,\\
      Norwegian University of Life Sciences, {\AA}s, Norway
    \item[$^{3}$]
      RIKEN Brain Science Institute \&\\
      Brain and Neural Systems Team, RIKEN Computational Science Research Program,\\
      Wako, Japan
    \item[$^\#$] These authors contributed equally to this work.
    \item[$^\ast$] corresponding author
    \end{itemize}
  }
}
\date{}
\maketitle
\titlepage

\clearpage

\begin{abstract}{}%
  \pdfbookmark[1]{Abstract}{AbstractPage}%
  Correlations in spike-train ensembles can seriously impair the
  encoding of information by their spatio-temporal structure.  An
  inevitable source of correlation in finite neural networks is common
  presynaptic input to pairs of neurons. Recent studies demonstrate
  that spike correlations in recurrent neural networks are
  considerably smaller than expected based on the amount of shared
  presynaptic input.
  Here, we explain this observation by means of a linear network model
  and simulations of networks of leaky integrate-and-fire neurons.  We
  show that inhibitory feedback efficiently suppresses pairwise
  correlations and, hence, population-rate fluctuations, thereby
  assigning inhibitory neurons the new role of active decorrelation.
  We quantify this decorrelation by comparing the responses of the
  intact recurrent network (feedback system) and systems where the
  statistics of the feedback channel is perturbed (feedforward
  system).
  Manipulations of the feedback statistics can lead to a significant
  increase in the power and coherence of the population response. In
  particular, neglecting correlations within the ensemble of feedback
  channels or between the external stimulus and the feedback amplifies
  population-rate fluctuations by orders of magnitude.
  The fluctuation suppression in homogeneous inhibitory networks is
  explained by a negative feedback loop in the one-dimensional
  dynamics of the compound activity. Similarly, a change of
  coordinates exposes an effective negative feedback loop in the
  compound dynamics of stable excitatory-inhibitory networks. The
  suppression of input correlations in finite networks is explained by
  the population averaged correlations in the linear network model: In
  purely inhibitory networks, shared-input correlations are canceled
  by negative spike-train correlations. In excitatory-inhibitory
  networks, spike-train correlations are typically positive. Here, the
  suppression of input correlations is not a result of the mere
  existence of correlations between excitatory (E) and inhibitory (I)
  neurons, but a consequence of a particular structure of correlations
  among the three possible pairings (EE, EI, II).
  \subsection*{Author summary}
  The spatio-temporal activity pattern generated by a recurrent
  neuronal network can provide a rich dynamical basis which allows
  readout neurons to generate a variety of responses by tuning the
  synaptic weights of their inputs. The repertoire of possible
  responses and the response reliability become maximal if the spike
  trains of individual neurons are uncorrelated. Spike-train
  correlations in cortical networks can indeed be very small, even for
  neighboring neurons. This seems to be at odds with the finding that
  neighboring neurons receive a considerable fraction of inputs from
  identical presynaptic sources constituting an inevitable source of
  correlation.  In this article, we show that inhibitory feedback,
  abundant in biological neuronal networks, actively suppresses
  correlations.  The mechanism is generic: It does not depend on the
  details of the network nodes and decorrelates networks composed of
  excitatory and inhibitory neurons as well as purely inhibitory
  networks.  For the case of the leaky integrate-and-fire model, we
  derive the correlation structure analytically. The new toolbox of
  formal linearization and a basis transformation exposing the
  feedback component is applicable to a range of biological systems.
  We confirm our analytical results by direct simulations.
\end{abstract}
\pagestyle{fancy}

\section{Introduction}
\label{sec:intro}
Neurons generate signals by weighting and combining input spike trains
from presynaptic neuron populations.  The number of possible signals
which can be read out this way from a given spike-train ensemble is
maximal if these spike trains span an orthogonal basis, i.e.~if they
are uncorrelated \citep{Tripp07_1830}. If they are correlated, the
amount of information which can be encoded in the spatio-temporal
structure of these spike trains is limited.  In addition, correlations
impair the ability of readout neurons to decode information reliably
in the presence of noise.  This is often discussed in the context of
\emph{rate coding}: for $N$ uncorrelated spike trains, the
signal-to-noise ratio of the compound spike-count signal can be
enhanced by increasing the population size $N$. In the presence of
correlations, however, the signal-to-noise ratio is bounded
\citep{Zohary94_140, Shadlen98}.  The same reasoning holds for any
other linear combination of spike trains, also for those where exact
spike timing matters \citep[for example for the coding scheme
presented in][]{Tripp07_1830}. Thus, the robustness of neuronal
responses against noise critically depends on the level of correlated
activity within the presynaptic neuron population.  
\par
Several studies suggested that correlated neural activity could be
beneficial for information processing: Spike-train correlations can
modulate the gain of postsynaptic neurons and thereby constitute a
gating mechanism \citep[for a review, see][]{Salinas01}. Coherent
spiking activity might serve as a means to bind elementary
representations into more complex objects
\citep[][]{Malsburg81,Bienenstock95}. Information represented by
correlated firing can be reliably sustained and propagated through
feedforward subnetworks \citep['synfire
chains';][]{Abeles91,Diesmann99}. Whether correlated firing has to be
considered favorable or not largely depends on the underlying
hypothesis, the type of correlation (e.g.~the time scale or the
affected frequency band) or which subpopulations of neurons are
involved. Most ideas suggesting a functional benefit of correlated
activity rely on the existence of an asynchronous 'ground
state'. Spontaneously emerging correlations, i.e.~correlations which
are not triggered by internal or external events, would impose a
serious challenge to many of these hypotheses. Functionally relevant
synfire activity, for example, cannot be guaranteed in the presence of
correlated background input from the embedding network
\citep{Tetzlaff04_117}. It is therefore--from several
perspectives--important to understand the origin of uncorrelated
activity in neural networks.
\par
It has recently been shown that spike trains of neighboring cortical
neurons can indeed be highly uncorrelated \citep{Ecker10}. Similar
results have been obtained in several theoretical studies
\citep{Vreeswijk96,Brunel00_183,Gerstner00,Tetzlaff08_2133,Kriener08_2185,Hertz10_427,Renart10_587}.
From an anatomical point of view, this observation is puzzling: in
general, neurons in finite networks share a certain fraction of their
presynaptic sources. In particular for neighboring neurons, the
overlap between presynaptic neuron populations is expected to be
substantial. This feedforward picture suggests that such presynaptic
overlap gives rise to correlated synaptic input and, in turn, to
correlated response spike trains.
\par
A number of theoretical studies showed that shared-input correlations
are only weakly transferred to the output side as a consequence of the
nonlinearity of the spike-generation dynamics
\citep{Stroeve01,Tetzlaff02,Morenobote06_028101,DeLaRocha07_802,Kriener08_2185}.
Unreliable spike transmission due to synaptic failure can further
suppress the correlation gain \citep{Rosenbaum10_00116}. In
\citep{Tetzlaff04_117}, we demonstrated that spike-train correlations
in finite-size recurrent networks are even smaller than predicted by
the low correlation gain of pairs of neurons with nonlinear
spike-generation dynamics. We concluded that this suppression of
correlations must be a result of the recurrent network dynamics. In
this article, we compare correlations observed in feedforward networks
to correlations measured in systems with an intact feedback loop. We
refer to the reduction of correlations in the presence of feedback as
``decorrelation''.  Different mechanisms underlying such a dynamical
decorrelation have been suggested in the recent past. Asynchronous
states in recurrent neural networks are often attributed to chaotic
dynamics \citep{Battaglia07_238106, Monteforte10}. In fact, networks
of nonlinear units with random connectivity and balanced excitation
and inhibition typically exhibit chaos
\citep{Vreeswijk96,Jahnke09}. The high sensitivity to noise may
however question the functional relevance of such systems
(\citealp{Legenstein07_323,Jahnke08}; cf., however,
\citealp{Toyoizumi10}). \citet{Zillmer06} and \citet{Jahnke08}
demonstrated that asynchronous irregular firing can also emerge in
networks with stable dynamics. Employing an analytical framework of
correlations in recurrent networks of binary neurons
\citet{Ginzburg94}, the balance of excitation and inhibition has
recently been proposed as another decorrelation mechanism
\citet{Renart10_587}: In large networks, fluctuations of excitation
and inhibition are in phase. Positive correlations between excitatory
and inhibitory input spike trains lead to a negative component in the
net input correlation which can compensate positive correlations
caused by shared input.
\par
In the present study, we demonstrate that dynamical decorrelation is a
fundamental phenomenon in recurrent systems with negative feedback. We
show that negative feedback alone is sufficient to efficiently
suppress correlations. Even in purely inhibitory networks,
shared-input correlations are compensated by feedback. A balance of
excitation and inhibition is thus not required. The underlying
mechanism can be understood by means of a simple linear model. This
simplifies the theory and helps to gain intuition, but it also
confirms that low correlations can emerge in recurrent networks with
stable, non-chaotic dynamics.
\par
The suppression of pairwise spike-train correlations by inhibitory
feedback is reflected in a reduction of population-rate
fluctuations. The main effect described in this article can therefore
be understood by studying the dynamics of the macroscopic population
activity
(\prettyref{sec:pop_fluct_iaf}--\prettyref{sec:EI_poprate_model}). This
approach leads to a simple mathematical description and emphasizes
that the described decorrelation mechanism is a general phenomenon
which may occur not only in neural networks but also in other
(biological) systems with inhibitory feedback. In
\prettyref{sec:pop_fluct_iaf}, we first illustrate the decorrelation
effect for random networks of $N$ leaky integrate-and-fire (LIF)
neurons with inhibitory or excitatory-inhibitory coupling. By means of
simulations, we show that low-frequency spike-train correlations, and,
hence, population-rate fluctuations are considerably smaller than
expected given the amount of shared input. As shown in
\prettyref{sec:I_poprate_model}, the suppression of population-rate
fluctuations by inhibitory feedback can readily be understood in the
framework of a simple one-dimensional linear model with negative
feedback.  \prettyref{sec:EI_poprate_model} extends this result to a
two-population system with excitatory-inhibitory coupling. Here, a
simple coordinate transform exposes the inherent negative feedback
loop as the underlying cause of the fluctuation suppression in
inhibition-dominated networks. The population-rate models used in
\prettyref{sec:I_poprate_model} and \prettyref{sec:EI_poprate_model}
are sufficient to understand the basic mechanism underlying the
decorrelation. They do, however, not describe how feedback in cortical
networks affects the detailed structure of pairwise correlations. In
\prettyref{sec:corr_ndim}, we therefore compute self-consistent
population averaged correlations for a random network of $N$ linear
excitatory and inhibitory neurons.
By determining the parameters of the linear network analytically from
the LIF model, we show that the predictions of the linear model
are---for a wide and realistic range of parameters---in excellent
agreement with the results of the LIF network model.
In \prettyref{sec:feedback_perturbations}, we make clear that the
active decorrelation in random LIF networks relies on the feedback of
the (sub)population averaged activity but not on the precise
microscopic structure of the feedback signal. In
\prettyref{sec:discussion}, we discuss the consequences of this work
in a broader context and point out limitations and possible extensions
of the presented theory. \prettyref{sec:methods} contains details on
the LIF network model, the derivation of the linear model from the LIF
dynamics and the derivation of population-rate spectra and population
averaged correlations in the framework of the linear model.  It is
meant as a supplement; the basic ideas and the main results can be
extracted from \prettyref{sec:results}.

\section{Results}
\label{sec:results}
In a recurrent neural network of size $N$, each neuron $i\in[1,N]$
receives in general inputs from two different types of sources:
External inputs $\xi_i(t)$ representing the sum of afferents from
other brain areas, and local inputs resulting from the recurrent
connectivity within the network. Depending on their origin, external
inputs $\xi_i$ and $\xi_j$ to different neurons $i$ and $j$ can be
correlated or not.  Throughout this manuscript, we ignore correlations
between these external sources, thereby ensuring that correlations
within the network activity arise from the local connectivity alone
and are not imposed by external inputs \citep{Renart10_587}.  The
local inputs feed the network's spiking activity
$\vec{s}(t)=(s_1(t),\ldots,s_N(t))\transp$ back to the network (we
refer to spike train $s_i(t)$, the $i$th component of the column
vector $\vec{s}(t)$ [the superscript ``$\text{T}$'' denotes the
transpose], as a sum over delta-functions centered at the spike times
$t_i^k$:
\begin{math}
  s_i(t)=\sum_{k}\delta(t-t_i^k)
\end{math};
the abstract quantity `spike train' can be considered as being
derived from the observable quantity `spike count' $n_i^{\Delta
  t}(t)$, the number of spikes occurring in the time interval
$[t,t+\Delta t)$, by taking the limit $\Delta t\to{}0$:
\begin{math}
  s_i(t)=\lim_{\Delta t\to{}0}\frac{1}{\Delta t}n_i^{\Delta t}(t)
\end{math}).
The structure and weighting of this feedback can be described
by the network's connectivity matrix $\mat{J}$ (see
\prettyref{fig:intro}A).  In a finite network, the local connectivity
typically gives rise to overlapping presynaptic populations: in a
random (Erd\H{o}s-R\'{e}nyi) network with connection probability $\epsilon$, for example,
each pair of postsynaptic neurons shares, on average, $\epsilon^2 N$
presynaptic sources. For a network size of, say, $N=10^4$ and a
connection probability $\epsilon=0.1$, this corresponds to a fairly
large number of $100$ identical inputs. For other network structures,
the amount of shared input may be smaller or larger. Due to this
presynaptic overlap, each pair of neurons receives, to some extent,
correlated input (even if the external inputs are uncorrelated).  One
might therefore expect that the network responses
$s_1(t),\ldots,s_N(t)$ are correlated as well. In this article, we
show that, in the presence of negative feedback, the effect of shared
input caused by the structure of the network is compensated by its
recurrent dynamics.

\subsection{Suppression of population-rate fluctuations in LIF networks}
\label{sec:pop_fluct_iaf}
\begin{figure}[ht!]
  \centering
  \doublecolumnfigure{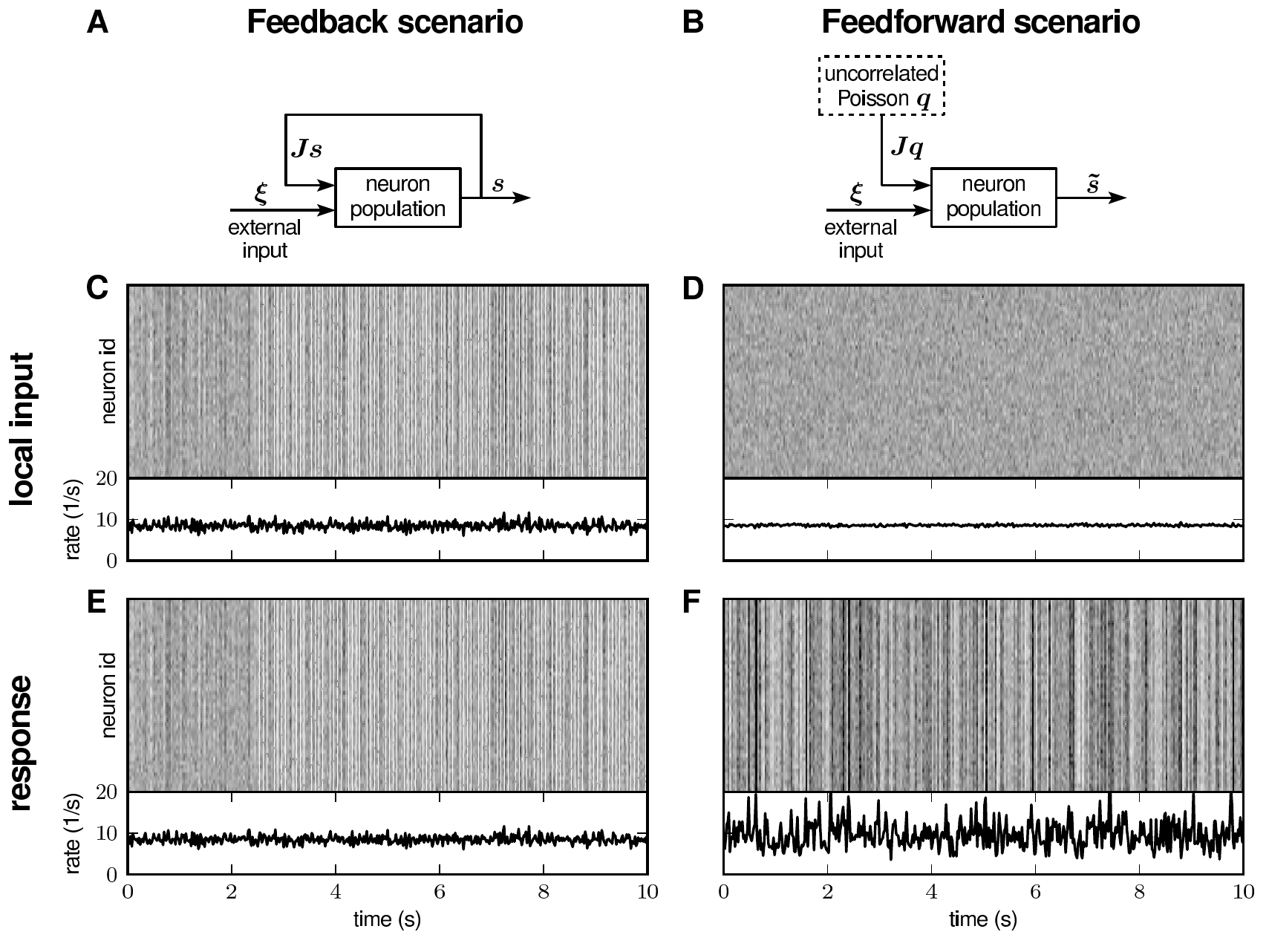}%
  \caption{%
    Spiking activity in excitatory-inhibitory LIF networks with intact
    (left column; \emph{feedback scenario}) and opened feedback loop
    (right column; \emph{feedforward scenario}).  
    {\bf A,B}: Network sketches for the feedback (A) and feedforward
    scenario (B).  {\bf C,D}: Spiking activity (top panels) and
    population averaged firing rate (bottom panels) of the local
    presynaptic populations.  {\bf E,F}: Response spiking activity
    (top panels) and population averaged response rate (bottom
    panels).  
    In the top panels of C--F, each pixel depicts the number of spikes
    (gray coded) of a subpopulation of $250$ neurons in a $10\,\ms$
    time interval.  
    In both the feedback and the feedforward scenario, 
    the neuron population $\{1,\ldots,N\}$
    is driven by the same realization  $\vec{\xi}(t)=(\xi_1(t),\ldots,\xi_N(t))\transp$ of an uncorrelated 
    white-noise ensemble; local input is fed to
    the population through the same connectivity matrix $\mat{J}$.
    The in-degrees, the synaptic
    weights and the shared-input statistics are thus exactly identical
    in the two scenarios.
    In the feedback case (A), local presynaptic spike-trains are provided by the network's response 
    $\vec{s}(t)=(s_1(t),\ldots,s_N(t))\transp$,
    i.e.~the pre- (C) and postsynaptic spike-train ensembles (E) are identical.
    In the feedforward scenario (B), the local presynaptic 
    spike-train population is replaced by an ensemble of $N$ independent realizations 
    $\vec{q}(t)=(q_1(t),\ldots,q_N(t))$ 
    of a Poisson point process (D). Its rate is identical to the time- and population-averaged firing rate 
    in the feedback case. 
    See \prettyref{tab:lif_network_model} and \prettyref{tab:lif_parameters} for details on 
    network models and parameters.
  }
   \label{fig:intro}
\end{figure}

To illustrate the effect of shared input and its suppression by the
recurrent dynamics, we compare the spike response
$\vec{s}(t)=(s_1(t),\ldots,s_N(t))\transp$ of a recurrent random
network (\emph{feedback scenario}; \prettyref{fig:intro}A,C,E) of $N$
LIF neurons to the case where the feedback is cut and replaced by a
spike-train ensemble $\vec{q}(t)=(q_1(t),\ldots,q_N(t))\transp$,
modeled by $N$ independent realizations of a stationary Poisson point
process (\emph{feedforward scenario}; \prettyref{fig:intro}B,D,F). The
rate of this Poisson process is identical to the time and population
averaged firing rate in the intact recurrent system.  In both the
feedback and the feedforward case, the (local) presynaptic spike
trains are fed to the postsynaptic population according to the same
connectivity matrix $\mat{J}$. Therefore, not only the in-degrees and
the synaptic weights but also the shared-input statistics are exactly
identical. 
\par
For realistic size $N$ and connectivity $\epsilon$,
asynchronous states of random neural networks
\citep{Brunel99,Brunel00_183} exhibit spike-train correlations which
are small but not zero \citep[compare raster displays in
\prettyref{fig:intro}C and D; see also][]{Kriener08_2185}.  Although
the presynaptic spike trains are, by construction, independent in the
feedforward case (\prettyref{fig:intro}D), the resulting response
correlations, and, hence, the population-rate fluctuations, are
substantially stronger than those observed in the feedback scenario
(compare \prettyref{fig:intro}F and E). In other words: A theory which
is exclusively based on the amount of shared input but neglects the
details of the presynaptic spike-train statistics can significantly
overestimate correlations and population-rate fluctuations in
recurrent neural networks.
\par
\begin{figure}[ht!]
  \centering
  \doublecolumnfigure{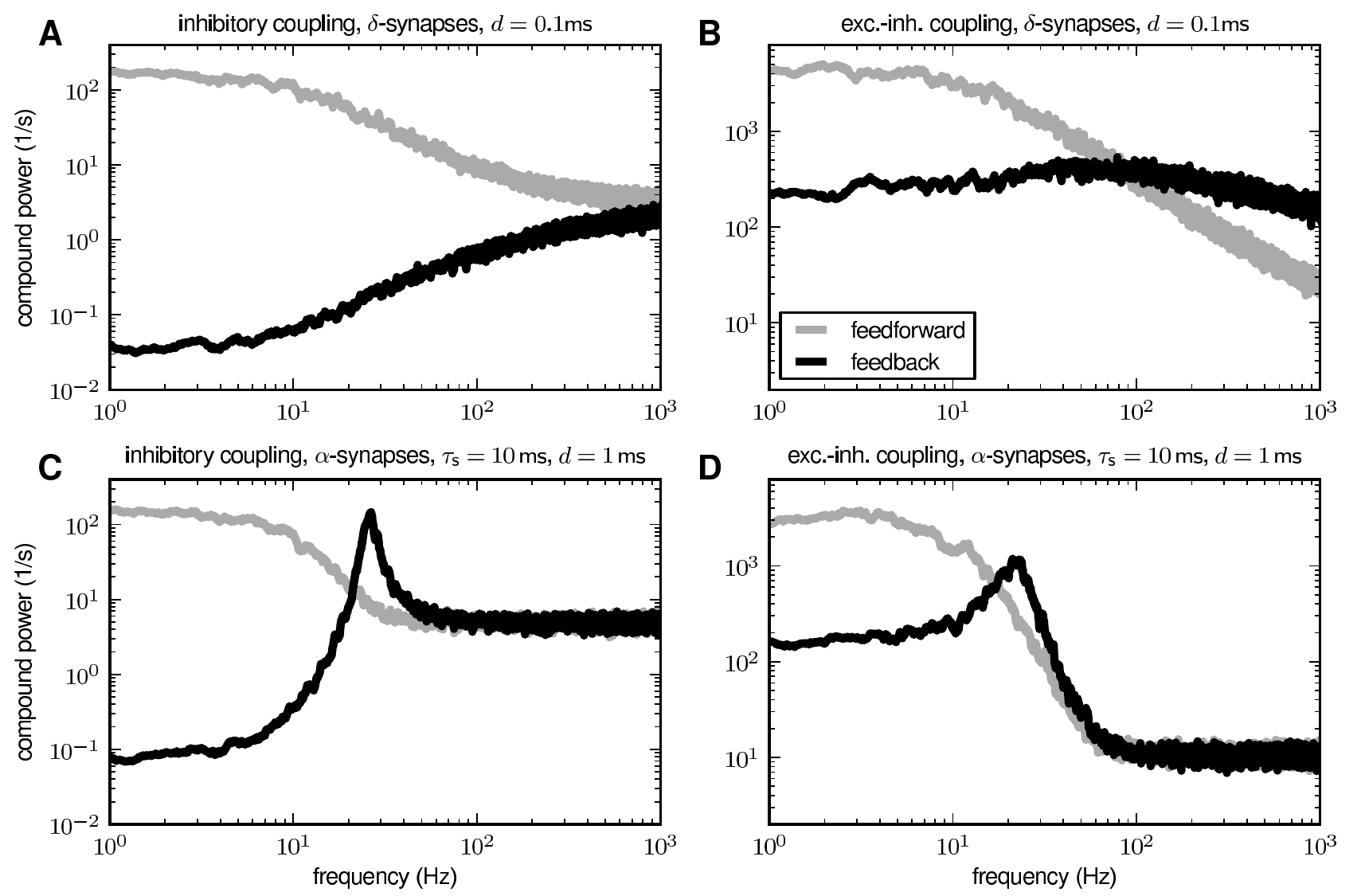}%
  \caption{Suppression of low-frequency fluctuations in recurrent LIF
    networks with purely inhibitory ({\bf A},{\bf C}) and mixed
    excitatory-inhibitory coupling ({\bf B},{\bf D}) for instantaneous
    synapses with delay $d=0.1\,\ms$ ({\bf A},{\bf B}) and low-pass
    synapses with $d=1\,\ms$ ({\bf C},{\bf D}). Power-spectra
    $N{}C_{SS}$ of population rates $s(t)$ for the feedback
    (black) and the feedforward case (gray;
    cf.~\prettyref{fig:intro}).  See \prettyref{tab:lif_network_model}
    and \prettyref{tab:lif_parameters} for details on network models
    and parameters.  In C and D, local synaptic inputs are modeled as
    currents $I_i(t)=\sum_j J_{ij} \sum_l \psc(t-t^j_l-d)$ with
    $\alpha$-function shaped kernel $\psc(t)=e t
    \tau_\syn^{-1}\exp(-t/\tau_\syn) \Theta(t)$ with time constant
    $\tau_\syn=10\,\ms$ ($\Theta(\cdot)$ denotes Heaviside
    function). (Excitatory) Synaptic weights are set to $J=1\,\pA$
    (see \prettyref{tab:lif_network_model} for details).  Simulation
    time $T=100\,\seconds$. Single-trial spectra smoothed by moving
    average (frame size $1\,\Hz$).}
  \label{fig:FB_vs_FF_spectra}
\end{figure}

The same effect can be observed in LIF networks with both purely
inhibitory and mixed excitatory-inhibitory coupling
(\prettyref{fig:FB_vs_FF_spectra}). To demonstrate this
quantitatively, we focus on the fluctuations of the population
averaged activity $s(t) =N^{-1} \sum_{i=1}^N s_i(t)$.  Its
power-spectrum (or auto-correlation, in the time domain)
\begin{eqnarray}
    \label{eq:spectrum_avg_activity}
  \begin{aligned}
    C_{SS}(\omega)&=|S(\omega)|^2 = |\Fourier{s(t)}{\omega}|^2
    =\frac{1}{N^2}\left[\sum_{i=1}^N A_i(\omega) + \sum_{i=1,j\ne{}i}^N C_{ij}(\omega)\right]
  \end{aligned}
\end{eqnarray}
is determined both by the power-spectra (auto-correlations)
$A_i(\omega)=|S_i(\omega)|^2$ of the individual spike trains and the
cross-spectra (cross-correlations)
$C_{ij}(\omega)=S_i(\omega)S_j(\omega)\cc$ ($i\ne{}j$) of pairs of spike trains
(throughout the article, we use capital letters to represent
quantities in frequency [Fourier] space;
\mbox{$S_k(\omega)=\Fourier{s_k(t)}{\omega}=\Int{s_k(t)e^{-i\omega{}t}}{t}{}{}$}
represents the Fourier transform of the spike train $s_k(t)$).  
We observe that the spike-train power-spectra $A_i(\omega)$ (and
auto-correlations) are barely distinguishable in the feedback and in
the feedforward case (not shown here; the main features of the
spike-train auto-correlation are determined by the average
single-neuron firing rate and the refractory mechanism; both are
identical in the feedback and the feedforward scenario).
The differences in the population-rate spectra $C_{SS}(\omega)$ are
therefore essentially due to differences in the spike-train
cross-spectra $C_{ij}(\omega)$. In other words, the fluctuations in
the population activity serve as a measure of pairwise spike-train
correlations \citep{Harris11_509}: small (large) population averaged
spike-train correlations are accompanied by small (large) fluctuations
in the population rate (see lower panels in
\prettyref{fig:intro}C--F). The power-spectra $C_{SS}(\omega)$ of the
population averaged activity reveal a feedback-induced suppression of
the population-rate variance at low frequencies up to several tens of
Hertz.  For the examples shown in \prettyref{fig:FB_vs_FF_spectra},
this suppression spans more than three orders of magnitude for the
inhibitory and more than one order of magnitude for the
excitatory-inhibitory network.
\par
The suppression of low-frequency fluctuations does not critically
depend on the details of the network model. As shown in
\prettyref{fig:FB_vs_FF_spectra}, it can, for example, be observed for
both networks with zero rise-time synapses ($\delta$-shaped synaptic
currents) and short delays and for networks with delayed low-pass
filtering synapses ($\alpha$-shaped synaptic currents).  In the latter
case, the suppression of fluctuations is slightly more restricted to
lower frequencies ($<10\,\Hz$). Here, the fluctuation suppression is
however similarly pronounced as in networks with instantaneous
synapses.
\par
In \prettyref{fig:FB_vs_FF_spectra}C,D, the power-spectra of the
population activity converge to the mean firing rate at high
frequencies. This indicates that the spike trains are uncorrelated on
short time scales.  For instantaneous $\delta$-synapses, neurons
exhibit an immediate response to excitatory input spikes
\citep{Helias10_1000929,Richardson10_178102}. This fast response
causes spike-train correlations on short time scales. Hence, the
compound power at high frequencies is increased. In a recurrent
system, this effect is amplified by reverberating simultaneous
excitatory spikes. Therefore, the high-frequency power of the compound
activity is larger in the feedback case
(\prettyref{fig:FB_vs_FF_spectra}B). Note that this high-frequency
effect is absent in networks with more realistic low-pass filtering
synapses (\prettyref{fig:FB_vs_FF_spectra}C,D) and in purely
inhibitory networks (\prettyref{fig:FB_vs_FF_spectra}A).
\par
Synaptic delays and slow synapses can promote oscillatory modes in
certain frequency bands \citep{Brunel99, Brunel00_183}, thereby
leading to peaks in the population-rate spectra in the feedback
scenario which exceed the power in the feedforward case (see peaks at
$\sim 25\,\Hz$ in \prettyref{fig:FB_vs_FF_spectra}C,D). Note that, in
the feedforward case, the local input was replaced by a stationary
Poisson process, whereas in the recurrent network (feedback case) the
presynaptic spike trains exhibit oscillatory modes. By replacing the
feedback by an inhomogeneous Poisson process with a time dependent
intensity which is identical to the population rate in the recurrent
network, we found that these oscillatory modes are neither suppressed
nor amplified by the recurrent dynamics, i.e.~the peaks in the
resulting power-spectra have the same amplitude in the feedback and in
the feedforward case (data not shown here). At low frequencies,
however, the results are identical to those obtained by replacing the
feedback by a homogeneous Poisson process (i.e.~to those shown in
\prettyref{fig:FB_vs_FF_spectra}; see
\prettyref{sec:feedback_perturbations}). In the present study, we
mainly focus on these low-frequency effects.
\par
The observation that the suppression of low-frequency fluctuations is
particularly pronounced in networks with purely inhibitory coupling
indicates that inhibitory feedback may play a key role for the
underlying mechanism. In the following subsection, we will demonstrate
by means of a one-dimensional linear population model that,
indeed, negative feedback alone leads to an efficient fluctuation
suppression.

\subsection{Suppression of population-activity fluctuations by negative feedback}
\label{sec:I_poprate_model}
Average pairwise correlations can be extracted from the spectrum
\prettyref{eq:spectrum_avg_activity} of the compound activity,
provided the single spike-train statistics (auto-correlations) is
known (see previous section). As the single spike-train statistics is
identical in the feedback and in the feedforward scenario, the
mechanism underlying the decorrelation in recurrent networks can be
understood by studying the dynamics of the population averaged
activity.
In this and in the next subsection (\prettyref{sec:EI_poprate_model}),
we will consider the linearized dynamics of random networks composed
of homogeneous subpopulations of LIF neurons. The high-dimensional
dynamics of such systems can be reduced to low-dimensional models
describing the dynamics of the compound activity (for details, see
\prettyref{sec:linear_model}). Note that this reduction is exact for
networks with homogeneous out-degree (number of outgoing
connections). For the networks studied here (random networks with
homogeneous in-degree), it serves as a sufficient approximation (in a
network of size $N$ where each connection is randomly and
independently realized with probability $\epsilon$
[Erd\H{o}s-R\'{e}nyi graph], the [binomial] in- and out-degree
distributions become very sharp for large $N$ [relative to the mean
in/out-degree]; both in- and out-degree are therefore approximately
constant across the population of neurons).
In this subsection, we
will first study networks with purely inhibitory coupling. In
\prettyref{sec:EI_poprate_model}, we will investigate the effect of
mixed excitatory-inhibitory connectivity.
\par
\begin{figure}[ht!]
  \centering
  \doublecolumnfigure{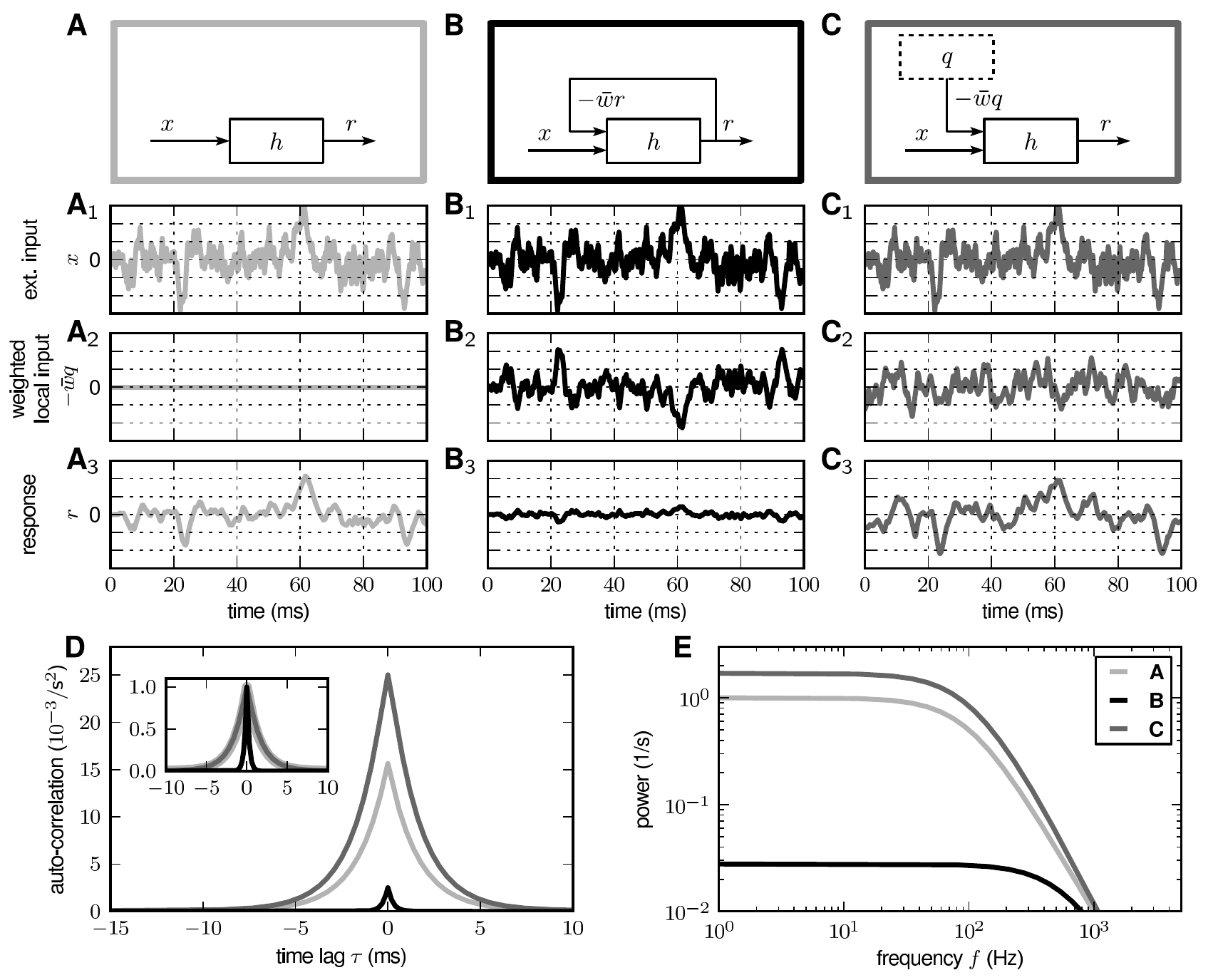}%
  \caption{Partial canceling of fluctuations in a linear system by inhibitory feedback.
    Response $r(t)$ of a linear system with impulse response $h(t)$ 
    (1st-order low-pass, cutoff frequency $100\,\Hz$) to Gaussian white noise input $x(t)$ with amplitude $\bar{\rho}=1$
    for three local-input scenarios.
    {\bf A} (light gray): No feedback (local input $q(t)=0$). 
    {\bf B} (black): Negative feedback ($q(t)=r(t)$) with strength $\w=5$.  
    The fluctuations of the weighted
    local input $-\w{}q(t)$ (B$_\mathsf{2}$) are anticorrelated to
    the external drive $x(t)$ (B$_\mathsf{1}$).
    {\bf C} (dark gray): Feedback in B is replaced by uncorrelated feedforward input $q(t)$ 
    with the same auto-statistics as the response $r(t)$ in B$_\mathsf{3}$.
    The local input $q(t)=\InvFourier{|R(\omega)|e^{i\xi(\omega)}}{t}$ is constructed by 
    assigning a random phase $\xi(\omega)$ to each Fourier component $R(\omega)=\Fourier{r(t)}{\omega}$ of the 
    response in B$_\mathsf{3}$. 
    Fluctuations in C$_\mathsf{2}$ and C$_\mathsf{1}$ are uncorrelated.
    {\bf A},{\bf B},{\bf C}: Network sketches.
    {\bf A$_\mathsf{1}$},{\bf B$_\mathsf{1}$},{\bf C$_\mathsf{1}$}: External input $x(t)$.
    {\bf A$_\mathsf{2}$},{\bf B$_\mathsf{2}$},{\bf C$_\mathsf{2}$}: Weighted local input $-\w q(t)$.
    {\bf A$_\mathsf{3}$},{\bf B$_\mathsf{3}$},{\bf C$_\mathsf{3}$}: Responses $r(t)$.
    {\bf D},{\bf E}:  Response auto-correlation functions 
    (D) and 
    power-spectra 
    (E) for the three cases shown in A,B,C
    (same gray coding as in A,B,C; inset in D: normalized auto-correlations). 
  } 
  \label{fig:intuition}
\end{figure}

Consider a random network of $N$ identical neurons with connection
probability $\epsilon$. Each neuron $i=1,\ldots,N$ receives
$K=\epsilon{}N$ randomly chosen inputs from the local network with
synaptic weights $-J$. In addition, the neurons are driven by external
uncorrelated Gaussian white noise $\xi_i(t)$ with amplitude $\eta$,
i.e.~\mbox{$\EW[t]{\xi_i(t)}=0$} and
\mbox{$\EW[t]{\xi_i(t)\xi_j(t+\tau)}=\delta_{ij} \eta^2\delta(\tau)$}.
For small input fluctuations, the network dynamics can be
linearized. This linearization is based on the averaged response of a
single neuron to an incoming spike and describes the activity of an
individual neuron $i$ by an abstract fluctuating quantity $r_i(t)$
which is defined such that within the linear approximation its auto-
and cross-correlations fulfill the same linearized equation as the
spiking model in the low-frequency limit. Consequently, also the
low-frequency fluctuations of the population spike rate are captured
correctly by the reduced model up to linear order. This approach is
equivalent to the treatment of finite-size fluctuations in spiking
networks \citep[see, e.g.]{Brunel99}. For details see
\prettyref{sec:linear_model}.  For large $N$, the population averaged
activity \mbox{$r(t)=\EW[i]{r_i(t)}=N^{-1}\sum_{i=1}^N r_i(t)$} can
hence be described by a one-dimensional linear system
\begin{equation}
  \label{eq:linear_model_1D}
  r(t)=([-\w r + x]*h)(t)
\end{equation}
with linear kernel $h(t)$, effective coupling strength $\w=Kw$
and the population averaged noise $x(t)=\EW[i]{x_i(t)}$ (see
\prettyref{sec:linear_model} and \prettyref{fig:intuition}B). The
coupling strength $\w$ represents the integrated linear response of
the neuron population to a small perturbation in the input rate of a
single presynaptic neuron. For a population of LIF neurons, its
relation to the synaptic weight $J$ (PSP amplitude) is derived in
\prettyref{sec:linear_model} and \prettyref{sec:w_of_j}. The
normalized kernel $h(t)$ (with $\int_0^\infty \dt\,h(t)=1$) captures
the time course of the linear response. It is determined by the
single-neuron properties (e.g.~the spike-initiation dynamics
\citep{Fourcaud03_11640,Naundorf05_297}), the properties of the
synapses (e.g.~synaptic weights and time constants
\citep{Brunel01_2186,Nordlie10}) and the properties of the input
(e.g.~excitatory vs.~inhibitory input \citep{Pressley09_63}). For many
real and model neurons, the linear population-rate response exhibits
low-pass characteristics
\citep{Knight72b,Koendgen08_2086,Boucsein09_1006,Blomquist09,Knight72a,Gerstner00,Brunel01_2186,Lindner01_2934,Fourcaud02,Fourcaud03_11640,Naundorf05_297,Pressley09_63,Nordlie10,Richardson10_178102}.
For illustration (\prettyref{fig:intuition}), we consider a 1st-order
low-pass filter, i.e.~an exponential impulse response
$h(t)=\tau^{-1}\exp(-t/\tau)\Theta(t)$ with time constant $\tau$
(cutoff frequency $f_\text{c}=(2\pi\tau)^{-1}$; see
\prettyref{fig:intuition}A, light gray curve in E). The results of our
analysis are however independent of the choice of the kernel $h(t)$.
The auto-correlation \mbox{$\EW[t]{x(t)x(t+\tau)} = \bar{\rho}^2\delta(\tau)$}
of the external noise is parametrized by the effective noise amplitude $\bar{\rho}=\rho/\sqrt{N}$.
\par
Given the simplified description \prettyref{eq:linear_model_1D}, the
suppression of response fluctuations by negative feedback can be
understood intuitively: Consider first the case where the neurons in
the local network are unconnected (\prettyref{fig:intuition}A; no
feedback, $\w=0$). Here, the response $r(t)$
(\prettyref{fig:intuition}A$_\mathsf{3}$) is simply a low-pass
filtered version of the external input $x(t)$
(\prettyref{fig:intuition}A$_\mathsf{1}$), resulting in an
exponentially decaying response auto-correlation
(\prettyref{fig:intuition}D; light gray curve) and a drop in the
response power-spectrum at the cutoff frequency $f_\text{c}$
(\prettyref{fig:intuition}E). At low frequencies, $r(t)$ and $x(t)$
are in phase; they are correlated.  In the presence of negative
feedback (\prettyref{fig:intuition}B), the local input $-\w r(t)$
(\prettyref{fig:intuition}B$_\mathsf{2}$) and the low-frequency
components of the external input $x(t)$
(\prettyref{fig:intuition}B$_\mathsf{1}$) are anticorrelated. They
partly cancel out, thereby reducing the response fluctuations $r(t)$
(\prettyref{fig:intuition}B$_\mathsf{3}$). The auto-correlation
function and the power-spectrum are suppressed
(\prettyref{fig:intuition}D,E; black curves). Due to the low-pass
characteristics of the system, mainly the low-frequency components of
the external drive $x(t)$ are transferred to the output side and, in
turn, become available for the feedback signal. Therefore, the
canceling of input fluctuations and the resulting suppression of
response fluctuations are most efficient at low
frequencies. Consequently, the auto-correlation function is sharpened
(see inset in \prettyref{fig:intuition}D). The cutoff frequency of the
system is increased (\prettyref{fig:intuition}E; black curve). This
effect of negative feedback is very general and well known in the
engineering literature. It is employed in the design of technical
devices, like, e.g., amplifiers \citep{Oppenheim96}.  As the
zero-frequency power is identical to the integrated auto-correlation
function, the suppression of low-frequency fluctuations is accompanied
by a reduction in the auto-correlation area
(\prettyref{fig:intuition}D; black curve).  Note that the suppression
of fluctuations in the feedback case is not merely a result of the
additional inhibitory noise source provided by the local input, but
follows from the precise temporal alignment of the local and the
external input. To illustrate this, let's consider the case where the
feedback channel is replaced by a feedforward input $q(t)$
(\prettyref{fig:intuition}C) which has the same auto-statistics as the
response $r(t)$ in the feedback case
(\prettyref{fig:intuition}B$_\mathsf{3}$) but is uncorrelated to the
external drive $x(t)$. In this case, external input fluctuations
(\prettyref{fig:intuition}C$_\mathsf{1}$) are not canceled by the
local input $-\w{}q(t)$
(\prettyref{fig:intuition}C$_\mathsf{2}$). Instead, the local
feedforward input acts as an additional noise source which leads to an
increase in the response fluctuations
(\prettyref{fig:intuition}C$_\mathsf{3}$). The response
auto-correlation and power-spectrum (\prettyref{fig:intuition}D,E;
dark gray curves) are increased. Compared to the unconnected case
(\prettyref{fig:intuition}E; light gray curve), the cutoff frequency
remains unchanged.
\par
\begin{figure}[ht!]
  \centering
  \doublecolumnfigure{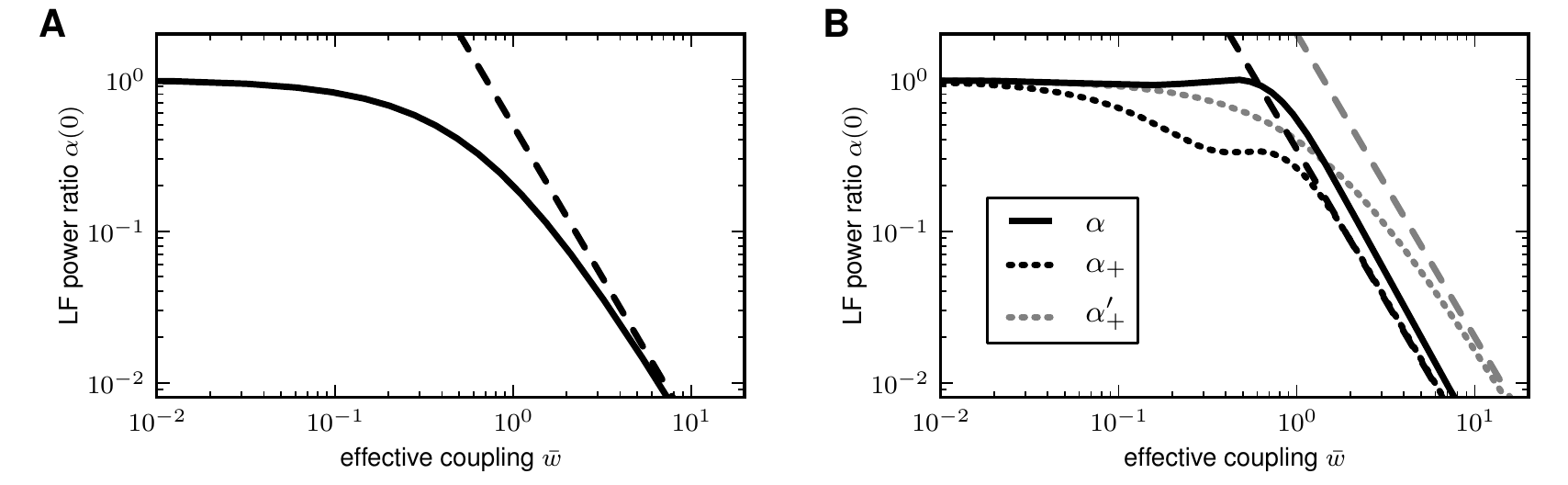}%
  \caption{
    Suppression of low-frequency (LF) population-rate fluctuations in linearized homogeneous random networks 
    with purely inhibitory ({\bf A}) and mixed excitatory-inhibitory coupling ({\bf B}).
    Dependence of the zero-frequency power ratio $\alpha(0)$ on the effective coupling strength $\w$     
    (solid curves: full solutions; dashed lines: strong-coupling approximations).
    The power ratio $\alpha(0)$ represents the ratio between the low-frequency population-rate power in 
    the recurrent networks ({\bf A}: \prettyref{fig:intuition}B; {\bf B}: \prettyref{fig:FB_vs_FF_schur}A,B) and 
    in networks where the feedback channels are replaced by uncorrelated feedforward input 
    ({\bf A}: \prettyref{fig:intuition}C;
    {\bf B}, black: \prettyref{fig:FB_vs_FF_schur}C,D;
    {\bf B}, gray: \prettyref{fig:FB_vs_FF_schur}D').
    Dotted curves in {\bf B} depict power ratio of the sum modes $r_+$ and $\tilde{r}_+$ (see text).
    B: Balance factor $\g=1.5$.
  }
  \label{fig:FB_vs_FF_lin_model}
\end{figure}

The feedback induced suppression of response fluctuations can be
quantified by comparing the response power-spectra
\begin{equation}
  \label{eq:spectr_1D_FB}
  C_{RR}(\omega) = \EW[x]{|R(\omega)|^2} = \frac{\bar{\rho}^2|H(\omega)|^2}{|1+\w{}H(\omega)|^2}
\end{equation}
and
\begin{equation}
  \label{eq:spectr_1D_FF}
  C_{\tilde{R}\tilde{R}}(\omega)= \EW[x]{ |\tilde{R}(\omega)|^2 } = |H(\omega)|^2(\w^2C_{RR}(\omega)+\bar{\rho}^2)
\end{equation}
in the feedback (\prettyref{fig:intuition}B) and the feedforward case
(\prettyref{fig:intuition}C), respectively (see
\prettyref{sec:spectr_1D}).  Here, $R(\omega)$ and $\tilde{R}(\omega)$
denote the Fourier transforms of the response fluctuations in the feedback and
the feedforward scenario, respectively, $H(\omega)$ the transfer
function (Fourier transform of the filter kernel $h(t)$) of the neuron
population, and $\EW[x]{}$ the average across noise realizations.  We
use the power ratio
\begin{eqnarray}
  \label{eq:power_ratio_1D}
  \alpha(\omega) &=& \frac{C_{RR}(\omega)}{C_{\tilde{R}\tilde{R}}(\omega)}=  
  \frac{1}{\w^2|H(\omega)|^2+|1+\w{}H(\omega)|^2}
\end{eqnarray}
as a measure of the relative fluctuation suppression caused by
feedback. For low frequencies ($\omega\to 0$) and strong effective
coupling $|\w|=|Kw|\gg{}1$, the power ratio
\prettyref{eq:power_ratio_1D} decays as $\w^{-2}$ (see
\prettyref{fig:FB_vs_FF_lin_model}A): the suppression of
population-rate fluctuations is promoted by strong negative feedback.
In line with the observations in \prettyref{sec:pop_fluct_iaf}, this
suppression is restricted to low frequencies; for high frequencies
($\omega\to\infty$, i.e.~$H(\omega)\to{}0$), the power ratio
$\alpha(\omega)$ approaches $1$. Note that the power ratio
\prettyref{eq:power_ratio_1D} is independent of the amplitude
$\bar{\rho}$ of the population averaged external input
$x(t)$. Therefore, even if we dropped the assumption of the external
inputs $x_i(t)$ being uncorrelated, i.e.~if
$\EW[t]{x_i(t)x_j(t+\tau)}\ne{}0$ for $i\ne{}j$, the power ratio
\prettyref{eq:power_ratio_1D} remained the same. For correlated
external input, the power $\bar{\rho}$ of the population average
$x(t)$ is different from $\rho/\sqrt{N}$. The suppression factor
$\alpha(\omega)$, however, is not affected by this. Moreover, it is
straightforward to show that the power ratio
\prettyref{eq:power_ratio_1D} is, in fact, independent of the shape of
the external-noise spectrum
$C_{XX}(\omega)=\EW[x]{|X(\omega)|^2}$. The same result
\prettyref{eq:power_ratio_1D} is obtained for any type of external
input (e.g.~colored noise or oscillating inputs).
%
\par
For low frequencies, the transfer function $H(\omega)$ approaches
unity ($\lim_{\omega\to 0} H(\omega)=1$); the exact shape of the
kernel $h(t)$ becomes irrelevant.  In particular, the cutoff
frequency (or time constant) of a low-pass kernel has no effect on the
zero-frequency power (integral correlation) and the zero-frequency
power ratio $\alpha(0)$ (\prettyref{fig:FB_vs_FF_lin_model}).
Therefore, the suppression of low-frequency fluctuations does not
critically depend on the exact choice of the neuron, synapse or input
model.  The same reasoning applies to synaptic delays: Replacing the
kernel $h(t)$ by a delayed kernel $h(t-d)$ leads to an additional
phase factor $e^{-i\omega d}$ in the transfer function
$H(\omega)$. For sufficiently small frequencies (long time scales),
this factor can be neglected ($\lim_{\omega\to 0}e^{-i\omega d}=1$).
\par
For networks with purely inhibitory feedback, the absolute power
\prettyref{eq:spectr_1D_FB} of the population rate decreases
monotonously with increasing coupling strength $\w$. As we will
demonstrate in \prettyref{sec:EI_poprate_model} and
\prettyref{sec:corr_ndim}, this is qualitatively different in networks
with mixed excitatory and inhibitory coupling $\w_\Ex=\w>0$ and
$\w_\In=-\g\w<0$, respectively: here, the fluctuations of the compound
activity increase with $\w$. The power ratio $\alpha(\omega)$, however,
still decreases with $\w$. 

\subsection{Population-activity fluctuations in excitatory-inhibitory networks}
\label{sec:EI_poprate_model}
In the foregoing subsection, we have shown that negative feedback
alone can efficiently suppress population-rate fluctuations and,
hence, spike-train correlations. So far, it is unclear whether the
same reasoning applies to networks with mixed excitatory and
inhibitory coupling. To clarify this, we now consider a random network
composed of a homogeneous excitatory and inhibitory subpopulation
$\Exc$ and $\Inh$ of size $N_\Ex=|\Exc|$ and $N_\In=|\Inh|=\gamma
N_\Ex$, respectively. Each neuron receives $K=\epsilon N_\Ex$
excitatory and $\gamma K=\epsilon N_\In$ inhibitory inputs from $\Exc$
and $\Inh$ with synaptic weights $w>0$ and $-gw<0$, respectively. In
addition, the neurons are driven by external Gaussian white noise. As
demonstrated in \prettyref{sec:linear_model}, linearization and
averaging across subpopulations leads to a two-dimensional system
\begin{equation}
  \label{eq:linearized_dynamics_2D}
  \vec{r}(t)=([\mat{W}\vec{r}+\vec{x}]*h)(t)
\end{equation}
describing the linearized dynamics of the subpopulation averaged
activity $\vec{r}(t)=(r_\Ex(t),r_\In(t))\transp$.  Here,
$\vec{x}(t)=(x_\Ex(t),x_\In(t))\transp$ denotes the
subpopulation averaged external uncorrelated white-noise input with
correlation functions
$\EW[x,t]{x_p(t),x_q(t+\tau)}=\bar{\rho}_p^2\delta_{pq}\delta(\tau)$
($\bar{\rho}_p=\rho/\sqrt{N_p}$, $p,q\in\{\Ex,\In\}$), and $h(t)$ a normalized linear kernel with
$\int_0^\infty \dt\, h(t)=1$. The excitatory and inhibitory
subpopulations are coupled through an effective connectivity matrix
\begin{equation}
  \label{eq:coupling_matrix_2D}
  \mat{W} = \w
  \begin{pmatrix}
  1 & -\g\\
  1 & -\g\\    
  \end{pmatrix}
\end{equation}
with effective weight $\w=Kw>0$ and balance parameter $\g=\gamma g>0$.
\par
\begin{figure}[ht!]
  \centering
    \oneandahalfcolumnfigure{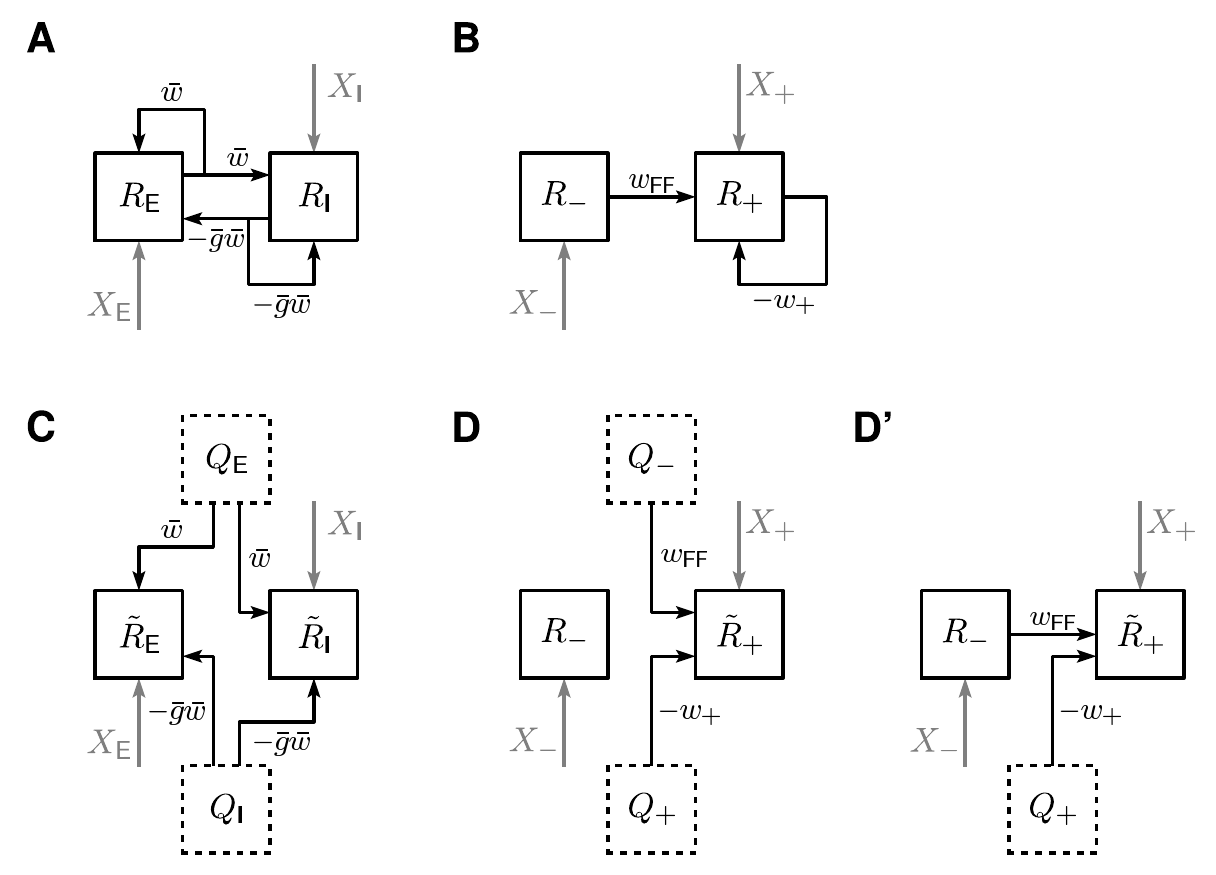}
    \caption{
      Sketch of the 2D (excitatory-inhibitory) model for the feedback
      (A,B) and the feedforward scenario (C,D) in normal (A,C) and
      Schur-basis representation (B,D). 
      {\bf A}: Original 2D recurrent system.
      {\bf B}: Schur-basis representation of the system shown in A.
      {\bf C}: Feedforward scenario: Excitatory and inhibitory feedback connections 
      of the original network (A) are replaced by feedforward input from 
      populations with rates $Q_\Ex$, $Q_\In$, respectively.
      {\bf D}: Schur-basis representation of the system shown in C.
      {\bf D'}: Alternative feedforward scenario: Here, the feedforward
      channel (weight $w_\FF$) of the original system in Schur basis
      (B) remains intact.  Only the inhibitory feedback (weight
      $-w_+$) is replaced by feedforward input $Q_+$.}
   \label{fig:FB_vs_FF_schur}
\end{figure}

The two-dimensional system
\prettyref{eq:linearized_dynamics_2D}/\prettyref{eq:coupling_matrix_2D}
represents a recurrent system with both positive and negative feedback
connections (\prettyref{fig:FB_vs_FF_schur}A). By introducing new
coordinates
\begin{equation}
  \label{eq:schur_basis}
  r_+(t)=(r_\Ex(t)+r_\In(t))/\sqrt{2}
  \,,\quad
  r_-(t)=(r_\Ex(t)-r_\In(t))/\sqrt{2}
\end{equation}
and $x_+(t)=(x_\Ex(t)+x_\In(t))/\sqrt{2}$, $x_-(t)=(x_\Ex(t)-x_\In(t))/\sqrt{2}$,
we obtain an equivalent representation of
\prettyref{eq:linearized_dynamics_2D}/\prettyref{eq:coupling_matrix_2D},
\begin{equation}
  \label{eq:schur_dynamics}
   \begin{pmatrix} 
     r_+(t)\\ r_-(t) 
   \end{pmatrix}
   = \left(\left[\mat{S} 
       \begin{pmatrix} 
         r_+\\ r_- 
       \end{pmatrix} +  
       \begin{pmatrix} 
         x_+\\ x_- 
       \end{pmatrix} 
     \right]*h \right)(t)
   \,,
\end{equation}
describing the dynamics of the sum and difference activity
$r_+(t)$ and $r_-(t)$, respectively, i.e.~the in- and anti-phase
components of the excitatory and inhibitory subpopulations
\citep[see][]{Troyer02_2741,Zhaoping04_198106,Murphy09_635}. The new
coupling matrix
\begin{equation}
  \label{eq:schur_coupling}
  \mat{S}=
  \begin{pmatrix}
    -w_+ & w_\FF\\ 0 & 0
  \end{pmatrix}
\end{equation}
reveals that the sum mode $r_+(t)$ is subject to self-feedback
($S_{11}=-w_+=\w(1-\g)$) and receives feedforward input from the
difference mode $r_-(t)$ ($S_{12}=w_\FF=\w(1+\g)$). All remaining
connections are absent ($S_{21}=S_{22}=0$) in the new representation
\prettyref{eq:schur_basis} (see \prettyref{fig:FB_vs_FF_schur}B). The
correlation functions of the external noise in the new coordinates are
given by $\EW[x,t]{x_p(t),x_q(t+\tau)}=\bar{\rho}_{pq}^2\delta(\tau)$
with $\bar{\rho}_{pq}=\rho^2/N_\Ex \left(\gamma^{-1}\delta_{pq}
  +(1-\gamma^{-1})/2\right)$ ($p,q\in\{+,-\}$).
\par
The feedforward coupling is positive
($w_\FF>0$): an excitation surplus ($r_-(t)>0$) will excite all
neurons in the network, an excitation deficit ($r_-(t)<0$) will lead
to global inhibition. In inhibition dominated regimes with $\g=\gamma
g>1$, the self-feedback of the sum activity $r_+(t)$ is
effectively negative ($-w_+<0$). The dynamics of the sum rate in
inhibition-dominated excitatory-inhibitory networks is therefore
qualitatively similar to the dynamics in purely inhibitory networks
(\prettyref{sec:I_poprate_model}). As shown below, the negative
feedback loop exposed by the transform \prettyref{eq:schur_basis}
leads to an efficient relative suppression of population-rate
fluctuations (if compared to the feedforward case).
\par
Mathematically, the coordinate transform \prettyref{eq:schur_basis}
corresponds to a \emph{Schur decomposition} of the dynamics: Any
recurrent system of type \prettyref{eq:linearized_dynamics_2D} (with
arbitrary coupling matrix $\mat{W}$) can be transformed to a system
with a triangular coupling matrix \citep[see e.g.][]{Murphy09_635}.
The resulting coupling between the different Schur modes can be
ordered so that there are only connections from modes with lower index
to modes with the same or larger index. In this sense, the resulting
system has been termed 'feedforward' \citep{Murphy09_635}.  The
original coupling matrix $\mat{W}$ is typically not normal,
i.e.~$\mat{W}\transp \mat{W} \neq \mat{W}\mat{W}\transp$. Its
eigenvectors do not form an orthogonal basis. By performing a
Gram-Schmidt orthonormalization of the eigenvectors, however, one can
obtain a (normalized) orthogonal basis, a Schur basis.  Our new
coordinates \prettyref{eq:schur_basis} correspond to the amplitudes
(the time evolution) of two orthogonal Schur modes.
\par
The spectra $C_{R_\Ex R_\Ex}(\omega)$, $C_{R_\In R_\In}(\omega)$,
$C_{R_\Ex R_\In}(\omega)$ and $C_{R_+ R_+}(\omega)$ of the
subpopulation averaged rates $r_\Ex$, $r_\In$ and the sum mode $r_+$,
respectively, are derived in \prettyref{sec:spectr_2D}.  In contrast
to the purely inhibitory network (see
\prettyref{sec:I_poprate_model}), the population-rate fluctuations of
the excitatory-inhibitory network increase monotonously with
increasing coupling strength $\w$. For strong coupling, $C_{R_+
  R_+}(\omega)$ approaches
\begin{equation}
  \label{eq:spectr_2D_FB_strong_coupling}
  \lim_{\w\to\infty}C_{R_+ R_+}(\omega)
  =|H(\omega)|^2\rho^2 \frac{1+\gamma^{-1}}{2N_\Ex} \frac{w_\FF^2}{w_+^2} 
\end{equation}
from below with $w_\FF/w_+=(\g+1)/(\g-1)$. Close to the critical point
($\g \simeq 1$), the rate fluctuations become very large;
\prettyref{eq:spectr_2D_FB_strong_coupling} diverges. Increasing the
amount of inhibition by increasing $\g$, however, leads to a
suppression of these fluctuations. In the limit $\g\to\infty$, $C_{R_+
  R_+}(\omega)$ and \prettyref{eq:spectr_2D_FB_strong_coupling}
approach the spectrum \mbox{$\lim_{\w\to 0}C_{R_+
    R_+}=|H|^2\rho^2(1+\gamma^{-1})/(2N_\Ex)$} of the unconnected
network.
For strong coupling ($\w\gg 1$), the ratio $C_{R_\Ex
  R_\Ex}(\omega)/C_{R_\In R_\In}(\omega)$ approaches $\g^2$: the
fluctuations of the population averaged excitatory firing rate exceed
those of the inhibitory population by a factor $\g^2$ (independently
of $H(\omega)$ and $\omega$).
\par
Similarly to the strategy we followed in the previous subsections, we
will now compare the population-rate fluctuations of the feedback
system \prettyref{eq:linearized_dynamics_2D}, or equivalently
\prettyref{eq:schur_dynamics}, to the case where the feedback channels
are replaced by feedforward input with identical auto-statistics. A
straight-forward implementation of this is illustrated in
\prettyref{fig:FB_vs_FF_schur}C: Here, the excitatory and inhibitory
feedback channels $R_\Ex$ and $R_\In$ are replaced by uncorrelated
feedforward inputs $Q_\Ex$ and $Q_\In$, respectively.  The Schur
representation of this scenario is depicted in
\prettyref{fig:FB_vs_FF_schur}D. According to
\prettyref{eq:linearized_dynamics_2D}, the Fourier transforms of the
response fluctuations of this system read
\begin{equation}
  \label{eq:linearized_dynamics_2D_FF}
  \begin{pmatrix}\tilde{R}_\Ex(\omega)\\ \tilde{R}_\In(\omega)\end{pmatrix}=
  H(\omega)\left[
  \mat{W}
  \begin{pmatrix} Q_\Ex(\omega)\\ Q_\In(\omega)\end{pmatrix}+
  \begin{pmatrix} X_\Ex(\omega)\\ X_\In(\omega)\end{pmatrix}
  \right]
  \,.
\end{equation}
With $\tilde{R}_+=(\tilde{R}_\Ex+\tilde{R}_\In)/\sqrt{2}$, and
using $C_{Q_\Ex Q_\Ex}=C_{R_\Ex R_\Ex}$, $C_{Q_\In Q_\In}=C_{R_\In
  R_\In}$, $C_{Q_\Ex Q_\In}=C_{Q_\Ex X_\Ex}=C_{Q_\Ex X_\In}=C_{Q_\In
  X_\Ex}=C_{Q_\In X_\In}=0$, we can express the spectrum
$C_{\tilde{R}_+\tilde{R}_+}(\omega)$ of the sum activity in the
feedforward case in terms of the spectra $C_{R_\Ex R_\Ex}(\omega)$ and
$C_{R_\In R_\In}(\omega)$ of the feedback system (see
eq.~\prettyref{eq:C_tilde_R_plus}). For strong coupling ($\w\gg 1$),
the zero-frequency component ($H(0)=1$) becomes
\begin{equation}
  C_{\tilde{R}_+\tilde{R}_+}(0) \simeq 
  \w^2\rho^2\frac{1+\gamma^{-1}}{N_\Ex}\frac{4\g^2}{(\g-1)^2}
  \,.
\end{equation}
Thus, for strong coupling, the zero-frequency power ratio
\begin{equation}
  \label{eq:power_ratio_2D_a}
  \alpha_+(0)=\frac{C_{R_+R_+}(0)}{C_{\tilde{R}_+\tilde{R}_+}(0)} \simeq
  \frac{(\g+1)^2}{8\w^2\g^2}
\end{equation}
reveals a relative suppression of the population-rate fluctuations in
the feedback system which is proportional to $1/\w^2$ (see
\prettyref{fig:FB_vs_FF_lin_model}B; black dashed line). The power
ratio $\alpha_+(0)$ for arbitrary weights $\w$ is depicted in
\prettyref{fig:FB_vs_FF_lin_model}B (black dotted curve). For a network
at the transition point $\g=1$, \prettyref{eq:power_ratio_2D_a} equals
$1/(2\w^2)$. Increasing the level of inhibition by increasing $\g$
leads to a decrease in the power ratio: in the limit $\g\to\infty$,
\prettyref{eq:power_ratio_2D_a} approaches $1/(8\w^2)$ monotonously.
\par
Above, we suggested that the negative self-feedback of the sum mode
$R_+$, weighted by $-w_+$ (\prettyref{fig:FB_vs_FF_schur}B), is
responsible for the fluctuation suppression in the recurrent
excitatory-inhibitory system. Here, we test this by considering the
case where this feedback loop is opened and replaced by uncorrelated
feedforward input $Q_+$, weighted by $-w_+$, while the feedforward
input from the difference mode $R_-$, weighted by $w_\FF$, is left
intact (see \prettyref{fig:FB_vs_FF_schur}D').  As before, we assume
that the auto-statistics of $Q_+$ is identical to the auto-statistics
of $R_+$ as obtained in the feedback case, i.e.~$C_{Q_+
  Q_+}(\omega)=C_{R_+ R_+}(\omega)$. According to the Schur
representation of the population dynamics
\prettyref{eq:schur_dynamics}/\prettyref{eq:schur_coupling}, the
Fourier transform of the sum mode of this modified system is
given by
\begin{equation}
  \tilde{R}_+(\omega)=H(\omega)\left(
    -w_+ Q_+(\omega)+w_\FF \tilde{R}_-(\omega) +X_+(\omega)
  \right)
  \,.
\end{equation}
With $C_{ \tilde{R}_+ \tilde{R}_+}(\omega)$ given in
\prettyref{eq:C_tilde_R_plus_FF2} and $C_{R_+{}R_+}(\omega)$, we obtain
the power ratio
\begin{equation}
  \label{eq:power_ratio_2D_b}
  \alpha^\prime_+(\omega)
  =\frac{C_{R_+R_+}(\omega)}{C_{\tilde{R}_+\tilde{R}_+}(\omega)}
  =\frac{1}{w_+^2|H(\omega)|^2+|1+w_+H(\omega)|^2}
  \,.
\end{equation}
Its zero-frequency component $\alpha_+^\prime(0)$ is shown in
\prettyref{fig:FB_vs_FF_lin_model}B (gray dotted curve). For strong
coupling, the power ratio decays as $1/(2 w_+^2)$ (gray dashed line in
\prettyref{fig:FB_vs_FF_lin_model}B). Thus, the (relative)
power in the recurrent system is reduced by strengthening the negative
self-feedback loop, i.e.~by increasing $w_+$.
\par
So far, we have presented results for the subpopulation averaged
firing rates $r_\Ex(t)$ and $r_\In(t)$ and the sum mode $r_+(t)$.  The
spectrum of the compound rate $r(t)=N^{-1}\sum_{i=1}^N
r_i(t)=N^{-1}[N_\Ex r_\Ex(t)+N_\In r_\In(t)]$, i.e.~the activity
averaged across the entire population, reads
\begin{equation}
  \label{eq:compound_spectrum_EI_FB}
  C_{RR}(\omega)=N^{-2}\left(N_\Ex^2 C_{R_\Ex{}R_\Ex}(\omega) + N_\In^2 C_{R_\In{}R_\In}(\omega)
    + N_\Ex{}N_\In [C_{R_\Ex{}R_\In}(\omega)+C_{R_\Ex{}R_\In}(\omega)\cc]\right) 
  \,.
\end{equation}
In the feedforward scenario depicted in \prettyref{fig:FB_vs_FF_schur}C, the spectrum 
of the compound rate $\tilde{R}=H(\w{}Q_\Ex+\w\g{}Q_\In+X)$ (with $X=N^{-1}\sum_{i=1}^N X_i$) is given by
\begin{equation}
  \label{eq:compoun_spectrum_EI_FF}
  C_{\tilde{R}\tilde{R}}(\omega)=|H(\omega)|^2\left(\w^2C_{R_\Ex{}R_\Ex}+\w^2\g^2C_{R_\In{}R_\In}+\rho^2/N\right)
  \,.
\end{equation}
For strong coupling, the corresponding low-frequency power ratio
$\alpha(0)=C_{RR}(0)/C_{\tilde{R}\tilde{R}}(0)$ (black solid curve in
\prettyref{fig:FB_vs_FF_lin_model}B) exhibits qualitatively the same
decrease $\propto\w^{-2}$ as the sum mode.
\par
To summarize the results of this subsection: the population dynamics
of a recurrent network with mixed excitatory and inhibitory coupling
can be mapped to a two-dimensional system describing the dynamics of
the sum and the difference of the excitatory and inhibitory
subpopulation activities. This equivalent representation uncovers
that, in inhibition dominated networks ($\g>1$), the sum activity is
subject to negative self-feedback. Thus, the dynamics of the sum
activity in excitatory-inhibitory networks is qualitatively similar to
the population dynamics of purely inhibitory networks (see
\prettyref{sec:I_poprate_model}). Indeed, the comparison of the
compound power-spectra of the intact recurrent network and networks
where the feedback channels are replaced by feedforward input reveals
that the (effective) negative feedback in excitatory-inhibitory
networks leads to an efficient suppression of population-rate
fluctuations.

\subsection{Population averaged correlations in cortical networks}
\label{sec:corr_ndim}
The results presented in the previous subsections describe the
fluctuations of the compound activity. Pairwise correlations
$c_{ij}(t) = \EW[t^\prime]{\bar s_i(t+t^\prime) \bar s_j(t^\prime)}$
between the (centralized) spike trains $\bar s_i(t) = s_i(t) -
\EW[t^\prime]{s_i(t^\prime)}$ are outside the scope of such a
description.  In this subsection, we consider the same
excitatory-inhibitory network as in \prettyref{sec:EI_poprate_model}
and present a theory for the population averaged spike-train
cross-correlations.  In general, this is a hard problem.  To
understand the structure of cross-correlations, it is however
sufficient to derive a relationship between the cross- and
auto-covariances in the network, because the latter can, to good
approximation, be understood in mean-field theory.  The integral of
the auto-covariance function of spiking LIF neurons can be calculated
by Fokker-Planck formalism \citep{Brunel99, Brunel00_183,
  MorenoBote08}.  To determine the relation between the
cross-covariance and the auto-covariance, we replace the spiking
dynamics by a reduced linear model whose covariances, to linear order,
obey the same relation. We present the full derivation in
\prettyref{sec:linear_model}.  There, we first derive an approximate
linear relation between the auto- and cross-covariance functions
$\mat{a}(\tau)$ and $\mat{c}(\tau)$, respectively, of the LIF
network. A direct solution of this equation is difficult. In the
second step, we therefore show that there exists a linear stochastic
system with activity $\mat{u}$ whose correlations $\mat{a}^u(\tau)$
and $\mat{c}^u(\tau)$ fulfill the same equation as the original LIF
model. This reduced model can be solved in the frequency domain by
standard Fourier methods.  Its solution allows us, by construction, to
determine the relation between the integral cross-covariances
$\mat{C}(0)$ and the integral auto-covariances $\mat{A}(0)$ up to
linear order.
\par
\begin{figure}[ht!]
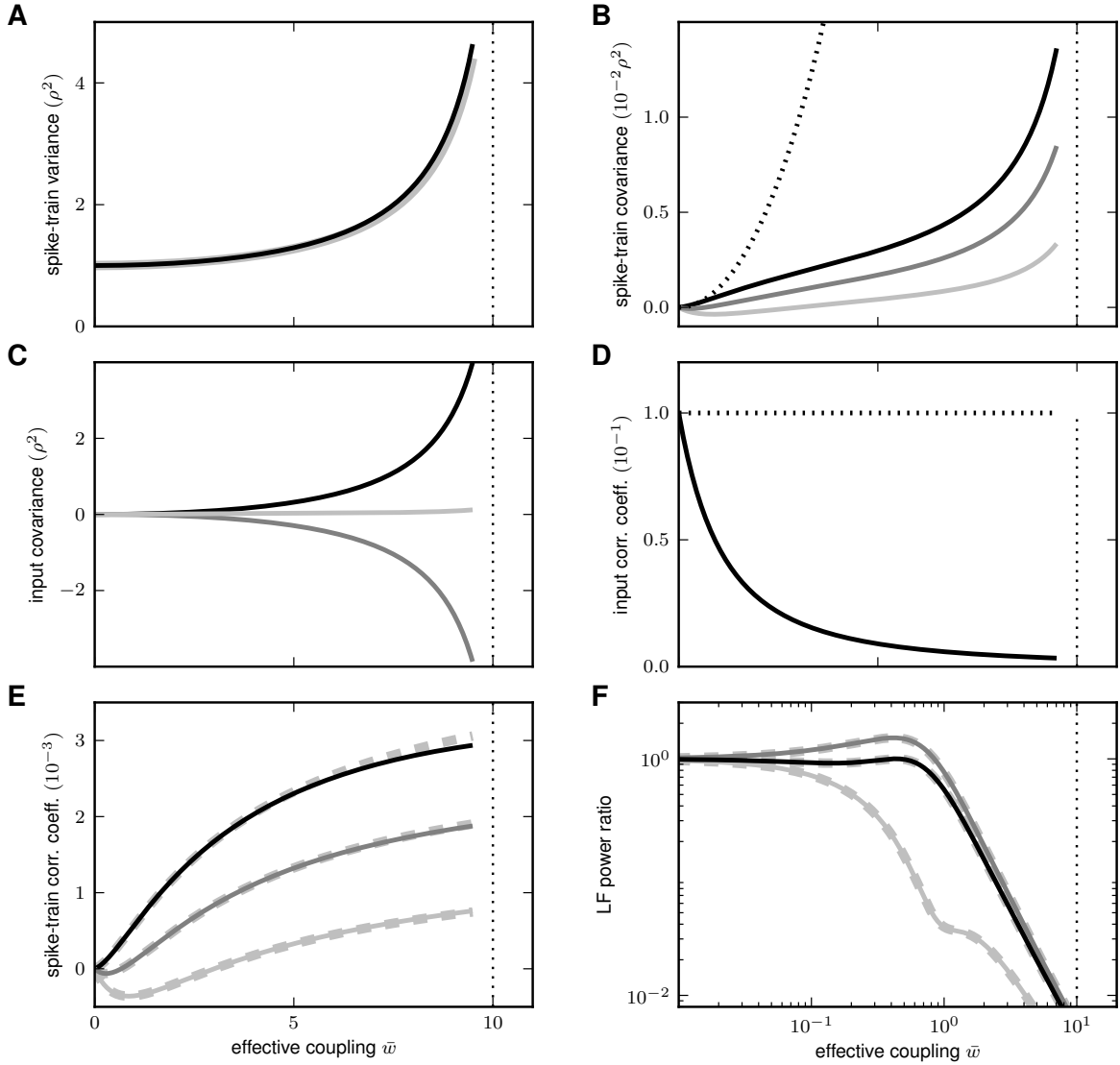

  \centering
  \doublecolumnfigure{\figpath/cov_ndim_w_dep_avg}%
  \caption{
    Dependence of population averaged correlations and population-rate fluctuations 
    on the effective coupling $\w = Kw$ in
    a linearized homogeneous network with excitatory-inhibitory coupling.
    {\bf A}: Spike-train variances $A_\Ex$ (black) and $A_\In$ (gray) of 
    excitatory and inhibitory neurons.
    {\bf B}: Spike-train covariances $C_{\Ex\Ex}$ (black solid), $C_{\Ex\In}$ (dark gray solid) and $C_{\In\In}$ (light gray solid) for excitatory-excitatory, excitatory-inhibitory and inhibitory-inhibitory neuron pairs in the recurrent network, respectively, and shared-input contribution $C^\inp_\text{shared}$ (black dotted curve; 'feedforward case').
    {\bf C}: Decomposition of the total input covariance $C^\inp$ (light gray) into 
    shared-input covariance $C^\inp_\text{shared}$ (black) and weighted spike-train 
    covariance $C^\inp_\text{corr}$ (dark gray).
    Covariances in A, B and C are given in units of the noise variance $\rho^2$.
    {\bf D}: Input-correlation coefficient $C^\inp/A^\inp$ in the recurrent network (black solid curve). In the feedforward case, the input-correlation coefficient is identical to the network connectivity $\epsilon$ (horizontal dotted line).
    {\bf E}: Spike-train correlation coefficients  $C_{\Ex\Ex}/A_\Ex$ (black), 
    $C_{\Ex\In}/\sqrt{A_\Ex A_\In}$ (dark gray) and $C_{\In\In}/A_\In$ (solid light gray curve) 
    for excitatory-excitatory, excitatory-inhibitory and inhibitory-inhibitory neuron pairs, respectively. 
    Thick dashed curves represent approximate solutions assuming $A_\Ex = A_\In$.
    {\bf F}: Low-frequency (LF) power ratios $\alpha$ (black), $\alpha_\Ex$ (dark gray), $\alpha_\In$ (solid light gray) for the population rate $r(t)$ and the excitatory and inhibitory subpopulation rates $r_\Ex(t)$ and $r_\In(t)$, respectively. The LF power ratio represents the ratio between the LF spectra in the recurrent network and for the case where the feedback channels are replaced by feedforward input with $C_{\Ex\In}=0$ (cf.~\prettyref{fig:FB_vs_FF_schur}C). Thick dashed curves in F show power ratios obtained by assuming that the auto-correlations are identical in the feedback and the feedforward scenario (see main text).
    Vertical dotted lines mark the stability limit of the linear model (see \prettyref{sec:linear_model}).
    A--F: $K=1000$, $\epsilon=0.1$, $\gamma=1/4$, $g=6$, $\g = \gamma g = 3/2$, $N=K(1+\gamma)/\epsilon=12500$.
  }
  \label{fig:cov_ndim_wdep}
 \end{figure}

As we are interested in the covariances averaged over many pairs of
neurons, we average the resulting set of $N$ linear self-consistency
equations \prettyref{eq:int_corr_lin} for the covariance matrix in the
frequency domain $\mat{C}(\omega)$ over statistically identical pairs
of neurons and many realizations of the random connectivity (see
\prettyref{sec:mean_field_corr}). This yields a four-dimensional
linear system \prettyref{eq:cov_eff_4dim} describing the population
averaged variances $A_\Ex$ and $A_\In$ of the excitatory and
inhibitory subpopulations, and the covariances $C_{\Ex \nlra \Ex}$ and
$C_{\In \nlra \In}$ for unconnected excitatory-excitatory and
inhibitory-inhibitory neuron pairs, respectively (note that we use the
terms ``variance'' and ``covariance'' to describe the \emph{integral}
of the auto- and cross-correlation function, respectively; in many
other studies, they refer to the zero-lag correlation functions
instead).
The dependence of the
variances and covariances on the coupling strength $\w$, obtained by
numerically solving \prettyref{eq:cov_eff_4dim}, is shown in
\prettyref{fig:cov_ndim_wdep}. 
We observe that the variances $A_\Ex$ and $A_\In$ of excitatory
and inhibitory neurons are barely distinguishable (\prettyref{fig:cov_ndim_wdep}A).
With the approximation $A:=A_\Ex = A_\In$, explicit expressions can be obtained for
the covariances (thick dashed curves \prettyref{fig:cov_ndim_wdep}E):
\begin{eqnarray}
C_{\Ex \Ex/\In \In} &=& \underbrace{\frac{1}{(1 - \w (1 - \g))^2}}_{\mathbf{I}} \; C^\inp_\text{shared} \label{eq:C_EE_C_II_ana_maintext} \\ \nonumber
  &+& \underbrace{\frac{1}{1 - \w (1-\g)}}_{\mathbf{II}} \underbrace{2 \w A \begin{cases}
      \frac{1}{N_\Ex} & \text{for} \; \Ex \Ex\\  
      \frac{-\g}{N_\In} & \text{for} \; \In \In
    \end{cases}}_{\mathbf{III}}
  \,,\\ \nonumber
  C_{\Ex \In} &=& \frac{1}{2}(C_{\Ex \Ex} + C_{\In \In}) \\ \nonumber
  \text{with}\quad
  C^\inp_\text{shared} &=& \w^2 \left(\frac{1}{N_\Ex} + \frac{\g^2}{N_\In} \right) A. \nonumber
\end{eqnarray}
The deviations from the full solutions (thin solid curves in
\prettyref{fig:cov_ndim_wdep}E), i.e.~for $A_\Ex \neq A_\In$, are
small. In the reduced model, both the external input and the spiking
of individual neurons contribute to an effective noise.  As the
fluctuations in the reduced model depend linearly on the amplitude
$\rho$ of this noise, the variances $A$ and covariances $C_{pq}$
($p,q\in\{\Ex,\In\}$) can be expressed in units of the noise variance
$\rho^2$. Consequently, the correlation coefficients $C_{pq}/A$ are
independent of $\rho^2$ (see \prettyref{fig:cov_ndim_wdep}).
\par 
The analytical form \prettyref{eq:C_EE_C_II_ana_maintext} of the
result shows that the correlations are smaller than expected given the
amount of shared input a pair of neurons receives: The quantity
$C^\inp_\text{shared}$ in the first line is the contribution of shared
input to the covariance.  For strong coupling $\w \gg 1$, the
prefactor $\mathbf{I}$ causes a suppression of this contribution. Its
structure is typical for a feedback system, similar to the solution
\prettyref{eq:spectr_1D_FB} of the one-population or the solution
\prettyref{eq:power_spectrum_EI} of the two-population model.  The
term $\w(1 - \g)$ in the denominator represents the negative feedback
of the compound rate.  The prefactor $\mathbf{II}$ in the second line
of \prettyref{eq:C_EE_C_II_ana_maintext} is again due to the feedback
and suppresses the contribution of the factor $\mathbf{III}$, which
represents the effect of direct connections between neurons. 
\par
Our results are consistent with a previous study of the decorrelation
mechanism: In \citep{Renart10_587}, the authors considered how
correlations scale with the size $N$ of the network where the synaptic
weights are chosen as $J \propto{}1/\sqrt{N}$.  As a result, the
covariance $C^\inp_\text{shared}$ in
\prettyref{eq:C_EE_C_II_ana_maintext} caused by shared input is
independent of the network size, while the feedback
$\w(1-\g)\propto\epsilon{}N(1-\g)\left(J + O(J^2)\right)$ scales---to
leading order---as $\sqrt{N}$ (see \prettyref{eq:w_of_J}).
Consequently, the first line in \prettyref{eq:C_EE_C_II_ana_maintext}
scales as $1/N$. The same scaling holds for the second line in
\prettyref{eq:C_EE_C_II_ana_maintext}, explaining the decay of
correlations as $1/N$ found in \citep{Renart10_587}.
\par
The first line in \prettyref{eq:C_EE_C_II_ana_maintext} is identical
for any pair of neurons.  The second line is positive for a pair of
excitatory neurons and negative for a pair of inhibitory neurons. In
other words, excitatory neurons are more correlated than inhibitory
ones.  Together with the third line in
\prettyref{eq:C_EE_C_II_ana_maintext}, this reveals a peculiar
correlation structure: $C_{\Ex \Ex} > C_{\Ex \In} > C_{\In \In}$
(\prettyref{fig:cov_ndim_wdep}B,E).  For strong coupling $\w \gg 1$,
the difference between the excitatory and inhibitory covariance is
\mbox{$C_{\Ex\Ex} - C_{\In \In} \simeq \frac{2}{\g - 1}
  (\frac{1}{N_\Ex} + \frac{\g}{N_\In}) A$}. The difference decreases as the level
$\g$ of inhibition is increased, i.e.~the further the network is in
the inhibition dominated regime, away from the critical point $\g=1$.

%
\par
To understand the suppression of shared-input correlations in
recurrent excitatory-inhibitory networks, consider the correlation
between the local inputs $I_{k/l}= [\mat{W} \vec{r}]_{k/l}$ of a pair
of neurons $k$, $l$. The input-correlation coefficient
$C^\inp/A^\inp=\Cov{I_k,I_l}/\sqrt{\Var{I_k}\Var{I_l}}$ can be
expressed in terms of the averaged spike-train covariances:
\begin{eqnarray}
  \label{eq:input_cov}
  \begin{aligned}
    C^\inp &= \Cov{I_k, I_l} = C^\inp_\text{shared} + C^\inp_\text{corr}\\
    A^\inp &= \Var{I_k} = \epsilon^{-1} \w^2 \left( \frac{1}{N_\Ex} + \frac{\g^2}{N_\In} \right) A + C^\inp_\text{corr}\\
    \text{with}\quad&
    C^\inp_\text{corr} = \w^2 (C_{\Ex \Ex} - 2 \g C_{\Ex \In} + \g^2 C_{\In \In})
  \end{aligned}
\end{eqnarray}
(see \prettyref{sec:mean_field_corr}: The input covariance $C^\inp$
equals the average quantity $C_{xy,\text{B}}$ given in
\prettyref{eq:C_avg_term_III}, the input variance $A^\inp$ is given by
\prettyref{eq:A_avg_term_III} as $A_{x,\text{B}}$). The term
$C^\inp_\text{shared}$ represents the contribution due to the
spike-train variances of the shared presynaptic neurons (see
\prettyref{eq:C_EE_C_II_ana_maintext}). This contribution is always
positive (provided the network architecture is consistent with Dale's
law; see \citep{Kriener08_2185}).
In a purely feedforward scenario with uncorrelated presynaptic
sources, $C^\inp_\text{shared}$ is the only contribution to the input
covariance of postsynaptic neurons. The resulting response correlation
for this feedforward case is much larger than in the feedback system
(\prettyref{fig:cov_ndim_wdep}B, black dotted curve). The correlation
coefficient between inputs to a pair of neurons in the feedforward
case is identical to the network connectivity $\epsilon$
\citep[horizontal dotted curve in \prettyref{fig:cov_ndim_wdep}D;
see][]{Kriener08_2185}.
In an inhibition dominated recurrent network, spike-train correlations
between pairs of different source neurons contribute the additional term $C^\inp_\text{corr}$,
which is negative and of similar absolute value as the shared-input contribution
$C^\inp_\text{shared}$. Thus, the two terms $C^\inp_\text{shared}$ and
$C^\inp_\text{corr}$ partly cancel each other (see
\prettyref{fig:cov_ndim_wdep}C).  In consequence, the resulting input
correlation coefficient $C^\inp/A^\inp$ is smaller than $\epsilon$
(see \prettyref{fig:cov_ndim_wdep}D; here: $\epsilon=0.1$).
\par
The correlations in a purely inhibitory network can be
obtained from \prettyref{eq:C_EE_C_II_ana_maintext} by replacing
$N_\Ex \rightarrow N$, taking into account the negative sign of $w$
in $\w = -Kw$ and setting $g=0$ and $\gamma = 0$:
\begin{eqnarray}
  C &=& \left( -1 + \frac{1}{(1+\w)^2} \right) \frac{A}{N}
  \,. 
  \label{eq:C_inh_ana_maintext}
\end{eqnarray}
For finite coupling strength $\w > 0$, this expression is
negative.  The contributions of shared input and spike-train
correlations to the input correlation are given by
$C^\text{in}_\text{shared} = \w^2 \frac{A}{N} > 0$ and
$C^\text{in}_\text{corr} = \w^2 C$, respectively (see
\prettyref{eq:C_EE_C_II_ana_maintext} and \prettyref{eq:input_cov}).
Using \prettyref{eq:C_inh_ana_maintext}, we can directly verify that
$C^\text{in}_\text{corr} < 0$, because pairwise correlations $C$ are
negative, leading to a partial cancellation $C^\text{in}_\text{shared}
+ C^\text{in}_\text{corr} = \w^2 \frac{1}{(1+\w)^2} \frac{A}{N}$: the
right hand side is smaller in magnitude by a factor of $\simeq
\frac{1}{\w^2}$ compared to each individual contribution.
Hence, as in the network with excitation and inhibition, shared-input
correlations are partly canceled by the contribution due to
presynaptic pairwise spike-train correlations.  In the feedforward
scenario with zero presynaptic spike-train correlations, in contrast,
the response correlations are determined by shared input alone and are
therefore increased. The suppression of shared-input correlations in
the feedback case is what we call 'decorrelation' in the current
work. In purely inhibitory networks, this decorrelation is caused by
weakly negative pairwise correlations
\prettyref{eq:C_inh_ana_maintext}. For sufficiently strong negative
feedback, correlations are smaller in absolute value as compared to
the feedforward case. The absolute value of these anti-correlations is
bounded by $A/N$.
\par
The similarity in the results obtained for purely inhibitory networks
and excitatory-inhibitory networks demonstrates that the suppression
of pairwise correlations and population-activity fluctuations is a generic
phenomenon in systems with negative feedback. It does not rely on an
internal balance between excitation and inhibition.
\par
As discussed in \prettyref{sec:pop_fluct_iaf}, the suppression of
correlations in the recurrent network is accompanied by a reduction of
population-activity fluctuations.  With the population averaged
correlations \prettyref{eq:C_EE_C_II_ana_maintext}, the power
\prettyref{eq:spectrum_avg_activity} of the population activity $r(t)$
reads
\begin{equation}
  \label{eq:spectrum_avg_activity_EI_FB}
  C_{RR}=\frac{1}{N^2}\left[N_\Ex A_\Ex + N_\In A_\In + N_\Ex (N_\Ex-1)C_{\Ex\Ex} + N_\In (N_\In-1)C_{\In\In}+2N_\Ex{}N_\In{}C_{\Ex\In}\right]
  \,.
\end{equation}
In \prettyref{sec:EI_poprate_model}, we showed that the
population-activity fluctuations are amplified if the local input in the
recurrent system is replaced by feedforward input from independent
excitatory and inhibitory populations (see
\prettyref{fig:FB_vs_FF_schur}C). This manipulation corresponds to a
neglect of correlations $C_{\Ex\In}$ between excitatory and inhibitory
neurons. All remaining correlations ($A_\Ex$, $A_\In$, $C_{\Ex\Ex}$,
$C_{\In\In}$) are preserved. With the resulting response auto- and
cross-correlations $\tilde{A}$ and $\tilde{C}$ given by
\prettyref{eq:A_C_FF}, the power \prettyref{eq:spectrum_avg_activity}
of the population activity becomes
\begin{equation}
  \label{eq:spectrum_avg_activity_EI_FF}
  C_{\tilde{R}\tilde{R}}=\frac{1}{N}\tilde{A}+\left(1-\frac{1}{N}\right)\tilde{C}
  \,.
\end{equation}
For large effective coupling $\w$, the power ratio
$\alpha=C_{RR}/C_{\tilde{R}\tilde{R}}$ decays as $1/\w^2$ (black curve
in \prettyref{fig:cov_ndim_wdep}F). Note that the power ratio $\alpha$
derived here is indistinguishable from the one we obtained in the
framework of the population model in
\prettyref{sec:EI_poprate_model} (black solid curve in
\prettyref{fig:FB_vs_FF_lin_model}B). Although the derivation of the
macroscopic model in \prettyref{sec:EI_poprate_model} is qualitatively
different from the one leading to the population averaged correlations
described here, the two models are consistent: They describe one and
the same system and lead to identical power ratios.
\par
The fluctuation suppression is not only observed at the level of the
entire network, i.e.~for the population activity $r(t)$, but also for each
individual subpopulation $\Exc$ and $\Inh$, i.e.~for the
subpopulation averaged activities $r_\Ex(t)$ and $r_\In(t)$. The derivation
of the corresponding power ratios $\alpha_\Ex$ and $\alpha_\In$ is
analog to the one described above. As a result of the correlation
structure $C_{\Ex\Ex}>C_{\In\In}$  in the feedback system (see
\prettyref{fig:cov_ndim_wdep}B), the power of the inhibitory
population activity is smaller than the power of the excitatory population
activity. In consequence, $\alpha_\Ex>\alpha_\In$ (gray curves in
\prettyref{fig:cov_ndim_wdep}F).
\par
In \prettyref{eq:spectrum_avg_activity_EI_FB} and
\prettyref{eq:spectrum_avg_activity_EI_FF}, the auto-correlations are
scaled by $1/N$, while the cross-correlations enter with a prefactor
of order unity. For large $N$, one may therefore expect that the
suppression of population-activity fluctuations is essentially mediated by
pairwise correlations. In the recurrent system, however, the
cross-correlations $C_{xy}$ ($x,y\in\{\Ex,\In\}$) are of order $A/N$
(see \prettyref{fig:cov_ndim_wdep} and
\prettyref{eq:C_EE_C_II_ana_maintext}). It is therefore a priori not
clear whether the fluctuation suppression is indeed dominated by
pairwise correlations. In our framework, one can explicitly show that
the auto-correlation is irrelevant: Replacing the auto-correlation
$\tilde{A}$ in \prettyref{eq:spectrum_avg_activity_EI_FF} by the
average auto-correlation $(N_\Ex A_\Ex + N_\In A_\In)/N$ of the
intact feedback system has no visible effect on the resulting power ratio
(dashed curves in \prettyref{fig:cov_ndim_wdep}F). The difference in
the spectra of the population activities $C_{RR}$ and $C_{\tilde{R}\tilde{R}}$ is
therefore essentially caused by the cross-correlations.
\par
The absolute population-activity fluctuations in purely inhibitory and
in excitatory-inhibitory networks show a qualitatively different
dependence on the synaptic coupling $\w$, in agreement with the
previous sections. In networks with excitation and inhibition, the
correlation coefficient increases with increasing synaptic coupling
(see \prettyref{fig:cov_ndim_wdep}E). Hence, the population-activity
fluctuations grow with increasing coupling strength. In purely
inhibitory networks, in contrast, the pairwise spike-train correlation
decreases monotonously with increasing magnitude of the coupling
strength $\w$, see \prettyref{eq:C_inh_ana_maintext}.  In consequence,
the population-activity fluctuations decrease.  The underlying reason
is that, in the inhibitory network, the power of the population
activity is directly proportional to the covariance of the input
currents, which is actively suppressed, as shown above. For
excitatory-inhibitory networks, these two quantities are not
proportional (compare \prettyref{eq:input_cov} and
\prettyref{eq:spectrum_avg_activity}) due to the different synaptic
weights appearing in the input covariance.
\par
\begin{figure}[ht!]
  \centering
  \doublecolumnfigure{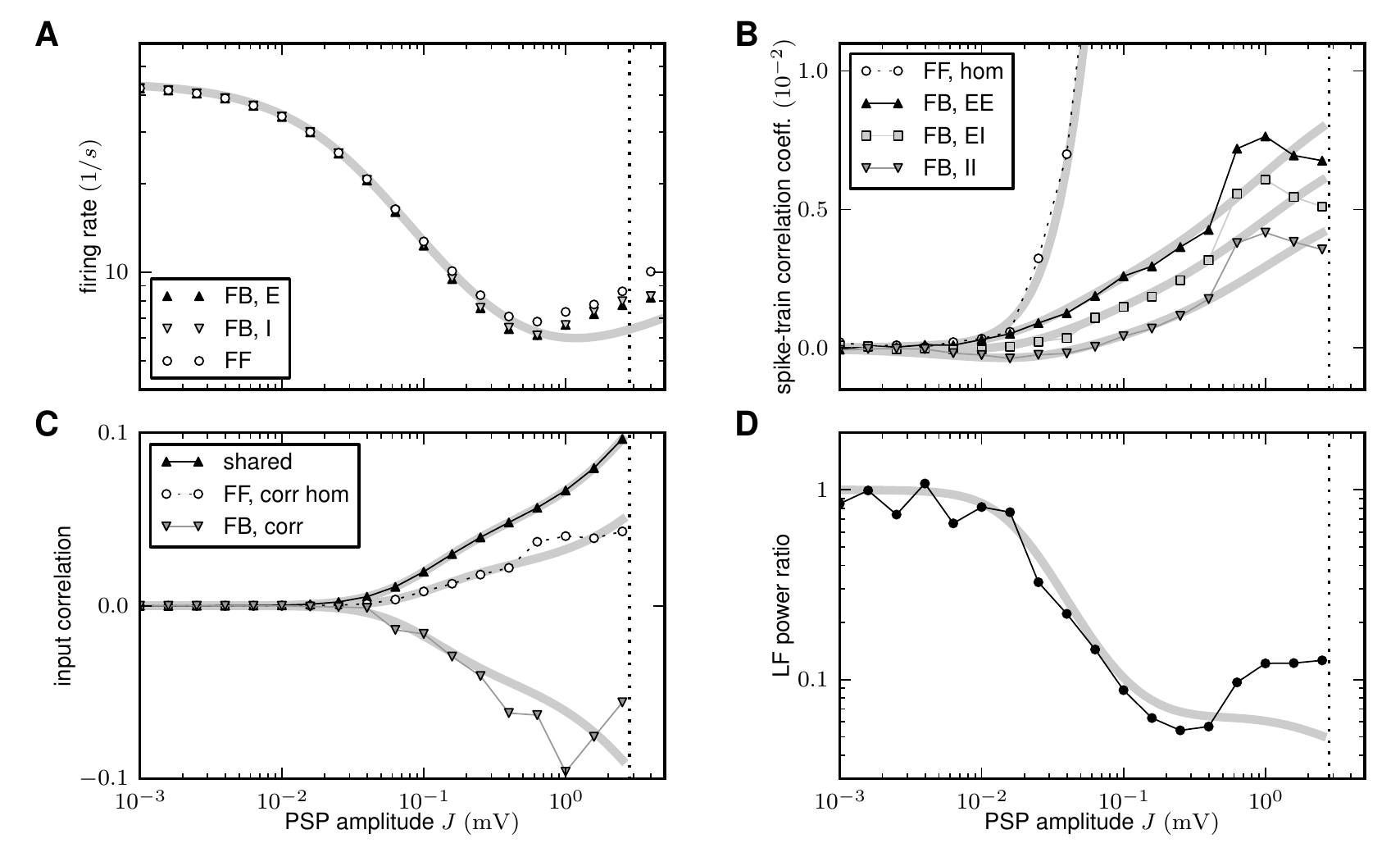}%
  \caption{%
    Comparison between predictions of the linear theory (thick gray curves) and direct simulation of the LIF-network model (symbols and thin lines).
    Dependence of the spike-train and population-rate statistics on the
    synaptic weight $J$ (PSP amplitude) in a recurrent
    excitatory-inhibitory network ('feedback system', 'FB') and in a
    population of unconnected neurons receiving randomized feedforward
    input ('feedforward system', 'FF') from neurons in the recurrent
    network.  Average presynaptic firing rates and shared-input structure
    are identical in the two systems.  In the FF case, the average
    correlations between presynaptic spike-trains are homogenized
    (i.e.~$C_{\Ex \Ex} =C_{\Ex \In}=C_{\In \In}$) as a result of the
    random reassignment of presynaptic neuron types. The mapping of the
    LIF dynamics to the linear reduced dynamics (\prettyref{sec:w_of_j})
    relates the PSP amplitude $J$ to the effective coupling strength $w(J)$ by
    \prettyref{eq:w_of_J}, as shown in \prettyref{fig:w_of_J}B.
    {\bf A}: Average firing rates $\nu_0$ in the FB
    (black up-triangles: excitatory neurons; gray down-triangles: inhibitory neurons) 
    and in the FF system (open circles). Analytical prediction
    \prettyref{eq:siegert} (gray curve).
    {\bf B}: Spike-train correlation coefficients
    $C_{\Ex \Ex}/A_\Ex$ (black up-triangles), 
    $C_{\Ex \In}/\sqrt{A_\Ex A_\In}$ (gray squares) and
    $C_{\In \In}/A_\In$ (gray down-triangles)
    for excitatory-excitatory, excitatory-inhibitory, and inhibitory-inhibitory neuron pairs, respectively, 
    in the FB system.
    Analytical prediction \prettyref{eq:C_EE_C_II_ana_maintext} (gray curves).
    Spike-train correlation coefficient $\tilde{C}/\tilde{A}$ (open circles) in the FF system with homogenized presynaptic spike-train correlations.
    Analytical prediction \prettyref{eq:A_C_FF_hom_ndim} (underlying gray curve).
    {\bf C}: 
    Shared-input ($C^\inp_\text{shared}$; black up-triangles) and 
    spike-correlation contribution $C^\inp_\text{corr}$ 
    (FB: gray down-triangles; FF: open circles) to the input correlation $C^\inp$ (normalized by $\sqrt{A_\Ex A_\In}$).
    Analytical predictions \prettyref{eq:input_cov}.
    {\bf D}: Low-frequency (LF) power ratio of the compound activity.
    Vertical dotted lines in A--D mark the stability limit of the linear model 
    (see \prettyref{sec:linear_model}).
    $N=12500$, $K=1000$, $\gamma=1/4$, $g=6$. Size of postsynaptic population in the FF case: $M=2000$. 
    Simulation time: $T=100\,\seconds$.
  }
  \label{fig:cov_ndim_iaf_wdep}
\end{figure}

To compare our theory to simulations of spiking LIF networks, we need
to determine the effect of a synaptic input on the response activity
of the neuron model.  To this end, we employ the Fokker-Planck theory
of the LIF model (see \prettyref{sec:w_of_j}).
In this context, the steady state of the recurrent network is
characterized by the mean $\mu$ and the standard deviation $\sigma$ of
the total synaptic input. Both $\mu$ and $\sigma$ depend on the
steady-state firing rate in the network.  The steady-state firing rate
can be determined in a self-consistent manner \cite{Brunel00_183} as
the fixed point of the firing rate approximation
\prettyref{eq:siegert}.  The approximation predicts the firing rate to
sufficient accuracy of about $\pm 1\,\sps$ (see
\prettyref{fig:cov_ndim_iaf_wdep}A).
We then obtain an analytical expression of the low-frequency transfer
which relates the fluctuation $\nu_j(t) = \bar \nu + \epsilon
\delta(t)$ of a synaptic input to neuron $i$ to the fluctuation of
neuron $i$'s response firing rate to linear order, so that
$\int_0^\infty\delta \nu_i(t) \; dt = \epsilon w(J_{ij})$.  This
relates the postsynaptic potential $J_{ij}$ in the LIF model to the
effective linear coupling $w_{ij} = w(J_{ij})$ in our linear theory.
The functional relation $w(J)$ can be derived in analytical form by
linearization of \prettyref{eq:siegert} about the steady-state
working point. Note that $w(J)$ depends on $\mu$ and $\sigma$ and,
hence, on the steady-state firing rate in the network. The derivation
outlined in \prettyref{sec:w_of_j} constitutes an extension of earlier
work \citep{DeLaRocha07_802, Helias10_1000929} to quadratic order in
$J$. The results agree well with those obtained by direct simulation
for a large range of synaptic amplitudes (see \prettyref{fig:w_of_J}).
\par
\prettyref{fig:cov_ndim_iaf_wdep}B compares the population averaged
correlation coefficients $C/A$ obtained from the linear reduced model,
see \prettyref{eq:C_EE_C_II_ana_maintext}, and simulations of LIF
networks. Note that the absolute value of the noise amplitude $\rho$
in the reduced model does not influence the correlation coefficient
$C/A$, as both quantities $C$ and $A$ depend linearly on $\rho^2$.
Theory and simulation agree well for synaptic weights up to
$J\approx{}1\,\mV$.  For larger synaptic amplitudes, the approximation
of the effective linear transfer for a single neuron obtained from the
Fokker-Planck theory deviates from its actual value (see
\prettyref{fig:w_of_J}B). \prettyref{fig:cov_ndim_iaf_wdep}C shows
that the cancellation of the input covariance in the LIF network is
well explained by the theory.
\par
Previous work \citep{Renart10_587} suggested that positive
correlations between excitatory and inhibitory inputs lead to a
negative component in the input correlation which, in turn, suppresses
shared-input correlations. The mere existence of positive correlations
between excitatory and inhibitory inputs is however not sufficient. To
explain the effect, it is necessary to take the particular correlation
structure $C_{\Ex\Ex}>C_{\Ex \In}>C_{\In \In}$ into account. To
illustrate this, consider the case where the correlation structure is
destroyed by replacing all pairwise correlations in the input
spike-train ensemble by the overall population average
\mbox{$C=(N_\Ex{}C_{\Ex \Ex}+N_\In{}C_{\In\In})/(N_\Ex+N_\In)>0$}
(homogenization of correlations). The resulting response correlations
(upper gray curve in \prettyref{fig:cov_ndim_iaf_wdep}B) are derived
in \prettyref{sec:mean_field_corr},
eq.~\prettyref{eq:A_C_FF_hom_ndim}. In simulations of LIF networks, we
study the effect of homogenized spike-train correlations by first
recording the activity of the intact recurrent network, randomly
reassigning the neuron type ($\Ex$ or $\In$) to each recorded spike
train, and feeding this activity into a second population of neurons.
Compared to the intact recurrent network, the response correlations
are significantly larger (\prettyref{fig:cov_ndim_iaf_wdep}B). The
contribution of homogenized spike-train correlations to the input
covariance $C^\inp$ (see \prettyref{eq:input_cov}) is given by
\mbox{$C^\inp_\text{corr,hom} = \w^2 (1 - \g)^2 C\ge{}0$}. For
positive spike-train correlations $C>0$, this contribution is greater
or equal zero (zero for $\g=1$). Hence, it cannot compensate the
(positive) shared-input contribution $C^\inp_\text{shared}$ (see
\prettyref{fig:cov_ndim_iaf_wdep}C). In consequence, input
correlations, output correlations and, in turn, population-rate
fluctuations (\prettyref{fig:cov_ndim_iaf_wdep}D) cannot be suppressed
by homogeneous positive correlations in the input spike-train
ensemble. Canceling of shared-input correlations requires either
negative spike-train correlations (as in purely inhibitory networks)
or a heterogeneity in correlations across different pairs of neurons
(e.g.~$C_{\Ex\Ex}>C_{\Ex \In}>C_{\In \In}$).


\subsection{Effect of feedback manipulations}
\label{sec:feedback_perturbations}
In the previous subsections, we quantified the suppression of
population-rate fluctuations in recurrent networks by comparing the
activity in the intact recurrent system (feedback scenario) to the
case where the feedback is replaced by feedforward input with some
predefined statistics (feedforward scenario). We particularly studied
the effect of neglecting the auto-statistics of the compound feedback,
(the structure of) correlations within the feedback ensemble and/or
correlations between the feedback and the external input. In all
cases, we observed a significant amplification of population-activity
fluctuations in the feedforward scenario. In this subsection, we
further investigate the role of different types of feedback
manipulations by means of simulations of LIF networks with
excitatory-inhibitory coupling. To this end, we record the spiking
activity of the recurrent network (feedback case), apply different
types of manipulations to this activity (described in detail below)
and feed this modified activity into a second population of identical
(unconnected) neurons (feedforward case). As before, the connectivity
structure (in-degrees, shared-input structure, synaptic weights) is
exactly identical in the feedback and the feedforward case.
\par
\begin{figure}[ht!]
  \centering
  \doublecolumnfigure{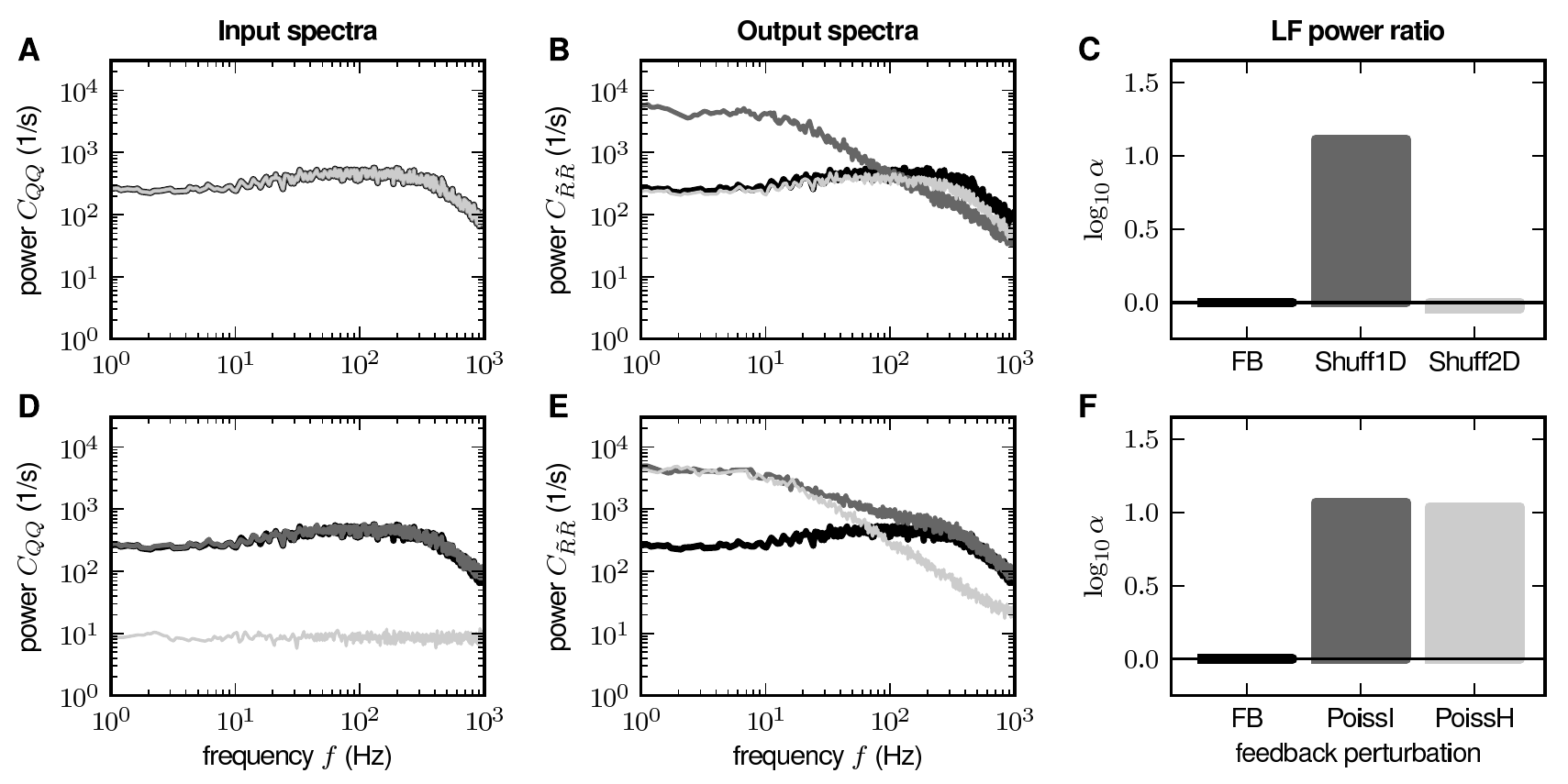}%
  \caption{
    Amplification of population-rate fluctuations by different types of feedback manipulations
    in a random network of excitatory and inhibitory LIF neurons (simulation results). 
    Top row ({\bf A}--{\bf C}): Unperturbed feedback (FB; black), 
    shuffling of spike-train senders across entire network (Shuff1D; dark gray) and 
    within each subpopulation (E,I) separately (Shuff2D; light gray).
    Bottom row ({\bf D}--{\bf F}): Unperturbed feedback (FB; black), replacement of spike trains by realizations of 
    inhomogeneous (PoissI; dark gray) and homogeneous Poisson processes (PoissH; light gray). 
    In the PoissI (PoissH) case, the (time averaged) subpopulation rates are approximately preserved.
    {\bf A},{\bf D}: Compound power-spectra $C_{QQ}$ of input spike-train ensembles.
    {\bf B},{\bf E}: Power-spectra $C_{\tilde{R}\tilde{R}}$ of population-response rates.
    {\bf C},{\bf F}: Low-frequency  (LF; $1$--$20\,\Hz$) power ratio $\alpha$ 
    (increase in LF power relative to the unperturbed case [FB]; logarithmic scaling).
    Note that in A, the compound-input spectra (FB, Shuff1D, Shuff2D) are identical. 
    In D, the input spectra for the intact recurrent network (FB) and the inhomogeneous-Poisson case (PoissI) 
    are barely distinguishable.
    See
    \prettyref{tab:lif_network_model} and
    \prettyref{tab:lif_parameters} for details on the network model and
    parameters. Simulation time $T=100\,\seconds$. Single-trial spectra
    smoothed by moving average (frame size $1\,\Hz$).
  }
  \label{fig:feedback_perturbations}
\end{figure}

In \prettyref{sec:linear_model}, we show that the low-frequency
fluctuations of the population rate $s(t)$ of the spiking model are
captured by the reduced model $r(t)$ presented in the previous
subsections.  To verify that the theory based on excitatory and
inhibitory population rates is indeed sufficient to explain the
decorrelation mechanism, we first consider the case where the sender
identities of the presynaptic spike train are randomly shuffled.
\prettyref{fig:feedback_perturbations}A shows the power-spectrum of
the population activity recorded in the original network (FB) as well
as the spectra obtained after shuffling spike-train identities within
the excitatory and inhibitory subpopulations separately (Shuff2D), or
across the entire network (Shuff1D). As shuffling of neuron identities
does not change the population rates, all three compound spectra are
identical.  \prettyref{fig:feedback_perturbations}B shows the response
power-spectra of the neuron population receiving the shuffled spike
trains. Shuffling within the subpopulations (Shuff2D) preserves the
population-specific fluctuations and average correlations. The effect
on the response fluctuations is negligible (compare black and light
gray curves in \prettyref{fig:feedback_perturbations}B). In
particular, the power of low-frequency fluctuations remains unchanged
(\prettyref{fig:feedback_perturbations}C).  This result confirms that
population models which take excitatory and inhibitory activity
separately into account are sufficient to explain the observations.
Shuffling of spike-train identities across subpopulations (Shuff1D),
in contrast, causes an increase in the population fluctuations by
about one order of magnitude
(\prettyref{fig:feedback_perturbations}B,C; dark gray).  This outcome
is in agreement with the result obtained by homogenizing pairwise
correlations (see \prettyref{fig:cov_ndim_iaf_wdep}) and demonstrates
that the excitatory and inhibitory subpopulation rates have to be
conserved to explain the observed fluctuation suppression.
\par
The shuffling experiments and the results of the linear model in the
previous subsections suggest that the precise temporal structure of
the \emph{population averaged} activities within homogeneous
subpopulations is essential for the suppression of population-rate
fluctuations. Preserving the exact structure of individual spike
trains is not required. This is confirmed by simulation experiments
where new sender identities were randomly reassigned for each
individual presynaptic spike (rather than for each spike train; data
not shown). This operation destroys the structure of individual spike
trains but preserves the compound activities. The results are similar
to those reported here.
\par
So far, it is unclear how sensitive the fluctuation-suppression
mechanism is to perturbations of the temporal structure of the
population rates.  To address this question, we replaced the
excitatory and inhibitory spike trains in the feedback ensemble by
independent realizations of inhomogeneous Poisson processes (PoissI)
with intensities given by the measured excitatory and inhibitory
population rates $s_\Ex(t)$ and $s_\In(t)$ of the recurrent network,
respectively.  Note that the compound rates of a single realization of
this new spike-train ensemble are similar but not identical to the
original population rates $s_\Ex(t)$, $s_\In(t)$ (in each time window
$[t+\Delta t)$, the resulting spike count is a random number drawn
from a Poisson distribution with mean and variance proportional to
$s_\Ex(t)$ and $s_\In(t)$, respectively). Although the compound
spectrum of the resulting local input is barely distinguishable from
the compound spectrum of the intact recurrent system
(\prettyref{fig:feedback_perturbations}D; black and dark gray curves),
the response spectra are very different: replacing the feedback
ensemble by inhomogeneous Poisson processes leads to a substantial
amplification of low-frequency fluctuations
(\prettyref{fig:feedback_perturbations}E; compare black and dark gray
curves). The effect is as strong as if the temporal structure of the
population rates was completely ignored, i.e.~if the feedback channels
were replaced by realizations of homogeneous Poisson processes with
constant rates (PoissH; light gray curves in
\prettyref{fig:feedback_perturbations}D,E).  This result indicates
that the precise temporal structure of the population rates is
essential and that even small deviations can significantly weaken the
fluctuation-suppression mechanism.  The results of the Poisson
experiments can be understood by considering the effect of the
additional noise caused by the stochastic realization of individual
spikes. 
Considering the auto-correlation, a Poisson spike-train ensemble with
rate profile $\nu(t)$ is equivalent to a sum of the rate profile and a
noise term resulting from the stochastic (Poissonian) realization of
spikes, $\vec{q}(t) = \nu(t)\vec{1} + \sqrt{\nu_0} \vec{z}(t)$. Here,
$\vec{z}(t)$ denotes a Gaussian white noise with auto-correlation
$\EW[t]{\vec{z}(t)\vec{z}(t+s)} = \vec{1} \delta(s)$ and
\mbox{$\nu_0=\EW[t]{\nu(t)}$} the mean firing rate.  The response
fluctuations of the population driven by the rate modulated Poisson
activity are, to linear approximation, given by $\tilde{\vec{R}} = H
(\mat{W}\vec{Q} + \vec{X})$.  Inserting $\vec{Q}$, we obtain an
additional noise term $\nu_0|H|^2\mat{W}\mat{W}^T$ in the spectrum
$\mat{C}_{\tilde{R}\tilde{R}} = \tilde{\vec{R}}\tilde{\vec{R}}\cc$
which explains the increase in power compared to the spectrum
$\mat{C}_{RR}$ of the recurrent network. As a generalization of the
Poisson model, one may replace the noise amplitude $\sqrt{\nu_0}$ by
some arbitrary prefactor $\eta$. In simulation experiments, we
observed a gradual amplification of the population-rate fluctuations
with increasing noise amplitude $\eta$ (data not shown).

\section{Discussion}\label{sec:discussion}
We have shown that negative feedback in recurrent neural networks
actively suppresses low-frequency fluctuations of the population
activity and pairwise correlations. This mechanism allows neurons to
fire more independently than expected given the amount of shared
presynaptic input.  We demonstrated that manipulations of the feedback
statistics, e.g.~replacing feedback by uncorrelated feedforward input,
can lead to a significant amplification of response correlations and
population-rate fluctuations.
\par
The suppression of correlations and population-rate fluctuations by
feedback can be observed in networks with both purely inhibitory and
mixed excitatory-inhibitory coupling. In purely inhibitory networks,
the effect can be understood by studying the role of the effective
negative feedback experienced by the compound activity. In networks of
excitatory and inhibitory neurons, a change of coordinates,
technically a Schur decomposition, exposes the underlying feedback
structure: the sum of the excitatory and inhibitory activity couples
negatively to itself if the network is in an inhibition dominated
regime (which is required for its stability; \citealp[see
e.g.][]{Brunel00_183}). This negative feedback suppresses fluctuations
in a similar way as in purely inhibitory networks. The fluctuation
suppression becomes more efficient the further the network is brought
into the inhibition dominated regime, away from the critical point of
equal recurrent excitation and inhibition ($\g = 1$). Having
identified negative feedback as the underlying cause of small
fluctuations and correlations, we can rule out previous explanations
based on a balance between (correlated) excitation and inhibition
\citep{Renart10_587}. We presented a self-consistent theory for the
average pairwise spike-train correlations which illuminates that the
suppression of population-rate fluctuations and the suppression of
pairwise correlations are two expressions of the same effect: as the
single spike-train auto-covariance is the same in the feedforward and
the feedback case, the suppression of population-rate fluctuations
implies smaller correlations.  Our theory enables us to identify the
cancellation of input correlations as a hallmark of small spike-train
correlations.
\par
In previous studies, shared presynaptic input has often been
considered a main source of correlation in recurrent networks
\citep[e.g.][]{Shadlen01_1916,Kriener08_2185}.  Recently,
\cite{Renart10_587} suspected that correlations between excitatory and
inhibitory neurons and the fast tracking of external input by the
excitatory and the inhibitory population are responsible for an active
decorrelation. We have demonstrated here that the mere fact that
excitatory and inhibitory neurons are correlated is not sufficient to
suppress shared-input correlations.  Rather, we find that the
spike-train correlation structure in networks of excitatory and
inhibitory networks arranges such that their overall contribution to
the covariance between the summed inputs to a pair of neurons becomes
negative, canceling partly the effect of shared inputs. This
cancellation becomes more precise the stronger the negative compound
feedback $Kw (1 - \gamma g)$ is.
In homogeneous networks where excitatory and inhibitory neurons
receive statistically identical input, the particular structure of
correlations is $C_{\Ex \Ex} > C_{\Ex \In} > C_{\In \In}$.  It can
further be shown that this structure of correlations is preserved in
the limit of large networks $N \rightarrow \infty$ ($K/N =
\text{const.}$).  For non-homogeneous synaptic connectivity, if the
synaptic amplitudes depend on the type of the target neuron
(i.e. $J_{\Ex \Ex} \neq J_{\In \Ex}$ or $J_{\Ex \In} \neq J_{\In
  \In}$), the structure of correlations may be different.  Still, the
correlation structure arranges such that shared input correlation is
effectively suppressed. Formally, this can be seen from a
self-consistency equation similar to our equation
\prettyref{eq:cov_unconnected_input_suppr}.
\par
The study by \citet{Renart10_587} has shown that correlations are
suppressed in the limit of infinitely large networks of binary neurons
receiving randomly drawn inputs from a common external population.
Its argument rests on the insight that the population-activity
fluctuations in a recurrent balanced network follow the fluctuations
of the external common population. An elegant scaling consideration
for infinitely large networks $N \rightarrow \infty$ with vanishing
synaptic efficacy $\propto 1/\sqrt{N}$ shows that this fast tracking
becomes perfect in the limit. This allows to determine the zero-lag
pairwise correlations caused by the external input.
The analysis methods and the recurrent networks presented here differ
in several respects from these previous results: We study networks of
a finite number of spiking model neurons. The neurons receive
uncorrelated external input, so that correlations are due to the local
recurrent connectivity among neurons, not due to tracking of the
common external input \citep{Renart10_587}.
Moreover, we consider homogeneous connectivity where synaptic weights
depend only on the type of the presynaptic neuron \citep[as, e.g.,
in][]{Brunel00_183}, resulting in a correlation structure $C_{\Ex \Ex}
> C_{\Ex \In} > C_{\In \In}$. For such connectivity, networks of
binary neurons with uncorrelated external input exhibit qualitatively
the same correlation structure as reported here (results not shown).
\par
In purely inhibitory networks, the decorrelation occurs in an analog
manner as in excitatory-inhibitory networks.  As only a single
population of neurons is available here, population averaged
spike-train correlations $C_{\In\In}$ are negative. This negative
contribution compensates the positive contribution of shared input.
\par
The structure of integrated spike-train covariances in networks
constitutes an experimentally testable prediction.  Note, however,
that the prediction \prettyref{eq:C_EE_C_II_ana_maintext} obtained in
the current work rests on two simplifying assumptions: identical
internal dynamics of excitatory and inhibitory neurons and homogeneous
connectivity (i.e.~$J_{\Ex \Ex}=J_{\In \Ex}$, $J_{\Ex
  \In}=J_{\In\In}$; see \prettyref{sec:EI_poprate_model}).  For such
networks, the structure of correlations is given by
$C_{\Ex\Ex}>C_{\Ex\In}>C_{\In \In}$.  Further, the relation between
subthreshold membrane-potential fluctuations and spike responses is
the same for both neuron types. Consequently, the above correlation
structure can be observed not only at the level of spike trains but
also for membrane potentials, provided the assumptions hold true.
A recent experimental study \citep{Gentet10_422} reports neuron-type
specific cross-correlation functions in the barrel cortex of behaving
mice, both for spike trains and membrane potentials. It is however
difficult to assess the integral correlations from the published
data. A direct test of our predictions requires either a reanalysis of
the data or a theory predicting the entire correlation functions. The
raw (unnormalized) II and EI spike-train correlations in
\citep{Gentet10_422} are much more pronounced than the EE correlations
\citep[Fig.\,6 in][]{Gentet10_422}. This seems to be in contradiction
to our results. Note, however, that the firing rates of excitatory and
inhibitory neurons are very different in \citep{Gentet10_422}. In our
study, in contrast, the average firing rates of excitatory and
inhibitory neurons are identical as a consequence of the assumed
network homogeneity. Future theoretical work is needed to generalize
our model to networks with heterogeneous firing rates and
non-homogeneous connectivity.  Recent results on the dependence of the
correlation structure on the connectivity may prove useful in this
endeavor \citep{Pernice11_e1002059, Pernice12_031916,
  Trousdale12_e1002408}.
\par
Correlations in spike-train ensembles play a crucial role for the en-
and decoding of information. A set of uncorrelated spike trains
provides a rich dynamical basis which allows readout neurons to
generate a variety of responses by tuning the strength and filter
properties of their synapses \citep{Tripp07_1830}. In the presence of
correlations, the number of possible readout signals is
limited. Moreover, spike-train correlations impair the precision of
such readout signals in the presence of noise. Consider, for example,
a linear combination $y(t)=\sum_{i=1}^{N} (s_i*h_i)(t)$ of $N$
presynaptic spike trains with arbitrary (linear) filter kernels
$h_i(t)$ (e.g.~synaptic filters). In a realistic scenario, the
individual spike trains $s_i(t)$ typically vary across trials
\citep{Softky93,Shadlen98}. To understand how robust the resulting
readout signal $y(t)$ is against this spike-train variability, let's
consider the variability of its Fourier transform
$Y(\omega)=\Fourier{y(t)}{\omega}=\sum_{i=1}^{N}S_i(\omega)H_i(\omega)$. Assuming
homogeneous spike-train statistics,
\begin{eqnarray}
  \label{eq:homogeneity}
  \begin{aligned}
    S(\omega)&:=\EW[i]{S_i(\omega)} && \text{(mean)}\\
    V(\omega)&:=\EW[i]{|S_i(\omega)-S(\omega)|^2} && \text{(variance)}\\
    C(\omega)&:=\EW[i \neq j]{\left(S_i(\omega)-S(\omega)\right)\left(S_j(\omega)-S(\omega)\right)\cc} && \text{(covariance)}
    \,,
  \end{aligned}
\end{eqnarray}
the (squared) signal-to-noise ratio of the readout
signal $Y(\omega)$ is given by
\begin{equation}
  \label{eq:sn2}
  \SN^2(\omega):=\frac{|\EW{Y(\omega)}|^2}{\EW{|Y(\omega)-\EW{Y(\omega)}|^2}}
  =\frac{|S|^2 |\bar{H}_1|^2}{N^{-1}(1-\kappa) V \bar{H}_2 
    + \kappa V |\bar{H}_1|^2}
  \,.
\end{equation}
Here, $\kappa(\omega)=C(\omega)/V(\omega)$ denotes the spike-train
coherence. The coefficients $\bar{H}_1:= \EW[i]{H_i}$ and $\bar{H}_2:=
\EW[i]{|H_i|^2}$ represent the 1st- and 2nd-order filter
statistics. For uncorrelated spike trains, i.e.~$\kappa(\omega)=0$,
and $S(\omega)\ne 0$, the signal-to-noise ratio $\SN^2$ grows
unbounded with the population size $N$. Thus, even for noisy spike
trains ($V>0$), the compound signal $y(t)$ can be highly reliable if
the population size $N$ is sufficiently large. In the presence of
correlations, $\kappa(\omega)\ne 0$, however, $\SN^2$ converges
towards a constant value $\kappa^{-1}|S|^2V^{-1}$ as $N$
grows. Even for large populations, the readout signal remains prone to
noise.
These findings constitute a generalization of the results reported for
population-rate coding, i.e.~sums of unweighted spike counts
\citep[see e.g.][]{Zohary94_140,Shadlen98}. The above arguments
illustrate that the same reasoning applies to coding schemes which are
based on the spatio-temporal structure of spike patterns.
\par
\begin{figure}[ht!]
  \centering
  \doublecolumnfigure{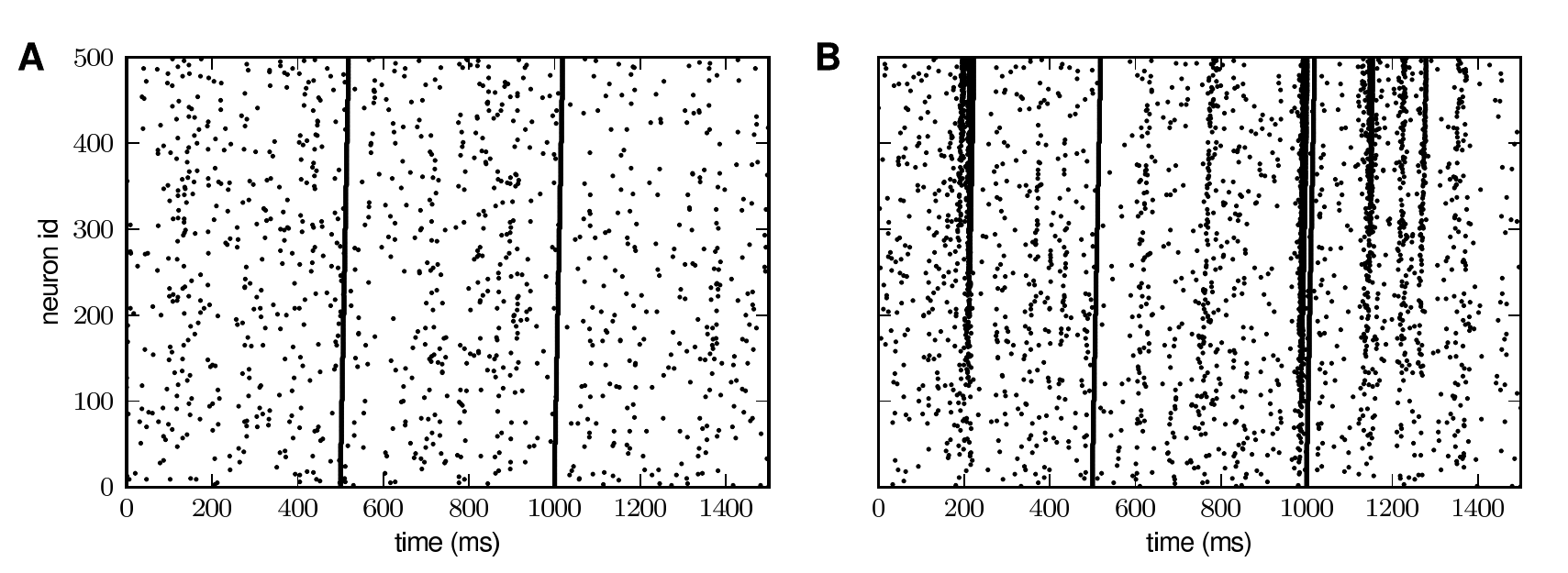}%
  \caption{Recurrent network dynamics stabilizes dynamics of embedded
    synfire chains.  Spiking activity in a synfire chain ($L=10$
    layers, layer width $b=50$) receiving background input from an
    excitatory-inhibitory network ({\bf A}, cf.~\prettyref{fig:intro}C) or from a finite pool of
    excitatory and inhibitory Poisson processes ({\bf B}, cf.~\prettyref{fig:intro}D). Average
    input firing rates, in-degrees and amount of shared input are
    identical in both cases. Neurons of the first synfire layer
    (neuron ids $1,\ldots,b$) are stimulated by current pulses at
    times $t=500$ and $1000\,\ms$.  Each neuron in layer $k\in[2,L]$ receives
    inputs from all $b$ neurons in the preceding layer $k-1$ (synaptic
    weights $J_\sfc=0.8\,\mV$, spike transmission delays
    $d_\sfc=2\,\ms$), and $K-b$ and $\gamma K$ excitatory and
    inhibitory background inputs, respectively, randomly drawn from the presynaptic populations.
    Neurons in the first layer $k=1$ receive $K$ and $\gamma K$ excitatory and inhibitory 
    background inputs, respectively. 
    Note that there is no feedback from the synfire chain to the embedding network.
    See \prettyref{tab:lif_parameters} for network parameters.
  }
  \label{fig:synfire_embedding}
\end{figure}

In a previous study \citep{Tetzlaff04_117}, we demonstrated that
active decorrelation in recurrent networks is a necessary
prerequisite for a controlled propagation of synchronous volleys of
spikes in embedded feedforward subnetworks ('synfire chains';
\prettyref{fig:synfire_embedding}): A synfire chain receiving
background input from a finite population of independent Poisson
sources amplifies the resulting shared-input correlations, thereby
leading to spontaneous synchronization within the chain
(\prettyref{fig:synfire_embedding}B). A distinction between these
spurious synchronous events and those triggered by an external
stimulus is impossible. The synfire chain loses its asynchronous
ground state \citep{Tetzlaff02_673}. A synfire chain receiving background inputs from a
recurrent network, in contrast, is much more robust. Here,
shared-input correlations are actively suppressed by the
recurrent-network dynamics. Synchronous events can be triggered by
external stimuli in a controlled manner
(\prettyref{fig:synfire_embedding}A).
Apart from the spontaneous synchronization illustrated in
\prettyref{fig:synfire_embedding}, decorrelation by inhibition might
solve another problem arising in embedded synfire structures: In the
presence of feedback connections between the synfire chain and the
embedding background network, synchronous spike volleys can excite
(high-frequency) oscillatory modes in the background network which, in
turn, interfere with the synfire dynamics and prevent a robust
propagation of synchronous activity within the chain \citep['synfire
explosion', see][]{Mehring01a,Aviel03a}. The decorrelation mechanism
we refer to in our work is efficient only at low frequencies. It
cannot prevent the build-up of these oscillations.
\citet{Aviel05_691} demonstrated that the 'synfire explosion' can be
suppressed by adding inhibitory neurons to each synfire layer ('shadow
inhibition') which diffusely project to neurons in the embedding
network, thereby weakening the impact of synfire activity on the
embedding network.
\par
In the present work we focus on the integral of
the correlation function, nurtured by our interest in the low-frequency
fluctuations. An analog treatment can however easily be performed for the
zero-lag correlations.  In contrast to infinite networks with sparse
connectivity ($N\rightarrow\infty$, $K=\text{const}$), in the case of
finite networks, pairs of neurons must be distinguished according to
whether they are synaptically connected or not in order to arrive at a
self-consistent theory for the averaged correlations.  Providing
explicit expressions for correlations between connected and
unconnected neurons, the current work provides the tools to relate
experimentally observed spiking correlations to the underlying
synaptic connectivity.
\par
The quantification of pairwise correlations is a necessary
prerequisite to understand how correlation sensitive synaptic
plasticity rules, like spike-timing dependent plasticity \citep{Bi98},
interact with the recurrent network dynamics
\citep{Gilson09_1}. Existing theories quantifying correlations employ
stochastic neuron models and are limited to purely excitatory networks
\citep{Burkitt07_533, Gilson09_1, Pfister10_22}.
Here, we provide an analytical equivalence relation between a reduced
linear model and spiking integrate-and-fire neurons describing
fluctuations correctly up to linear order. A formally similar approach
has been employed earlier to study delayed cumulative inhibition in
spiking networks \citep{Lindner05_061919}. We show that the
correlations observed in recurrent networks in the asynchronous
irregular regime are quantitatively captured for realistic synaptic
coupling with postsynaptic potentials of up to about $1\,\mV$.
The success of this approach can be explained by the linearization of
the neural threshold units by the afferent noise experienced in the
asynchronous regime. For linear neural dynamics, the second-order
description of fluctuations is closed \citep{Buice09_377}.  We exploit
this finding by applying perturbation theory to the Fokker-Planck
description of the integrate-and-fire neuron to obtain the linear
input-output transfer at low frequencies \citep{Helias10_1000929},
thereby determining the effective coupling in our linear model.
\par
The scope of the theory presented in the current work is limited
mainly by three assumptions. The first is the use of a linear theory
which exhibits an instability as soon as a single eigenvalue of the
effective connectivity matrix assumes a positive real part.  This
ultimately happens when increasing the synaptic coupling strength,
because the eigenvalues of the random connectivity matrix are located
in a circle centered in the left half of the complex plain with a
radius given by the square root of the variance of the matrix elements
\citep{Sommers88, Rajan06}. Nonlinearities, like those imposed by
strictly positive firing rates, prevent such unbounded growth (or
decay) by saturation.
For nonlinear rate models with sigmoidal transfer functions it has
been shown that the activity of recurrent random networks of such
units makes a transition to chaos at the point where the linearized
dynamics would loose stability \citep{Sompolinsky88_259}.  However,
this point of transition is sharp only in the limit of infinitely
large networks. From the population averaged firing rate and the
pairwise correlations averaged over pairs of neurons considered in
\prettyref{fig:cov_ndim_iaf_wdep} we cannot conclude whether or not a
transition to chaos occurs in the spiking network.  In simulations and
in the linearized reduced model, we could however observe that the
distribution of pairwise correlations broadens when approaching the
point of instability.  Future work needs to examine this question in
detail, e.g.~by considering measures related to the Lyapunov exponent.
Recently developed semi-analytical theories accounting for nonlinear
neural features \citep{Toyoizumi09} may be helpful to answer this
question.
The second limiting factor of the current theory is the use of a
perturbative approach to quantify the response of the
integrate-and-fire model. Although the steady-state firing rate of the
network is found as the fixed point of the nonlinear self-consistency
equation, the response to a synaptic fluctuation is determined up to
linear order in the amplitude of the afferent rate fluctuation, which
is only valid for sufficiently small fluctuations.  For larger input
fluctuations, nonlinear contributions to the neural response can
become more important \citep{Helias10_1000929}.  Also for strong
synaptic coupling, deviations from our theory are to be expected.
Thirdly, the employment of Fokker-Planck theory to determine the
steady-state firing rate and the response to incoming fluctuations
assumes uncorrelated presynaptic firing with Poisson statistics and
synaptic amplitudes which are vanishingly small compared to the
distance between reset and threshold. For larger synaptic amplitudes,
the Fokker-Planck theory becomes approximate and deviations are
expected \citep{sirovich00_2009, Jacobsen07_1330, Richardson10_178102,
  Helias10_1000929}. This can be observed in
\prettyref{fig:cov_ndim_iaf_wdep}A, showing a deviation between the
self-consistent firing rate and the analytical prediction at about
$J\simeq{1}\,\mV$.
In this work, we obtained a sufficiently precise self-consistent
approximation of the correlation coefficient $C/A$ by relating the
random recurrent network of spiking neurons in the asynchronous
irregular state to a reduced linear model which obeys the same
relation between $C$ and $A$ up to linear order.  This reduced linear
model, however, does not predict the absolute values of the variance
$A$ and covariance $C$. The variance $A$ of the LIF model, for
example, is dominated by nonlinear effects, such as the reset
mechanism after each action potential. Previous work \citep{Brunel99,
  Brunel00_183} has shown that the single spike-train statistics can
be approximated in the diffusion approximation if the recurrent firing
rate in the network is determined by mean-field theory.  One may
therefore extend our approach and determine the integral
auto-correlation function as $A = \nu\mathrm{FF}$ with the Fano factor
$\mathrm{FF}$ \citep[see][]{MorenoBote08}. For a renewal process and
long observation times, the Fano factor is given by
$\mathrm{FF}=\mathrm{CV}^2$ \citep{Cox62, Nawrot08_374}. The
coefficient of variation $\mathrm{CV}$ can be obtained from the
diffusion approximation of the membrane-potential dynamics \citep[][
App.\,A.1]{Brunel00_183}. The covariance $C$ can then be determined by
\prettyref{eq:C_EE_C_II_ana_maintext}.
Another possibility is the use of a refractory-density approach
\citep{Chizhov08_011910, Meyer02}.
\par
The spike-train correlation as a function of the time lag
is an experimentally accessible measure.  Future theoretical work
should therefore also focus on the temporal structure of correlations
in recurrent networks, going beyond zero-lag correlations
\citep{Kriener08_2185, Renart10_587} and the integral measures studied
in the current work. This would allow to compare the theoretical
predictions to direct experimental observations in a more detailed
manner.  Moreover, the relative spike timing between pairs of neurons
is a decisive property for Hebbian learning \citep{Hebb49} in
recurrent networks, as implemented by spike timing-dependent
plasticity \citep{Bi98}, and suspected to play a role for synapse
formation and elimination \citep{Helias08_7}.
\par
The simulation experiments performed in this work revealed that the
suppression of correlations is vulnerable to certain types of manipulations of the
feedback loop. One particular biological source of additional
variability in the feedback loop is probabilistic vesicle release at
synapses \citep{Loebel09}.  In feedforward networks, such unreliable
synaptic transmission has been shown to decrease the transmission of
correlations by pairs of neurons \citep{Rosenbaum10_00116}.  Stochastic
synaptic release is very similar to the replacement of the population
activity in the feedback branch by a rate modulated Poisson processes
that conserves the population rate. In these simulations we observed
an increase of correlations due to the additional noise caused by the
stochastic Poisson realization.  Future work should investigate more
carefully which of the two opposing effects of probabilistic release
on correlations dominates in recurrent networks.
\par
The results of our study do not only shed light on the decorrelation
of spiking activity in recurrent neural networks. They also
demonstrate that a standard modeling approach in theoretical
neuroscience is problematic: When studying the dynamics of a local
neural network (e.g.~a ``cortical column''), it is a common strategy
to replace external inputs to this neural population $\mathcal{P}$ by
spike-train ensembles with some predefined statistics, e.g.~by
stationary Poisson processes. Most neural systems, however, exhibit a
high degree of recurrence. Nonlocal input to the population
$\mathcal{P}$, i.e.~input from other brain areas, therefore has to be
expected to be shaped by the activity within $\mathcal{P}$. The
omission of these feedback loops can lead to qualitatively wrong
predictions of the population statistics.  The analytical results for
the correlation structure of recurrent networks presented in this
study provide the means to a more realistic specification of such
external activity.

\section{Methods}
\label{sec:methods}
\subsection{LIF network model}
\label{sec:lif_network}
In the present study, we consider two types of sparsely connected
random networks: networks with purely inhibitory coupling (``I
networks'') and networks with both excitatory and inhibitory
interactions (``EI networks''). To illustrate the main findings of
this study and to test the predictions of the linear model described
in \prettyref{sec:linear_model}, both architectures were implemented
as networks of leaky integrate-and-fire (LIF) neurons. The model
details and parameters are reported in
\prettyref{tab:lif_network_model} and \prettyref{tab:lif_parameters},
respectively. All network simulations were carried out with NEST
(\href{http://www.nest-initiative.org}{www.nest-initiative.org}).
\def\tabspace{0.5ex}
\begin{table}[ht!]

  \begin{tabularx}{\linewidth}{|p{0.15\linewidth}|X|}
    \hline\modelhdr{2}{A}{Model summary}\\\hline
    \bf Populations & one (inhibitory network) or two (excitatory-inhibitory network)\\\hline
    \bf Connectivity & random, fixed in-degrees\\\hline
    \bf Neuron & leaky integrate-and-fire (LIF)\\\hline
    \bf Synapse &current based, delta-shaped postsyn.~currents with constant amplitudes\\\hline
    \bf Input & uncorrelated Gaussian white noise currents\\
    \hline
  \end{tabularx}\vspace{\tabspace}

  \begin{tabularx}{\linewidth}{|p{0.15\linewidth}|p{0.15\linewidth}|X|}
    \hline\modelhdr{3}{B}{Populations}\\
    \hline\nettypehdr{3}{}{Inhibitory network}\\\hline
    \bf Name & \bf Elements & \bf Size\\\hline
    $\In$    & LIF & $N=K/\epsilon$\\
    \hline\nettypehdr{3}{}{Excitatory-inhibitory network}\\\hline
    \bf Name & \bf Elements & \bf Size\\\hline
    $\Ex$    & LIF & $N_\Ex=K/\epsilon$\\
    $\In$    & LIF & $N_\In=\gamma N_\Ex=\gamma K/\epsilon$\\    
    \hline
  \end{tabularx}\vspace{\tabspace}

  \begin{tabularx}{\linewidth}{|p{0.15\linewidth}|p{0.15\linewidth}|X|}
    \hline\modelhdr{3}{C}{Connectivity}\\
    \hline\nettypehdr{3}{}{Inhibitory network}\\\hline
    \bf Source & \bf Target & \bf Pattern\\\hline
    $\In$      & $\In$      & random convergent $K\to 1$, delay $d$, weight $-J$\\
    \hline\nettypehdr{3}{}{Excitatory-inhibitory network}\\\hline
    \bf Source & \bf Target & \bf Pattern\\\hline
    $\Ex$      & $\Ex$      & random convergent $K\to 1$, delay $d$, weight $J$\\
    $\Ex$      & $\In$      & random convergent $K\to 1$, delay $d$, weight $J$\\
    $\In$      & $\Ex$      & random convergent $\gamma K\to 1$, delay $d$, weight $-gJ$\\
    $\In$      & $\In$      & random convergent $\gamma K\to 1$, delay $d$, weight $-gJ$\\
    \hline
  \end{tabularx}\vspace{\tabspace}

  \begin{tabularx}{\linewidth}{|p{0.15\linewidth}|X|}
    \hline\modelhdr{2}{D}{Neuron}\\\hline
    \bf Type & Leaky integrate-and-fire \citep[LIF;][]{Tuckwell88a}\\
    \bf Description & 
    Dynamics of membrane potential $V_i(t)$ ($i\in[1,N]$):
    \begin{itemize}
      \setlength{\itemsep}{0pt}
    \item[-] Spike emission at times $t^i_k$ with $V_i(t^i_k)\ge\threshold{}$
    \item[-] Subthreshold dynamics:
    \item[]
      $\tauM\dot{V_i}=-V_i+\RM I_i(t)$ \quad if
      $\forall k:\,t\notin(t^i_k,t^i_k+\tauR]$
    \item[-] Reset + refractoriness: 
      $V_i(t)=\resetpot$ \quad if $\forall k:\,t\in(t^i_k,t^i_k+\tauR]$
    \end{itemize}\\    
    & Exact integration \citep{Rotter99a} with temporal resolution $\dt$\\
    & Initial membrane-potential distribution at $t=0$: uniform between $0$ and $\threshold$\\
    \hline
  \end{tabularx}\vspace{\tabspace}

  \begin{tabularx}{\linewidth}{|p{0.15\linewidth}|X|}
    \hline\modelhdr{2}{E}{Synapse}\\\hline
    \bf Type  & Current synapse with $\delta$-shaped
    postsyn.~currents (PSCs)\\
    \bf Description &
    Input current of neuron $i$:
    \begin{math}
      I_i(t) = C_m \sum_j J_{ij}\sum_l \delta(t-t^j_l-d) + I_{i,\ext}(t)
    \end{math}\\
    & Static synaptic weights $J_{ij}$ (see Connectivity)\\
    \hline
  \end{tabularx}\vspace{\tabspace}

  \begin{tabularx}{\linewidth}{|p{0.15\linewidth}|X|}
    \hline\modelhdr{2}{F}{Input}\\\hline
    \bf Type  & uncorrelated Gaussian white noise $\RM I_{i,\ext}(t) = \mu_\ext + \sqrt{\tauM} \xi_i(t)$ (for $i\in[1,N]$)\\
    \bf Description & mean $\mu_\ext = \RM \EW[t]{I_{i,\ext}(t)}$, auto-correlation $\RM ^2 \EW[t]{I_{i,\ext}(t) I_{j,\ext}(t+\tau)} = \mu_\ext^2 + \eta^2 \tauM \delta_{ij} \delta(\tau)$\\
    & in discrete time $t\in\{n\cdot\dt|n\in\mathbb{N}\}$, $\xi(n \cdot \dt)$ piecewise constant within time interval $\dt$, value drawn independently for each time point from a normal distribution with zero mean and standard deviation $1/\sqrt{\dt}$
    \\
    \hline
  \end{tabularx}\vspace{\tabspace}

   \caption{LIF network: Model overview}
  \label{tab:lif_network_model}
\end{table}

\begin{table}[ht!]
  \def\tabspace{0.5ex}
  \begin{tabularx}{\textwidth}{|p{0.15\linewidth}|p{0.3\linewidth}|X|}
    \hline\parameterhdr{3}{A}{Connectivity}\\\hline
    \bf Name & \bf Value & \bf Description\\\hline
    $K$ & $1250$ (inhibitory network)& in-degree \\
        & $1000$ (E-I network) & excitatory in-degree \\
    $\epsilon$ & $0.1$ & network connectivity \\
    $\gamma=N_\In/N_\Ex$ & $1/4$ (E-I network) & relative size of inhibitory subpopulation \\
    \hline
  \end{tabularx}\vspace{\tabspace}
  \begin{tabularx}{\textwidth}{|p{0.15\linewidth}|p{0.3\linewidth}|X|}
    \hline\parameterhdr{3}{B}{Neuron}\\\hline
    \bf Name & \bf Value & \bf Description\\\hline
    $\RM$ & $80\,\MOhm$ & membrane resistance\\
    $\tauM$ & $20\,\ms$ & membrane time constant\\
    $\tauR$ & $2\,\ms$ & refractory period \\
    $\resetpot$ & $0\,\mV$ & reset potential \\
    $\threshold$ & $15\,\mV$ & spike threshold \\
    \hline
  \end{tabularx}\vspace{\tabspace}
  \begin{tabularx}{\textwidth}{|p{0.15\linewidth}|p{0.3\linewidth}|X|}
    \hline\parameterhdr{3}{C}{Synapse}\\\hline
    \bf Name & \bf Value & \bf Description\\\hline
    $J$ & $0.2 \,\mV$ & EPSP amplitude\\
    $g$ & $6$ (E-I network) & relative IPSP amplitude \\
    $d$ & $0.1\,\ms$ & synaptic delay \\
    \hline
  \end{tabularx}\vspace{\tabspace}
  \begin{tabularx}{\textwidth}{|p{0.15\linewidth}|p{0.3\linewidth}|X|}
    \hline\parameterhdr{3}{D}{Input}\\\hline
    \bf Name & \bf Value & \bf Description\\\hline
    $\mu_\ext$ & $1.5 \, \threshold$ & mean external GWN input\\
    $\eta$ & $0.3\, \threshold$ & SD of external GWN input\\
    \hline
  \end{tabularx}\vspace{\tabspace}
  \begin{tabularx}{\textwidth}{|p{0.15\linewidth}|p{0.3\linewidth}|X|}
    \hline\parameterhdr{3}{E}{Simulation}\\\hline
    \bf Name & \bf Value & \bf Description\\\hline
    $T$ & $10$ or $100\,\seconds$ & simulation time \\
    $\dt$ & $0.1\,\ms$ & time resolution \\
    \hline
  \end{tabularx}
  \caption{LIF network: Parameters (default values)}
  \label{tab:lif_parameters}
\end{table}


\subsection{Linearized network model}
\label{sec:linear_model}
In this section we show how the dynamics of the spiking network can be
reduced to an effective linear model whose fluctuations, by
construction, fulfill the same relationship as the original system up
to linear order. We first outline the major steps of this reduction,
and then provide the formal derivation.
\par
We make use of the observation that the effect of a single synaptic
impulse on the output activity of a neuron is typically small. Writing
the response spike train of a neuron as a functional of the history of
all incoming impulses therefore allows us to perform a linearization
with respect to each of the afferent spike trains.  Formally, this
corresponds to a Volterra expansion up to linear order, the
generalization of a Taylor series to functionals.
In \prettyref{sec:w_of_j}, we perform this linearization explicitly
for the example of the LIF model. This determines how the linear
response kernel depends on the parameters of the LIF model.
The linear dependence on the input leads to an approximate convolution
equation \prettyref{eq:convolution_equation_C_A} linearly connecting
the auto- and the cross-correlation functions in the network. As this
equation is complicated to solve directly, we introduce a reduced
linear model \prettyref{eq:linearized_dynamics} obeying the same
convolution equation. The reduced linear model can be solved by
standard Fourier methods and yields an explicit form for the
covariance matrix in the frequency domain
\prettyref{eq:covariance_matrix}.
The diagonal and off-diagonal elements of the $N=N_\Ex+N_\In$
dimensional covariance matrix $\mat{C}(\omega)$ in
\prettyref{eq:int_corr_lin} correspond to the power-spectra of
individual neurons and the cross-spectra of individual neuron pairs,
respectively. As, in this linear approximation, both the auto- and the
cross-covariances are proportional to the variance of the driving
noise, the resulting correlation coefficients are independent of the
noise amplitude (see \prettyref{sec:mean_field_corr}).
As shown in \prettyref{sec:I_poprate_model} and
\prettyref{sec:EI_poprate_model}, the suppression of fluctuations in
recurrent networks is most pronounced at low frequencies. It is
therefore sufficient to restrict the discussion to the zero-frequency
limit $\omega \rightarrow 0$. Note that the zero-frequency variances
and covariances correspond to the integrals of the auto- and
cross-correlation functions in the time domain.
In this limit, we may combine the two different sources of
fluctuations caused by the spiking of the neurons and by external
input to the network into a single source of white noise
\prettyref{eq:linearized_dynamics_lowomega} with variance $\rho^2$.
\par
In general, the spiking activity $s_i(t)$ of neuron $i$ at time $t$ is
determined by the entire history $\{ \s(\tpr)| \tpr < t \}$ of the
activity of all neurons $\s = (s_1, \ldots, s_N)$ in the network up to
time $t$.
Formally, this dependence can be expressed by a functional
\begin{equation}
s_{i}(t) = G^i_{t}[\s(t^{\prime})].
\end{equation}
The subscript $t$ in $G^i_{t}$ indicates that $t^\prime<{}t$
(causality).  In the following, we will use the abbreviation
$G^i_{t}[\s]\equiv{}G^i_{t}[\s(t^{\prime})]$.
The effect of a single synaptic input on the state of a neuron is
typically small. We therefore approximate the influence of an incoming
spike train on the activity of the target neuron up to linear
order. The sensitivity of neuron $i$'s activity to the input from
neuron $k$ can be expressed by the functional derivative of $G^i_t$
with respect to input spike train $s_k$:
\begin{equation}
  \label{eq:functional_derivative}
  \frac{\delta G_t^i[\s]}{\delta s_k(\tppr)} 
  =\lim_{\epsilon \rightarrow 0} \frac{1}{\epsilon} 
  \left(G_t^i[\s+\epsilon\, \delta(\circ-\tppr)\, \e_{k}]-G_{t}^{i}[\s]\right).
\end{equation}
It represents the response of the functional to a single
$\delta$-shaped perturbation in input channel $k$ at time $\tppr$,
normalized by the perturbation amplitude $\epsilon$. In
\prettyref{eq:functional_derivative},
$\e_{k}=(0,\ldots,0,1,0,\ldots,0)$ denotes the unity vector with
elements $e_{k_k}=1$ and $e_{k_i}=0$ for all $i\ne{}k$.  By
introducing the vector $\s_{\hat{k}}(t)=(s_1(t), \ldots,s_{k-1}(t),0,
s_{k+1}(t), \ldots, s_N(t))$ of spike trains with the $k$-th component
set to zero, $G^i_{t}[\s]$ can be approximated by
\begin{equation}
  \label{eq:linearized_functional}
G_{t}^{i}[\s] 
\simeq\sum_{k=1}^{N}\int_{-\infty}^{t}\frac{\delta G_t^i[\s_{\hat{k}}]}{\delta s_{k}(\tppr)} \, s_k(\tppr)\, d\tppr.
\end{equation}
Eq.~\prettyref{eq:linearized_functional} is a Volterra expansion up to
linear order, the formal extension of a Taylor expansion of a function
of $N$ variables to a functional, truncated after the linear term.
With the linearized dynamics \prettyref{eq:linearized_functional}, the
pairwise spike-train cross-correlation function between two neurons
$i$ and $j\ne{}i$ is given by
\begin{eqnarray}
  \label{eq:correlation}
  \begin{aligned}
    c_{ij}(\tau) & = \langle s_i(t+\tau)\,\tilde{s}_j(t)\rangle_{\s}\\
    & = \left\langle G_{t+\tau}^i[\s]\,\tilde{s}_j(t)\right\rangle _{\s}\\
    & = \sum_{k=1}^N \int_{-\infty}^{t+\tau} \left\langle \frac{\delta
        G_{t+\tau}^i [\s_{\hat{k}}]} {\delta s_k(\tppr)} \,
      \left\langle s_k (\tppr) \, \tilde{s}_j(t) \right\rangle_{s_k}
    \right\rangle _{\s \backslash s_k} \, d\tppr
    \qquad(\forall{}\tau>0).
  \end{aligned}
\end{eqnarray}
Note that \prettyref{eq:correlation} is valid only for positive time
lags $\tau>0$, because for $\tau < 0$ a possible causal influence of
$s_i$ on $s_j$ is not expressed by the functional.  Here,
$\langle\cdot\rangle_{\s}$ denotes the average across the ensemble of
realizations of spike trains in the stationary state of the network
(e.g.~the ensemble resulting from different initial conditions), and
$\tilde{\s}(t)=\s(t)-\langle\s\rangle_{\s}$ the centralized (zero
mean) spike train. In the last line in \prettyref{eq:correlation}, the
average
$\langle\cdot\rangle_{\s}=\langle\langle\cdot\rangle_{s_k}\rangle_{\s
  \backslash s_k}$ is split into the average $\langle\cdot\rangle_{\s
  \backslash s_k}$ across all realizations of spike trains excluding
$s_k$ and the average $\langle\cdot\rangle_{s_k}$ across all
realizations of $s_k$. Note that the latter does not affect the
functional derivative because it is, by construction, independent of
the actual realization of $s_{k}$.
A consistent approximation up to linear order is equivalent to the
assumption that for all $j$ the linear dependence of the functional on
$s_j$ is completely contained in the respective derivative with
respect to $s_j$ \prettyref{eq:linearized_functional}.  Dependencies
beyond linear order include higher-order derivatives and are neglected
in this approximation. This is equivalent to neglecting the dependence
of $\frac{\delta G_{t+\tau}^{i}[\s_{\hat{k}}]}{\delta s_k(\tppr)}$ on
$s_{j}$ for any $j\neq k$. Hence, we can average the inner term over
$s_{k}$ and $s_{j}$ separately. In the stationary state, this
correlation can only depend on $\tppr-t$ and equals the auto- or the
cross-correlation function:
\begin{eqnarray*}
\left\langle s_k(\tppr) \, \tilde{s}_j(t) \right\rangle_{s_k,s_j} & = & \begin{cases}
a_k(\tppr-t) & \text{for } k=j\\
c_{kj}(\tppr-t) & \text{for } k\neq j.
\end{cases}
\end{eqnarray*}
The pairwise spike-train correlation function is therefore given by
\begin{eqnarray*}
c_{ij}(\tau) & = & \sum_{k=1}^N \int_{-\infty}^{t+\tau}\, d\tppr \left\langle \frac{\delta G_{t+\tau}^i[\s_{\hat{k}}]}{\delta s_k(\tppr)} \right\rangle _\s \,\begin{cases}
a_k(\tppr-t) & \text{for } k=j\\
c_{kj}(\tppr-t) & \text{for } k\neq j
\end{cases}
\qquad(\forall{}\tau>0),
\end{eqnarray*}
where we used the fact that $\langle f[s_{\hat{k}}] \rangle_{\s
  \backslash s_k} = \langle f[s_{\hat{k}}] \rangle_\s$ for any
functional $f$ that does not depend on $s_{k}$.  The average of the
functional derivative has the intuitive meaning of a response kernel
with respect to a $\delta$-shaped perturbation of input $s_{k}$ at
time $\tppr$. Averaged over the realizations of the stationary network
activity this response can only depend on the relative time
$t+\tau-\tppr$. In a homogeneous random network, the input statistics
(number of synaptic inputs and synaptic weights) and the parameters of
the internal dynamics are identical for each cell, so that the
temporal shape $h(t)$ of the response kernel can be assumed to be the
same for all neurons. The synaptic coupling strength from neuron $k$
to neuron $i$ determines the prefactor $w_{ik}$:
\begin{eqnarray}
w_{ik} h(t+\tau-\tppr) & \equiv & \left\langle \frac{\delta G^i_{t+\tau}[\s_{\hat{k}}]}{\delta s_k(\tppr)} \right\rangle_\s. \label{eq:functional_derivative_kernel}
\end{eqnarray}
In this notation, the linear equation connecting the auto-correlations
$a_{k}$ and the cross-correlations $c_{ij}$ takes the form
\begin{equation}
  \label{eq:convolution_equation_C_A}
  c_{ij}(\tau) = \sum_{k=1}^{N}w_{ik}\int_{-\infty}^{\tau}dt\, h(\tau-t)\,
  \begin{cases}
    a_{k}(t) & \text{for }k=j\\
    c_{kj}(t) & \text{for }k\neq j
  \end{cases}
  \qquad(\forall{}\tau>0).
\end{equation}
\par
Our aim is to find a simpler model which is equivalent to the LIF
dynamics in the sense that it fulfills the same equation
\prettyref{eq:convolution_equation_C_A}.  Let's $\vec{u}(t)$ denote
the vector of dynamic variables of this reduced model. Analog to the
original model, we define the cross-correlation for $i\neq j$ and
$\tau>0$ as
\begin{eqnarray}
c_{ij}^u(\tau) & = & \langle u_i(t+\tau)\,\tilde{u}_j(t)\rangle_{\vec{u}} \nonumber \\
 & = & \left\langle L_{t+\tau}^i[\vec{u}] \, \tilde{u}_j(t) \right\rangle_{\vec{u}}.\label{eq:correlation_r}
\end{eqnarray}
The simplest functional $L_{t}^{i}[\vec{u}]$ consistent with equation
\prettyref{eq:convolution_equation_C_A} is linear in $\vec{u}$. Since
we require equivalence only with respect to the ensemble averaged
quantities, i.e. $c_{ij}^{u}(\tau)=c_{ij}(\tau)$, the reduced activity
and therefore $L_{t}^{i}[\vec{u}]$ can contain a stochastic element
which would disappear after averaging. The linear functional
\begin{eqnarray}
u_i(t) = L_t^i[\vec{u}] & = & \sum_{k=1}^N w_{ik} \int_{-\infty}^t h(t-\tpr) \, u_k(\tpr) \, d\tpr + z_i (t) \label{eq:effective_reduced_model}
\end{eqnarray}
with a pairwise uncorrelated, centralized white noise $z_i(t)$
($\langle z_i(t+\tau)z_k(t) \rangle_\bz = \rho_z^2 \delta_{ij}
\delta(\tau)$) fulfills the requirement, since for $\tau>0$ and $i\neq
j$
\begin{eqnarray*}
c_{ij}^u(\tau) = \langle u_i(t+\tau) \, \tilde{u}_j(t) \rangle_{\vec{z}} & = & \sum_{k=1}^N w_{ik} \int_{-\infty}^{t+\tau} h(t+\tau-\tpr) \left\langle u_k(\tpr) \, \tilde{u}_j(t) \right\rangle_{\vec{z}} \, d\tpr + \underbrace{\langle z_i(t+\tau) \, \tilde{u}_j(t)\rangle_{\vec{z}}}_{=0}\\
 & = & \sum_{k=1}^N w_{ik} \int_{-\infty}^{\tau} dt\, h(\tau-t)
\begin{cases}
  a_k^u(t) & \text{for } k=j\\
  c_{kj}^u(t) & \text{for } k\neq j
\end{cases}. \label{eq:convoution_equation_c_a_u}
\end{eqnarray*}
This equation has the same form as
\prettyref{eq:convolution_equation_C_A}, so both models, within the
linear approximation, exhibit an identical relationship between the
auto- and cross-covariances.
The physical meaning of the noise $\vec{z}(t)$ is the variance caused
by the spiking of the neurons. The auto-correlation function of a
spike train of rate $\nu$ has a $\delta$-peak of weight $\nu$.  The
reduced model \prettyref{eq:effective_reduced_model} exhibits such a
$\delta$-peak if we set $\rho_z^{2}=\nu$. A related approach has been
pursued before \citep[see Sec.~3.5 in][]{Brunel99} to determine the
auto-correlation of the population averaged firing rate. This
similarity will be discussed in detail below.
\par
So far, we considered a network without external drive, i.e. all spike
trains $\s(t)$ originated from within the network.  If the network is
driven by external input, each neuron receives, in addition, synaptic
input $y_i$ from neurons outside the network. We assume uncorrelated
external drive $\langle y_i(t+\tau) \rangle y_j(t) \rangle = \rho^2_y
\delta_{ij} \delta(\tau)$.  In the reduced model, this input
constitutes a separate source of noise:
\begin{eqnarray}
u_i(t) = L_t^i[\vec{u}, \vec{y}] & = & \sum_{k=1}^N w_{ik} (u_k \ast h) + (y_i \ast h_{iy}) + z_i (t) \label{eq:effective_reduced_model_external}.
\end{eqnarray}
Here, $(f*g)(t)=\int_{-\infty}^{t}\dt'\,f(t')g(t-t')$ denotes the
convolution and $h_{iy}(t)$ the response kernel with respect to an
external input.  For simplicity, let's assume that the shape of these
kernels is identical for all pairs of pre- and postsynaptic sources,
i.e.~$h_{ix}(t)=h(t)$. If we further absorb the synaptic amplitude of
the external drive in the strength of the noise $\rho_y$, the
linearized dynamics \prettyref{eq:effective_reduced_model_external}
can be written in matrix notation
\begin{equation}
  \label{eq:linearized_dynamics}
  \vec{u}(t)=([\mat{W}\vec{u}+\vec{y}]*h)(t) + \vec{z}(t)
\end{equation}
with $\mat{W}=\{w_{ij}\}$.
The reduced model \prettyref{eq:linearized_dynamics} can be solved
directly by means of Fourier transform:
\begin{eqnarray}
\vec{U}(\omega) & = & [\mathbf{1}-\mat{W} H(\omega)]^{-1} (H(\omega) \vec{Y}(\omega) + \vec{Z}(\omega)). \label{eq:solution_U}
\end{eqnarray}
The full covariance matrix follows by averaging over the sources of
noise $\vec{Z}$ and $\vec{Y}$ as
\begin{eqnarray}
  \mat{C}^u(\omega) & = & \langle \vec{U}(\omega)\vec{U}^{T}(-\omega) \rangle_{\vec{Z}, \vec{Y}} \nonumber \\
  & = & (\rho_z^2 + |H(\omega)|^2 \rho_y^2) [\mathbf{1}-\mat{W} H(\omega)]^{-1}[\mathbf{1}-\mat{W}^T H(-\omega)]^{-1}.\label{eq:covariance_matrix}
\end{eqnarray}
The diagonal elements of $\mat{C}^u$ represent the auto-covariances,
the off-diagonal elements the cross-covariances. Both are proportional
to the driving noise $\rho_z^2 + |H(\omega)|^2 \rho_y^2$.  This is
consistent with \prettyref{eq:convolution_equation_C_A} which is a
linear relationship between the cross- and auto-covariances.
%
\par
For networks which can be decomposed into homogeneous subpopulations,
the $N$ dimensional system \prettyref{eq:linearized_dynamics} can be
further simplified by population averaging. Consider, for example, a
homogeneous random network with purely inhibitory coupling. Assume
that the neurons are randomly connected with probability $\epsilon$ and
coupling strength $-w<0$. The average number of in/outputs per neuron
(in/out-degree) is thus given by $K=\epsilon{}N$. By introducing the
population averaged external input $y(t)=\EW[i]{y_i(t)}$, the averaged
spiking noise $z(t) = \EW[i]{z_i(t)}$, and the effective coupling
strength $\w=K w$, the dynamics of the population averaged activity becomes
\begin{equation}
  \label{eq:mean_field_inh}
  u(t)
  =\EW[i]{u_i(t)}
  =\left(\left[\sum_j \EW[i]{w_{ij}} u_j+\EW[i]{y_i}
    \right]*h\right)(t) + \EW[i]{z_i(t)}
  =([-\w u+y]*h)(t) + z(t)
  \,.
\end{equation}
Here we assumed that $\EW[i]{w_{ij}}$ is independent of the
presynaptic neuron $j$ and can be replaced by
$-\epsilon{}w=-\w/N$. Note that this replacement is exact for networks
with homogeneous out-degree, i.e.~if the number of outgoing connections
is identical for each neuron $j$.  For large random networks with
binomially distributed out-degrees (e.g.~Erd\H{o}s-R\'{e}nyi networks
or random networks with constant in-degree),
\prettyref{eq:mean_field_inh} serves as an approximation.
\par
To relate our approach to the treatment of finite-size fluctuations in
\citep{Brunel99}, consider the population-averaged dynamics
\prettyref{eq:mean_field_inh} of a single population with mean firing
rate $\nu$.  We set $\rho_z^2 = \nu$ for all single neuron noises
$z_i$ in order for the reduced model's auto-covariances to reproduce
the $\delta$-peak of the spiking dynamics.  In the population averaged
dynamics, this leads to the variance of the noise $z(t)$ given by
$\langle z(t+\tau)\,
z(t)\rangle=\frac{1}{N}\rho_z^{2}\delta(\tau)=\frac{\nu}{N}\delta(\tau)$.
This agrees with the variance of the population rate in
\citep{Brunel99}.  Therefore, the dynamics of the population averaged
quantity $u$ in \prettyref{eq:mean_field_inh} agrees with the earlier
definition of a population averaged firing rate $s(t) = \frac{1}{N}
\sum_i s_i(t)$ for the spiking network \citep{Brunel99}.
\par
In equation \prettyref{eq:mean_field_inh}, two distinct sources of
noise appear: The noise due to external uncorrelated activity $y$ and
the noise $z$ which is required to obtain the $\delta$-peak of the
auto-correlation functions of the reduced model. The qualitative
results of \prettyref{sec:I_poprate_model} and
\prettyref{sec:EI_poprate_model}, however can be understood with an
even simpler model.  As we are mainly concerned with the low-frequency
fluctuations, we only need a model that has the same limit $\omega
\rightarrow 0$.  As we normalized the kernel so that $H(0)=1$ we
can combine both sources of noise and require $X(0) \equiv Y(0) +
Z(0)$ in \prettyref{eq:solution_U} in the zero frequency limit. Hence,
in \prettyref{sec:I_poprate_model} and
\prettyref{sec:EI_poprate_model}, we consider the model
\begin{eqnarray}
  \vec{r}(t) &=& ([-\mat{W} \vec{r} + \vec{x}] \ast h)(t) \label{eq:linearized_dynamics_lowomega}
\end{eqnarray}
with a pairwise uncorrelated centralized white noise
$\EW[x]{x_i(t+\tau) x_j(t)} = \rho^2 \delta_{ij} \delta(\tau)$ to
explain the suppression of fluctuations at low frequencies.
%
%
\par
As a second example, consider a random network composed of an
excitatory and an inhibitory subpopulation $\Exc$ and $\Inh$ with
population sizes $N_\Ex=|\Exc|$ and $N_\In=|\Inh|=\gamma N_\Ex$,
respectively. Assume that each neuron receives excitatory and
inhibitory inputs from $\Exc$ and $\Inh$ with coupling strengths $w>0$
and $-gw<0$, respectively, and probability $\epsilon$, such that the
average excitatory and inhibitory in/out-degrees are given by
$K=\epsilon{}N$ and $\gamma{}K$, respectively. The dynamics of the
subpopulation averaged activities
$\vec{u}(t)=(u_\Ex(t),u_\In(t))\transp$ is given by
\prettyref{eq:linearized_dynamics} with subpopulation averaged noise
$\vec{y}(t)=(y_\Ex(t),y_\In(t))\transp$ and
$\vec{z}(t)=(z_\Ex(t),z_\In(t))\transp$ and effective coupling
\begin{equation}
  \label{eq:mean_field_exc_inh}
  \mat{W}=\w
  \begin{pmatrix}
    1 & -\g\\
    1 & -\g
  \end{pmatrix}
  \,.
\end{equation}
Here, $\w=Kw$ denotes the effective coupling strength, $\g=\gamma g$
the effective balance parameter and $y_{\Ex/\In}(t)=\EW[i\in
  \Exc/\Inh]{y_i(t)}$ and $z_{\Ex/\In}(t)=\EW[i\in \Exc/\Inh]{z_i(t)}$
the (sub)population averaged external and spiking sources of noise,
respectively. Again, the reduction of the $N$-dimensional linear
dynamics to the two-dimensional dynamics
\prettyref{eq:mean_field_exc_inh} is exact if the out-degrees are
constant within each subpopulation.
As before, both sources of noise can be combined into a single source
of noise, if the we are only interested in the low-frequency behavior
of the model, leading to the dynamics
\prettyref{eq:linearized_dynamics_lowomega} with the effective
coupling \prettyref{eq:mean_field_exc_inh}.

\par
The linear theory is only valid in the domain of its stability, which
is determined by the eigenvalue spectrum of the effective coupling
matrix $\mat{W}$. For random coupling matrices, the eigenvalues are
located within a circle with a radius equal to the square root of the
variance of the matrix entries \citep{Rajan06} $\sqrt{\Var{w_{ij}}} =
w \sqrt{N\epsilon (1-\epsilon) (1+\gamma g^2)}$.  Writing the effective
dynamics for the exponential kernel as a differential equation $\tau
\frac{\partial\vec{r}}{\partial t} = (\mat{W}-\mat{1}) \vec{r} +
\vec{x}(t)$, the eigenvalues of the right hand side matrix $\mat{W} -
\mat{1}$ are confined to a circle centered at $-1$ in the complex
plain with radius $\sqrt{\Var{w_{ij}}}$.
%
%
Given $\Var{w_{ij}} > 1$, eigenvalues might exist which have a
positive real part, leading to unstable dynamics.  This condition is
indicated by the vertical dotted lines in
\prettyref{fig:cov_ndim_wdep}A-F and
\prettyref{fig:cov_ndim_iaf_wdep}B-D near $J=2.8\,\mV$. Beyond this
line, the linear model predicts an explosive growth of
fluctuations. In the LIF-network model, an unbounded growth is avoided
by the nonlinearities of the single-neuron dynamics.

\subsection{Response kernel of the LIF model}
\label{sec:w_of_j}
\begin{figure}[ht!]
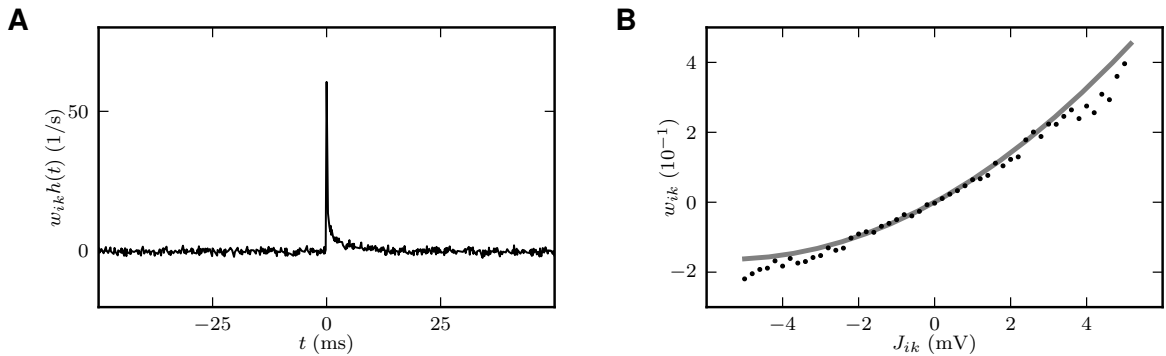

  \centering
  \doublecolumnfigure{\figpath/w_of_J}
  \caption{{\bf A}: Firing-rate deflection $w_{ik} h(t)$ of a LIF neuron
    caused by an incoming spike event of postsynaptic amplitude
    $J_{ik} = 0.6\,\mV$.  {\bf B}: Integral $w_{ik}= w_{ik} \int_0^\infty
    h(t) \; dt$ of the firing rate deflection shown in A as a
    function of the postsynaptic amplitude $J_{ik}$ (simulation: black
    dots; analytical approximation \prettyref{eq:w_of_J}: gray
    curve). The neuron receives constant synaptic background input
    with $J=0.1\,\mV$, $g=4$, and rates $\nu_\Ex = 5960 \, \frac{1}{\mathrm{s}}$,
    $\nu_\In = 1190 \, \frac{1}{\mathrm{s}}$ resulting in a first and second
    moment \prettyref{eq:siegert}
    $\bar{\mu}_i = 12 \,\mV$ and $\bar{\sigma}_i = 5 \,\mV$.
    Simulation results are obtained by averaging over $1000$ trials of $10 \, \seconds$
    duration each with $25000$ input impulses on average. For
    further parameters of the neuron model, see \prettyref{tab:lif_network_model} and
    \prettyref{tab:lif_parameters}.  }
  \label{fig:w_of_J}
\end{figure}

We now perform the formal linearization
\prettyref{eq:functional_derivative_kernel} for a network of $N$ LIF
neurons $i=1,\ldots,N$. A similar approach has been employed in
previous studies to understand the population dynamics in these
networks \citep{Brunel99, Brunel00_183}.
We consider the input $\sum_j J_{ij} s_j(t)$ received by neuron $i$
from the local network, where $s_j$ denotes the spike train of the
neuron $j$ projecting to neuron $i$ with synaptic weight $J_{ij}$.
Given the time dependent firing rate $\nu_j(t)$ of each afferent, and
assuming small correlations and small synaptic weights, the total
input to neuron $i$ can be replaced by a Gaussian white noise with
mean $\mu_i(t)$ and variance $\sigma_i^2(t)$,
\begin{eqnarray}
  \mu_i(t) &=& \tau_m \sum_j J_{ij} \nu_j(t) \label{eq:mu_sigma} \nonumber \\
  \sigma_i^2(t) &=& \tau_m \sum_j J_{ij}^2 \nu_j(t),
\end{eqnarray}
where $j$ sums over all synaptic inputs. $J_{ij} \in \{J, -gJ\}$
denotes the amplitude of the postsynaptic potential evoked by synapse
$j\to{}i$. $\tau_m$ is the membrane time constant of the model.  In
the stationary state, the firing rate of each afferent is well
described by the constant time average $\bar \nu_j = \EW[t]{\nu_j}$.
The working point at which we perform the linearization of the
neural response \prettyref{eq:functional_derivative_kernel} is then
given by analog equations as \prettyref{eq:mu_sigma}, resulting in a
constant mean $\bar \mu_i = \tau_m \sum_j J_{ij} \bar \nu_j$ and
variance $\bar \sigma^2_i = \tau_m \sum_j J_{ij}^2 \bar \nu_j$.  If
the amplitude of each postsynaptic potential is small compared to the
distance of the membrane potential to threshold, the dynamics of the
LIF model can be approximated by a diffusion process, employing
Fokker-Planck theory \citep{Risken96}.  The stationary firing rate of
the neuron is then given by \citep{Siegert51, Brunel99, Brunel00_183}
\begin{eqnarray}
  \bar \nu_i^{-1}(\bar \mu_i, \bar \sigma_i) &=& \tau_\text{ref} + \sqrt{\pi} \tau_m (F(y_{\theta,i}) - F(y_{r,i})) 
  \label{eq:siegert} \nonumber \\
  \text{with}\nonumber && \\
  F(y) &=& \int^{y} f(y) \; dy \quad f(y) = e^{y^2} (\text{erf}(y) + 1) \nonumber \\
  y_{\theta,i} &=& \frac{\theta - \bar \mu_i}{\bar \sigma_i} \quad y_{r,i} = \frac{V_\text{reset} - \bar \mu_i}{\bar \sigma_i},
\end{eqnarray}
with the reset voltage $V_\text{reset}$, the threshold voltage
$\theta$ and the refractory time $\tau_\text{ref}$.  In homogeneous
random networks, the stationary rate
(\prettyref{fig:cov_ndim_iaf_wdep}A) is the same for all neurons. It
is determined in a self-consistent manner \cite{Brunel00_183} as the
fixed point of \prettyref{eq:siegert}.  The stationary mean $\bar
\mu_i$ and variance $\bar \sigma_i^2$ are determined by the stationary
rate.
To determine the kernel \prettyref{eq:functional_derivative_kernel} we
need to consider how a $\delta$-shaped deflection in the input to this
neuron at time point $t^\prime$ affects its output up to linear order
in the amplitude of the fluctuation. In the stationary state, we may
set $t^\prime = 0$. It is therefore sufficient to focus
on the effect of a single fluctuation
\begin{equation}
  s_k(t) = a \delta(t). \label{eq:rate_deflection}
\end{equation}
We therefore ask how the density of spikes per time 
$\nu_i(t) = \langle G^i_{t}[\s] \rangle_{\s \backslash s_k}$ of neuron $i$, averaged
over different realizations of the remaining inputs to neuron $i$,
changes in response to the fluctuation \prettyref{eq:rate_deflection}
of the presynaptic neuron $k$ in the limit of vanishing amplitude $a$.
This kernel $w_{ik} h$ \prettyref{eq:functional_derivative_kernel} is identical
to the impulse response of the neuron and can directly be measured in simulation
by trial averaging over many responses to the given $\delta$-deflection
\prettyref{eq:rate_deflection} in the input (see \prettyref{fig:w_of_J}A).
For the theory of low-frequency fluctuations, we only need the
integral of the kernel, also known as the DC susceptibility,
\begin{eqnarray}
  w_{ik} &=& w_{ik} \int_0^\infty h(t) \; dt \label{eq:effective_w} \\ \nonumber
  &=& \lim_{a \rightarrow 0} \frac{\bar \nu_i(\bar \mu_i + \delta \mu_i, \bar \sigma_i + \delta \sigma_i) - \bar \nu_i(\bar \mu_i, \bar \sigma_i)}{a} \\ \nonumber
  &=& \frac{\partial \bar \nu_i}{\partial \bar \mu_i} \tau_m J_{ik} + \frac{\partial \bar \nu_i}{\partial \bar \sigma_i} \frac{\tau_m}{2 \bar \sigma_i} J_{ik}^2 + O(a^2).
\end{eqnarray}
The second equality follows from the equivalence of the integral of the
impulse response and the step response in linear approximation
\citep{DeLaRocha07_802, Helias10_1000929}. Following from \prettyref{eq:mu_sigma}, both mean
and variance are perturbed as $\delta \mu_i = a \tau_m J_{ik}$
and $\delta \sigma_i^2 = a \tau_m J^2_{ik}$ in response to a step $a$
in the afferent rate $\nu_j$. Moreover, we used the chain rule
$\delta \sigma_i = \frac{1}{2\bar \sigma_i}\delta \sigma_i^2$.
The variation of the afferent firing rate hence co-modulates the mean
and the variance and both modulations need to be taken into account
to derive the neural response \citep{Brunel99}.
Although the finite amplitude of postsynaptic potentials has an effect
on the response properties \citep{Helias10_1000929, Richardson10_178102},
the integral response is rather insensitive to the granularity of the
noise \citep{Helias10_1000929}.  We therefore employ the diffusion
approximation to linearize the dynamics of the LIF neuron around its
working point characterized by the mean $\bar \mu_i$ and the variance
$\bar \sigma_i^2$ of the total synaptic input.
In \prettyref{eq:effective_w}, we evaluate the partial derivatives of
$\bar \nu_i$ with respect to $\bar \mu_i$ and $\bar \sigma^2_i$ using
\prettyref{eq:siegert}.  First, observe that by chain rule
$\frac{\partial \bar \nu_i}{\partial \bar \mu_i}=-\bar \nu_i^2
\frac{\partial \bar \nu_i^{-1}}{\partial \bar \mu_i}$.  We then again
make use of the chain rule $\frac{\partial \bar \nu_i^{-1}}{\partial
  \bar \mu_i} = \frac{\partial \bar \nu_i^{-1}}{\partial y_{\theta,i}}
\frac{\partial y_{\theta,i}}{\partial \bar \mu_i} + \frac{\partial
  \bar \nu_i^{-1}}{\partial y_{r,i}} \frac{\partial y_{r,i}}{\partial
  \bar \mu_i}$.  Analog expressions hold for the derivative with
respect to $\bar \sigma_i$.
The first derivative yields
$\frac{\partial \bar \nu_i^{-1}}{\partial y_{\theta,i}} = \sqrt{\pi} \tau_m f(y_{\theta,i})$,
the one with respect to $y_{r,i}$ follows analogously, but with a negative sign.
We further observe that $\frac{\partial y_A}{\partial \bar \mu_i} = \frac{-1}{\bar \sigma_i}$
and $\frac{\partial y_A}{\partial \bar \sigma_i} = \frac{-y_A}{\bar \sigma_i}$ with $y_A \in \{ y_{r,i}, y_{\theta,i} \}$.
Taken together, we obtain the explicit result for \prettyref{eq:effective_w} 
\begin{equation}
  w_{ik} = (\bar \nu_i \tau_m)^2 \sqrt{\pi} \frac{J_{ik}}{\bar \sigma_i} \left(f(y_{\theta,i}) (1 + \frac{J_{ik}}{2 \bar \sigma_i} y_{\theta,i}) - f(y_{r,i}) (1 + \frac{J_{ik}}{2 \bar \sigma_i} y_{r,i}) \right). \label{eq:w_of_J}
\end{equation}
Note that the modulation of $\mu_i$ results in a contribution to
$w_{ik}$ that is linear in $J_{ik}$, whereas the modulation of
$\sigma_i$ causes a quadratic dependence on $J_{ik}$. This expression
therefore presents an extension to the integral response presented in
\cite{DeLaRocha07_802, Helias10}.
%
\prettyref{fig:w_of_J}B shows the comparison of the analytical
expression \prettyref{eq:w_of_J} and direct simulation.  The agreement
is good over a large range of synaptic amplitudes $J_{ik} \in [-4, 4]
\,\mV$ in the case of constant background noise caused by small
synaptic amplitudes (here $0.1\,\mV$ for excitation and $-0.4\,\mV$
for inhibition). For background noise caused by stronger impulses, the
deviations are expected to grow \citep{Helias10_1000929}.
%


\subsection{Population-activity spectra in the linear model: feedback
  vs.~feedforward scenario}
\label{sec:linear_model_spectra}
The recurrent linear neural dynamics defined in the previous section
is conveniently solved in the Fourier domain.  The driving external
Gaussian white noise $\vec{X}$ is mapped to the response
$\vec{R}(\omega) = \mat{T}(\omega) \vec{X}(\omega)$ by means of the
transfer matrix $\mat{T}(\omega)$. According to
\prettyref{eq:linearized_dynamics_lowomega}, it is given by
$\mat{T}(\omega) = H(\omega)\left(\eye-H(\omega)\mat{W}\right)^{-1}$.
The covariance matrix in the frequency domain, the spectral matrix,
thus reads
\begin{equation}
  \mat{C}_{RR}(\omega) =  \EW[x] { \vec{R}(\omega)\vec{R}(\omega)\adj }= \mat{T}(\omega) \rho^2 \eye \mat{T}(\omega)\adj
  \,, \label{eq:rate_spectrum}
\end{equation}
where we used $\EW[x]{X X\adj} = \rho^2 \eye$ and
the expectation operator $\EW[x]{}$ represents an average over noise realizations.
%
%
To identify the effect of recurrence on the network dynamics, we
replace the local feedback input by a feedforward input $\vec{Q}$ with
spectral matrix $\mat{C}_{QQ}$. The resulting response firing rate is
given by $\tilde{\vec{R}} = H (\mat{W} \vec{Q} + \vec{X})$.
%
%
Assuming that the feedforward input $\vec{Q}$ is uncorrelated to the external noise source
$\vec{X}$ ($\mat{C}_{QX} = 0$) yields a response spectrum
\begin{equation}
  \mat{C}_{\tilde{R}\tilde{R}}  
  = \EW[x] { \tilde{\vec{R}}\tilde{\vec{R}}\adj }
  = |H|^2 \left(\mat{W}\mat{C}_{QQ}\mat{W}\adj + \rho^2 \eye \right)
  \,. 
  \label{eq:power_spectrum_feed_forward_uncorr}
\end{equation}

\subsection{Population-activity spectrum of the linear inhibitory network} 
\label{sec:spectr_1D}
In the Fourier domain, the solution of the mean-field dynamics
\prettyref{eq:mean_field_inh} of the inhibitory network is $R(\omega)
= H(\omega) X(\omega) / (1 + \w H(\omega))$.
The power-spectrum $C_{RR}(\omega) = \EW[x]{R(\omega) R(\omega)\cc}$ hence becomes
\begin{equation}
  C_{RR}(\omega) = \frac{|H(\omega)|^2}{|1 + \w H(\omega)|^2} \rho^2,
\end{equation}
using the spectrum of the noise $\EW[x]{X(\omega) X(\omega)\cc} = \rho^2$.
\par
We compare this power-spectrum to the case where the feedback loop is
opened, i.e.~where the recurrent input is replaced by feedforward
input with unchanged auto-statistics $C_{QQ}(\omega) =
C_{RR}(\omega)$, but which is uncorrelated to the external input
$C_{QX}(\omega)=0$. The resulting power-spectrum is given by
\prettyref{eq:power_spectrum_feed_forward_uncorr} as
$C_{\tilde{R}\tilde{R}} = |H|^2 \left(\w^2 C_{RR} + \rho^2 \right)$.

\subsection{Population-activity spectra of the linear excitatory-inhibitory network} 
\label{sec:spectr_2D}
%
%
%
In a homogeneous random network of excitatory and inhibitory neurons,
the population averaged activity
\prettyref{eq:mean_field_exc_inh} can be solved in the Schur basis
\prettyref{eq:schur_dynamics} introduced in
\prettyref{sec:EI_poprate_model}
\begin{eqnarray}
R_+(\omega) &=& H(\omega) \frac{H(\omega) w_\FF X_-(\omega) + X_+(\omega)}{1+H(\omega) w_+}\label{eq:schur_solution} \\ \nonumber
R_-(\omega) &=& H(\omega) X_-(\omega),
\end{eqnarray}
with $w_+ = -\w (1-\g)$ and $w_\FF = \w (1 + \g)$.  The power of the
population rate therefore is
\begin{eqnarray}
\label{eq:pop_power}
\begin{aligned}
C_{R_+R_+}(\omega) 
&=\frac{\rho^2|H(\omega)|^2}{2N_\Ex}\cdot\frac{|H(\omega)w_\FF+1|^2+\gamma^{-1}|H(\omega)w_\FF-1|^2}{ |1 + H(\omega) w_+|^2}\\
C_{R_-R_-}(\omega) 
&=\frac{\rho^2|H(\omega)|^2}{2N_\Ex}  (1+\gamma^{-1})
\,.
\end{aligned}
\end{eqnarray}
The fluctuations of the excitatory and the inhibitory population
follow as
\begin{eqnarray}
R_{\Ex/\In}(\omega) &=& \frac{1}{\sqrt{2}}\frac{H(\omega)}{1 + H(\omega) w_+} X_+(\omega) \label{eq:spectrum_EI} \\ \nonumber
 &+& \frac{1}{\sqrt{2}} \left( \frac{H(\omega) w_\FF}{1+H(\omega) w_+} \pm 1 \right) H(\omega) X_-(\omega).
\end{eqnarray}
So the power-spectra are
\begin{eqnarray}
  \label{eq:power_spectrum_EI} 
  \begin{aligned}
    C_{R_\Ex R_\Ex}(\omega)
    &=\frac{|H(\omega)|^2\rho^2}{N_\Ex}\cdot
    \frac{|1+\w\g{}H(\omega)|^2+\gamma^{-1}(\w\g)^2|H(\omega)|^2}{|1+H(\omega)w_+|^2}\\
    C_{R_\In R_\In}(\omega)
    &=\frac{|H(\omega)|^2\rho^2}{N_\Ex}\cdot
    \frac{\w^2|H(\omega)|^2+\gamma^{-1}|1-\w{}H(\omega)|^2}{|1+H(\omega)w_+|^2}\\
    C_{R_\Ex R_\In}(\omega)
    &=\frac{|H(\omega)|^2\rho^2}{N_\Ex}\cdot
    \w \frac{\w \g (1+\gamma^{-1})|H(\omega)|^2 + H^\ast(\omega) - \g \gamma^{-1} H(\omega)}{|1+H(\omega)w_+|^2}
    \,.
  \end{aligned}
\end{eqnarray}
\par
Replacing the recurrent input of the sum activity $R_+$ by activity
$Q_+$ with the same auto-statistics, but which is uncorrelated to the remaining input into
$R_+$ (\prettyref{fig:FB_vs_FF_schur}D') results in the fluctuations
\begin{eqnarray}
  \tilde{R}_+(\omega) &=& H(\omega) \left( -w_+ Q_+(\omega) + w_\FF \tilde{R}_-(\omega) + X_+(\omega) \right)\,,\\ \nonumber
  \tilde{R}_-(\omega) &=& H(\omega) X_-(\omega)\,.
\end{eqnarray}
The power-spectrum of the sum activity therefore becomes
\begin{eqnarray}
  \label{eq:C_tilde_R_plus_FF2}
  \begin{aligned}
  C_{\tilde{R}_+\tilde{R}_+}(\omega) 
  &= |H(\omega)|^2 \left[ w_+^2C_{R_+ R_+}+ \frac{\rho^2}{2N_\Ex}\left(|H(\omega)w_\FF+1|^2+\gamma^{-1}|H(\omega)w_\FF-1|^2\right)\right]\,.    
  \end{aligned}
\end{eqnarray}
\par
If, alternatively, the excitatory and the inhibitory feedback terms
$R_\Ex$ and $R_\In$ are replaced by uncorrelated feedforward input
$Q_\Ex$ and $Q_\In$ with power-spectra $C_{R_\Ex R_\Ex}$ and $C_{R_\In
  R_\In}$ (\prettyref{fig:FB_vs_FF_schur}C,D), 
the spectrum of the sum activity reads
\begin{eqnarray}
  \label{eq:C_tilde_R_plus}
  \begin{aligned}
    C_{\tilde{R}_+ \tilde{R}_+}(\omega) 
    &=|H(\omega)|^2\left[2 \w^2 \left(C_{R_\Ex R_\Ex}(\omega) + \g^2 C_{R_\In R_\In}(\omega)\right)+\frac{\rho^2}{2N_\Ex}(1+\gamma^{-1})\right].     
  \end{aligned}
\end{eqnarray}
The limit \prettyref{eq:power_ratio_2D_a} for inhibition dominated
networks with $\g > 1$ can be obtained from this and the former
expressions by taking $H(0) = 1$ and assuming strong coupling $\w \gg
1$.

\subsection{Population averaged correlations in the linear EI network} 
\label{sec:mean_field_corr}
In this subsection, we derive a self-consistency equation for the
covariances in a recurrent network.  We start from \prettyref{eq:covariance_matrix}
(we drop the superscript $u$ of $C^u$ for brevity) multiply by $\mathbf{1} - \mat{W} H(\omega)$
from left and its transpose from right to obtain
\begin{eqnarray}  
\mat{C}(\omega) &=& H(\omega)\mat{W} \mat{C}(\omega) + \mat{C}(\omega) H(-\omega)\mat{W}\transp \label{eq:int_corr_lin} \\ \nonumber
&-& |H(\omega)|^2\mat{W} \mat{C}(\omega) \mat{W}\transp\ + \eye (|H(\omega)|^2 \rho_y^2 + \rho_z^2) 
\end{eqnarray}
We assume a recurrent network of $N_\Ex$ excitatory and $N_\In$
inhibitory neurons, in which each neuron receives $K$ excitatory
inputs of weight $w$ and $\gamma K$ inhibitory inputs of weight $-gw$
drawn randomly from the presynaptic pool of neurons.  To obtain a
theory for the variances and covariances at zero frequency (with $H(0)
= 1$) we may abbreviate $\rho^2_z(0) + |H(0)|^2 \rho_y^2(0)$ by
$\rho^2(0)$.  For a population averaged theory, we need to replace in
\prettyref{eq:int_corr_lin} the variances $A_i$ of an individual
neuron by the population average and replace the covariance $C_{ij}$
for a given pair of neurons $(i,j)$ by the average over pairs that are
statistically equivalent to $(i, j)$.  For a pair $(i,j)$ of neurons
we will show that the set of equivalent pairs depends on the current
realization of the connectivity since unconnected pairs are not
equivalent to connected ones.  Therefore it is necessary to first
average the covariance matrix over statistically equivalent neuron
pairs given a fixed connectivity and to subsequently average over all
possible realizations of the connectivity.  The latter will be denoted
as $\EW[W]{}$.
For compactness of the notation, first we perform the averaging for
the general case, where neuron $i$ belongs to population $x$ and
neuron $j$ to population $y$.  We denote by $\X$, $\Y$ the sets of
neuron indices belonging to populations $x$ and $y$, respectively.
Subsequently replacing $x$ and $y$ by all possible combinations $x, y
\in \{\Ex, \In\}$, we obtain the averaged self-consistency equations
for the network.  We denote the number of incoming connections to a
neuron of type $x$ from the population of neurons of type $y$ as $K_{x
  y}$ and the strength of a synaptic coupling as $w_{x y}$.  Rewriting
the self-consistency equation \prettyref{eq:int_corr_lin} explicitly
with indices yields
\begin{equation}
  \mat{C}_{ij} = 
  \sum_k w_{ik} C_{kj} + \sum_k w_{jk} C_{ik} - \sum_{k,l} w_{ik} C_{kl} w_{jl} + \rho^2 \delta_{ij}
  \,. 
  \label{eq:C_ij_self_consistent}
\end{equation}
The last equation shows that for a connected pair $(i,j)$ of neurons
($w_{ij} \neq 0$ or $w_{ji} \neq 0$) either of the first two sums
contains a contribution $w_{ij}C_{jj}$ or $w_{ji}C_{ii}$ proportional
to the variance of the projecting neuron.  We therefore need to
perform the averaging separately for connected and for unconnected
pairs of neurons. We use the notation
\begin{equation}
  C_{x \la y} = 
  \EW[\mat{W}]{ \frac{1}{N_{\text{pairs}, x \la y}} \sum_{i \in \X, j \in \Y, i \la j} C_{ij} } 
  \label{eq:C_conn_avg}
\end{equation}
for the average covariance over pairs of neurons of types $x, y \in
\{\Ex,\In\}$ with a connection from neuron $j \in \Y$ to neuron $i \in
\X$, where $N_{\text{pairs}, x \la y}$ is the number neuron pairs
connected in this way. An arrow to the right, $i \ra j$, denotes a
connection from neuron $i$ to neuron $j$.  Note that we use the same
letter $C$ for the population averaged covariances and for the
covariances of individual pairs. The distinction can be made by the
indices: $i,j,k,l$ throughout indexes a single neuron, $u,v,x,y,z$
identifies one of the populations $\{\Ex, \In\}$.  We denote the
covariance averaged over unconnected pairs as
\begin{equation}
  C_{x \nlra y} = 
  \EW[\mat{W}] { \frac{1}{N_{\text{pairs}, x \nlra y}} \sum_{i \in \X, j \in \Y, i \nlra j} C_{ij} } 
  \,.
  \label{eq:C_unconn_avg}
\end{equation}
We further use
\begin{equation}
 A_x = \EW[\mat{W}] { \frac{1}{N_x} \sum_{i \in \X} C_{ii} } \label{eq:A_avg}
\end{equation}
for the integrated variance averaged over all neurons of type
$x$.  Connected and the unconnected averaged covariances differ by the
term proportional to the variance of the projecting neuron, as
mentioned above
\begin{eqnarray}
  C_{x \la y} &=& C_{x \nlra y} + w_{xy} A_y \label{eq:C_conn_unconn} \\ \nonumber
  C_{x \ra y} &=& C_{x \nlra y} + w_{yx} A_x
  \,.
\end{eqnarray}
As a consequence, we can express all quantities in terms of the
averaged variance \prettyref{eq:A_avg} and the covariance
averaged over unconnected pairs \prettyref{eq:C_unconn_avg}.  We now
proceed to average the integrated variance over population
$x$. Since there are no self-connections in the network, we do not
need to distinguish two cases here.  Replacing $C_{ii}$ on the right
hand side of \prettyref{eq:A_avg}, the first term of
\prettyref{eq:C_ij_self_consistent} contributes
\begin{eqnarray}
  A_{x,\text{A}} 
  &=& \EW[\mat{W}] { \frac{1}{N_x} \sum_{i \in \X} \sum_{k \in \Exc \vee \Inh} w_{ik} C_{ki} } 
  = \EW[\mat{W}] { \frac{1}{N_x} \sum_{i \in \X} \left( \sum_{k \in \Exc} w_{ik} C_{ki} + \sum_{k \in \Inh} w_{ik} C_{ki} \right) } 
  \label{eq:A_avg_term_I} 
  \\ \nonumber
  &=& \sum_{z \in \{\Ex, \In\}} K_{xz} w_{xz} C_{z \rightarrow x} 
  = \sum_{z \in \{\Ex, \In\}} K_{xz} w_{xz} (C_{z \nlra x} + w_{xz} A_{z})
  \,.
\end{eqnarray}
From the second to the third step we used that the sum over $k$ ($l$)
yields non-zero contributions only if neuron $k$ ($l$) connects to
neuron $i$. This happens in $K_{x\Ex}$ ($K_{x\In}$) cases with the
coupling weight $w_{x\Ex}$ ($w_{x\In}$). Therefore the covariance
averaged over connected pairs appears on the right hand side.  In the
last line we used the relation \prettyref{eq:C_conn_unconn} to express
the connected covariance in terms of the variance and the
covariance over unconnected pairs. The second term in
\prettyref{eq:A_avg} is identical because of the symmetry $C_{ik} =
C_{ki}$. Up to here, the structure of the network only entered in
terms of the in-degree of the neurons.  The contribution of the third
term follows from a similar calculation
\begin{eqnarray}
  && \EW[\mat{W}] { \frac{1}{N_x} \sum_{i \in \X} \sum_{k,l \in \Exc \vee \Inh} w_{ik} C_{kl} w_{il} } 
  \label{eq:A_avg_term_III} \\ \nonumber
  &=& \EW[\mat{W}] { \frac{1}{N_x} \sum_{i \in \X} \left( \sum_{k \neq l \in \Exc \vee \Inh} w_{ik} C_{kl} w_{il} + \sum_{k \in \Exc \vee \Inh} w_{ik}^2 C_{kk} \right) } \\ \nonumber
  &=& \sum_{u,v \in \{\Ex, \In\}} K_{xu} K_{xv} w_{xu} w_{xv} \left( \frac{K_{wu}}{N_u}(C_{v \la u} - C_{v \nlra u}) + \frac{K_{uv}}{N_v}(C_{v \ra u} - C_{v \nlra u}) + C_{v \nlra u} \right)\\ \nonumber 
  &+& \sum_{z \in \{\Ex, \In\}} K_{xz} w_{xz}^2 A_z \\ \nonumber
  &=& \sum_{u,v \in \{\Ex, \In\}} K_{xu} K_{xv} w_{xu} w_{xv} C_{vu} + \sum_{z \in \{\Ex, \In\}} K_{xz} w_{xz}^2 A_z
  \,.
\end{eqnarray}
From the second to the third step we assumed that among the $K_{xz}
K_{xw}$ pairs of neurons $k \in \Z, l \in \W$ projecting to neuron
$i$, the fraction $\frac{K_{wz}}{N_z}$ has a connection $k \ra
l$. These pairs contribute with the connected covariance. The
connections in opposite direction contribute the other term of similar
structure.  We ignore multiple and reciprocal connections here,
assuming the connection probability is low. We introduce the shorthand
$C_{xy}$ for the covariance averaged over all neuron pairs including
connected and unconnected pairs
\begin{equation}
  C_{xy} = C_{x \nlra y} + w_{xy} \frac{K_{xy}}{N_y} A_y + w_{yx} \frac{K_{yx}}{N_x} A_x. \label{eq:avg_cov}
\end{equation}
This is the covariance which is observed on average when picking a
pair of neurons of type $x$ and $y$ randomly.  In this step, beyond
the in-degree, the structure of the network entered through the
expected number of connections between two populations. Taken all
three terms together, we arrive at
\begin{eqnarray}
  A_x &=& \rho^2 + \sum_{z \in \{\Ex, \In\}} K_{xz} w_{xz} (2 C_{z \nlra x} + w_{xz} A_z) - C_{xx,\text{corr}} \\ \nonumber       
  C_{xy,\text{corr}} &\defeq& \sum_{u,v \in \{\Ex, \In\}} K_{xu} K_{yv} w_{xu} w_{yv} C_{uv}
  \,.
\end{eqnarray}
The averaged covariances follow by similar calculations. Here we
only need to calculate the average over unconnected pairs $(i,j)$
given by \prettyref{eq:C_unconn_avg}, because the connected covariance
follows from \prettyref{eq:C_conn_unconn}. The first sum in
\prettyref{eq:C_ij_self_consistent} contributes
\begin{eqnarray}
  C_{xy,\text{A}} &\defeq& \EW[\mat{W}] { \frac{1}{N_{\text{pairs}, x \nlra y}} \sum_{i \in \X, j \in \Y, i \neq j, i \nlra j} \sum_{k \in \Exc \vee \Inh} w_{ik} C_{kj} } 
  \label{eq:C_avg_term_I} \\ \nonumber
  &=& \sum_{z \in \{\Ex, \In\}} K_{xz} w_{xz} C_{zy},
\end{eqnarray}
where due to the absence of a direct connection between $i$ and $j$,
the term linear in the coupling and proportional to the
variance is absent. From the symmetry $C_{kl} = C_{lk}$ it
follows that the second term corresponds to an exchange of $x$ and $y$
in the last expression.  The third sum in
\prettyref{eq:C_ij_self_consistent} follows from an analog calculation
as before
\begin{eqnarray}
  C_{xy,\text{B}} &\defeq& \EW[\mat{W}] { \frac{1}{N_{\text{pairs}, x \nlra y}} \sum_{i \in \X, j \in \Y, i \neq j, i \nlra j} \sum_{k,l \in \Exc \vee \Inh} w_{ik} C_{kl} w_{jl} } \label{eq:C_avg_term_III} \\ \nonumber
  &=& \sum_{z \in \{\Ex, \In\}} w_{xz} w_{yz} \frac{K_{xz}K_{yz}}{N_z} A_z + C_{xy,\text{corr}}
  \,.
\end{eqnarray}
In summary, the contributions from \prettyref{eq:C_avg_term_I} and
\prettyref{eq:C_avg_term_III} together result in the self-consistency
equation for the covariance
\begin{eqnarray}
  C_{x \nlra y} &=& C_{xy,\text{A}} + C_{yx,\text{A}} - C_{xy,\text{B}}
  \,. 
  \label{eq:self_consistent_C}
\end{eqnarray}
We now simplify the expressions by assuming that the in-degree of a
neuron and the incoming synaptic amplitudes do not depend on the type
of the neuron, i.e. that excitatory and inhibitory neurons receive
statistically the same input.  Formally this means that we need to
replace $K_{xy}$ by $K_y$, the number of incoming connections from
population $y$ and $w_{xy}$ by $w_y$, the coupling strength of a
projection from a neuron of type $y$.  The covariance $C_{x
  \nlra y}$ then has two distinct contributions, $C_{xy,\text{sep}}$
that depends on the type of neurons $x,y$, and $C_\text{com}$ that
does not. In particular $C_{xy,\text{B}}$ and $C_{xy,\text{corr}}$ do
not depend on $x, y$ and we omit their subscripts in the
following. The variances fulfill
\begin{eqnarray}
  A_x &=& A_{x,\text{sep}} + A_\text{com} + \rho^2\\ \nonumber
  A_{x,\text{sep}} &=& \sum_{u \in \{\Ex, \In\}} 2 K_u w_u C_{u \nlra x} \\ \nonumber
  A_{\text{com}} &=& \sum_{u \in \{\Ex, \In\}} K_u w_u^2 A_u - C_\text{corr}
  \,,
\end{eqnarray}
the covariances satisfy
\begin{eqnarray}
  C_{x \nlra y} &=& C_{xy,\text{sep}} + C_\text{com}\\ \nonumber
  C_{xy,\text{sep}} &=& \sum_{u \in \{\Ex, \In\}} K_u w_u \left( \frac{K_x}{N_x} w_x A_x + \frac{K_y}{N_y} w_y A_y + C_{u \nlra y} + C_{u \nlra x} \right)\\ \nonumber
  C_\text{com} &=& \sum_{u \in \{\Ex, \In\}} \frac{K_u^2}{N_u} w_u^2 A_u - C_\text{corr} \\ \nonumber
  C_\text{corr} 
  &=& \sum_{u,v \in \{\Ex, \In\}} 2 K_v w_v \frac{K_u^2}{N_u} w_u^2 A_u +  K_u K_v w_u w_v C_{u \nlra v}
  \,.
\end{eqnarray}
The disjoint part $C_{xy,\text{sep}}$ determines the difference
between the covariances for pairs of neurons of different type. Using
the parameters $K_{\Ex} = K$, $K_\In = \gamma K$, $w_\Ex = w$, $w_\In
= -gw$, the explicit form is
\begin{eqnarray}
  C_{\Ex \Ex,\text{sep}} &=& 2 K w^2 (1 - \gamma g) \frac{K}{N_\Ex} A_\Ex + 2 K w C_{\Ex \nlra \Ex} - 2 \gamma g K w C_{\Ex \nlra \In} \label{eq:CEI_avg} \\ \nonumber
  C_{\In \In,\text{sep}} &=& -2 K g w^2 (1 - \gamma g) \frac{\gamma K}{N_\In} A_\In + 2 \gamma g K w C_{\In \nlra \In} + 2 K w C_{\Ex \nlra \In} \\ \nonumber
  C_{\Ex \In,\text{sep}} &=& \frac{1}{2} (C_{\Ex \Ex, \text{sep}} + C_{\In \In, \text{sep}})
  \,.
\end{eqnarray}
Therefore, also the covariances in the network obey the relation
\begin{equation}
  C_{\Ex \nlra \Ex} + C_{\In \nlra \In} = 2 C_{\Ex \nlra \In}, \label{eq:C_EI_arith}
\end{equation}
i.e. the mixed covariance can be eliminated and is given by the
arithmetic mean of the covariances between neurons of same type.
%
%
In matrix representation with the vector $\vec{Q} = (A_E, A_I, C_{\Ex
  \nlra \Ex}, C_{\In \nlra \In})$, the self-consistency equation is
\begin{eqnarray}
  \mat{M_\text{dis}} &=& K w \begin{pmatrix}
    0 & 0 & 2-\gamma g & -\gamma g\\
    0 & 0 & 1 & 1 - 2 \gamma g\\
    2w(1 - \gamma g)\frac{K}{N_\Ex} & 0 & 2-\gamma g & -\gamma g\\  
    0 & -2gw(1 - \gamma g)\frac{\gamma K}{N_\In} & 1 & 1 - 2 \gamma g 
  \end{pmatrix}\label{eq:Mdiss_matrix} \\
  \mat{M_\text{com}} &=& (Kw)^2 \begin{pmatrix}
    \frac{1}{K} & \frac{g^2\gamma}{K} & 0 & 0 \\ 
    \frac{1}{K} & \frac{g^2\gamma}{K} & 0 & 0 \\ 
    \frac{1}{N_\Ex} & \frac{(g\gamma)^2}{N_\In} & 0 & 0 \\ 
    \frac{1}{N_\Ex} & \frac{(g\gamma)^2}{N_\In} & 0 & 0 \\ 
  \end{pmatrix} - M_\FF\label{eq:Mcomm_matrix} \\
  \mat{M_\FF} &=& (Kw)^2 (1 - \gamma g) \begin{pmatrix}
    2 \frac{Kw}{N_\Ex} & 2 (\gamma g)^2 \frac{Kw}{N_\In} & 1 & -\gamma g\\ 
    2 \frac{Kw}{N_\Ex} & 2 (\gamma g)^2 \frac{Kw}{N_\In} & 1 & -\gamma g\\ 
    2 \frac{Kw}{N_\Ex} & 2 (\gamma g)^2 \frac{Kw}{N_\In} & 1 & -\gamma g\\ 
    2 \frac{Kw}{N_\Ex} & 2 (\gamma g)^2 \frac{Kw}{N_\In} & 1 & -\gamma g\\ 
  \end{pmatrix}
  \,.
  \label{eq:MFF_matrix}
\end{eqnarray}
The self consistent covariance can then be obtained by solving the
system of linear equations
\begin{equation}
  (\mat{I} - \mat{M_\text{dis}} - \mat{M_\text{com}}) \vec{Q} = \rho^2 (1,1,0,0)\transp
  \,. 
  \label{eq:cov_eff_4dim}
\end{equation}
The numerical solution shows that the variances for excitatory
and inhibitory neurons are approximately the same, as depicted in
\prettyref{fig:cov_ndim_wdep}A. In the following we therefore assume
$A_\Ex = A_\In = A$ and then solve \prettyref{eq:cov_eff_4dim} for the
covariances.  With the abbreviation $G = \left(\frac{1}{N_\Ex} +
  \frac{(\gamma g)^2}{N_\In} \right)$, the third and fourth line
yields the equation for the covariances
\begin{eqnarray}
  C_{\Ex \nlra \Ex/\In \nlra \In} &=& (Kw)^2 A \left[ G (1 - 2 Kw (1-\gamma g))
  + 2(1-\gamma g) \begin{cases}
      \frac{1}{N_\Ex} & \text{for} \; \Ex \Ex\\  
      \frac{-\gamma g}{N_\In} & \text{for} \; \In \In\\  
    \end{cases} \right] \label{eq:C_EX_IN} \\ \nonumber
  &+& Kw (C_{\Ex \nlra \Ex} - \gamma g C_{\In \nlra \In}) (1 - Kw (1-\gamma g))
  + Kw (1-\gamma g) \begin{cases}
     C_{\Ex \nlra \Ex} & \text{for} \; \Ex \Ex\\
     C_{\In \nlra \In} & \text{for} \; \In \In
  \end{cases}
\end{eqnarray}
The structure of the equation suggests to introduce the linear
combination $m = C_{\Ex \Ex} - \gamma g C_{\In \In}$ which satisfies
\begin{eqnarray}
  m &=& (Kw)^2 (1-\gamma g) G (3 - 2 Kw(1-\gamma g)) A \label{eq:def_m} \\ \nonumber
    &+& Kw (1 - \gamma g) (2 - Kw(1 - \gamma g)) m\\ \nonumber
  m &=& (Kw)^2 G (1-\gamma g) \frac{3 - 2 Kw(1-\gamma g)}{(1 - Kw (1 - \gamma g))^2} A
  \,.
\end{eqnarray}
We solve \prettyref{eq:C_EX_IN} for $C_{\Ex \nlra \Ex}$ and $C_{\In \nlra \In}$
and insert \prettyref{eq:def_m} for $m$ to obtain the covariances as
\begin{eqnarray}
  C_{\Ex \nlra \Ex/\In \nlra \In} &=& (Kw)^2 A \left[ G \frac{1-2 Kw (1-\gamma g)}{1 - Kw (1-\gamma g)} + 2 \frac{1-\gamma g}{1 - Kw (1-\gamma g)} \begin{cases} \label{eq:C_EX_IN_m}
      \frac{1}{N_\Ex} & \text{for} \; \Ex \Ex \label{eq:CEE_CII_ana} \\ \nonumber
      \frac{-\gamma g}{N_\In} & \text{for} \; \In \In
    \end{cases} \right] + Kw m\\
  &=& G \frac{(Kw)^2}{(1 - Kw (1 - \gamma g))^2} A\\ \nonumber
  &+& 2 \frac{Kw (1-\gamma g)}{1 - Kw (1-\gamma g)} A \begin{cases}
    \frac{Kw}{N_\Ex} & \text{for} \; \Ex \Ex\\  
    \frac{-K \gamma w g}{N_\In} & \text{for} \; \In \In
  \end{cases}
  \,.
\end{eqnarray}
The covariance $C_{x \nlra y}$ between unconnected neurons can be
related to the covariance between the incoming currents this pair of
neurons receives. Expressing the self-consistency
\prettyref{eq:self_consistent_C} in terms of the covariances averaged
over connected and unconnected pairs \prettyref{eq:avg_cov} uncovers
the connection
\begin{eqnarray}
  C_{x \nlra y} &=& \sum_{z \in \{\Ex, \In\}} K_z w_z \left( C_{zx} +  C_{zy} \right) 
  \label{eq:cov_unconnected_input_suppr} \\ \nonumber
  &-& \sum_{z \in \{\Ex, \In\}} w_z^2 \frac{K_z^2}{N_z} A_z - \sum_{u,v \in \{\Ex, \In\}} K_u K_v w_u w_v C_{uv} \\ \nonumber
  &=& Kw \left[ C_{\Ex x} + C_{\Ex y} - \gamma g \left( C_{\In x} + C_{\In y} \right) \right]\\ \nonumber
  &-& (Kw)^2 \left[ \frac{1}{N_\Ex} A_\Ex + \frac{(\gamma g)^2}{N_\In} A_\In + C_{\Ex \Ex} - 2 \gamma g C_{\Ex \In} + (\gamma g)^2 C_{\In \In} \right]
  \,.
\end{eqnarray}
This self-consistency equation yields the argument, why the
shared-input correlation $C^\text{in}_\text{shared}$
\prettyref{eq:C_EE_C_II_ana_maintext} cancels the contribution
$C^\inp_\text{corr}$ \prettyref{eq:input_cov} due to spike-train
correlations in the covariance to the input currents (see
\prettyref{fig:cov_ndim_wdep}C,D).  Rewriting
\prettyref{eq:cov_unconnected_input_suppr} in terms of these
quantities results in
\begin{eqnarray}
  && \frac{C_{x \nlra y}}{K w} - \left[ C_{\Ex x} + C_{\Ex y} - \gamma g ( C_{\In x} + C_{\In y} ) \right] 
  \label{eq:input_cov_suppression} \\ \nonumber
  &=&  K w \left[ C^\text{in}_\text{shared}/(Kw)^2 + C^\inp_\text{corr}/(Kw)^2 \right]
  \,.
\end{eqnarray}
If a self-consistent solution with small correlation $|C_{x \nlra y}|,
|C_{xy}| < \varepsilon$ exists, the right hand side of
\prettyref{eq:input_cov_suppression} must be of the same order of
magnitude. The right hand side of this equation has a prefactor $Kw$
which typically is $\gg 1$ (for the parameters in
\prettyref{fig:cov_ndim_wdep}, $Kw$ becomes larger than $1$ for $w >
10^{-3}$).  The first term in the bracket is proportional to the
contribution of shared input, the second term is due to correlations
among pairs of different neurons. Each of these terms is of order
$\varepsilon$. Due to the prefactor $Kw$, however, the sum of the two
terms needs to be of order $\varepsilon/(Kw)$ to fulfill the
equation. Hence, the terms must have different signs to cause the
mutual cancellation.
\par
To illustrate how the correlation structure is affected by feedback,
let us now consider the case where the feedback activity is perturbed
(``feedforward scenario'').  We start from
\prettyref{eq:power_spectrum_feed_forward_uncorr} and, again, only
consider the fluctuations at zero frequency,
\begin{equation}
  \mat{C}_{\tilde{R}\tilde{R}}(0) = \mat{W} \mat{C_{QQ}} \mat{W}\transp + \eye \rho^2
  \,.
\end{equation}
\par
First, we consider a manipulation that preserves the single-neuron
statistics $A_{\Ex}$, $A_{\In}$ and the pairwise correlations
$C_{\Ex\Ex}$, $C_{\In\In}$ within each subpopulation, but neglects
correlations $C_{\Ex\In}$ between excitatory and inhibitory neurons.
Formally, this corresponds to the block diagonal correlation matrix
\begin{equation}
  \mat{C_{QQ}}_{ij} = 
  \begin{cases}
    \delta_{ij}A_\Ex + (1-\delta_{ij})C_{\Ex\Ex} & i,j\in\Exc\\
    \delta_{ij}A_\In + (1-\delta_{ij})C_{\In\In} & i,j\in\Inh
    \,.
  \end{cases}
\end{equation}
Here, we have replaced the individual entries of the correlation
matrix by the corresponding subpopulation averaged correlations. The
calculation of the response auto- and cross-correlation
$\tilde{A}$ and $\tilde{C}$ is similar as for the expressions
\prettyref{eq:A_avg_term_III} and \prettyref{eq:C_avg_term_III}, with
the difference that terms containing $C_{\Ex\In}$ are absent:
\begin{eqnarray}
  \label{eq:A_C_FF}
  \begin{aligned}
    \tilde{A}&=Kw^2\left(A_\Ex + \gamma g^2 A_\In \right) +  c + \rho^2\\
    \tilde{C}&=(Kw)^2\left(\frac{A_\Ex}{N_\Ex}+(\gamma{}g)^2\frac{A_\In}{N_\In}\right) + c\\
    \text{with} &\quad
    c=(Kw)^2\left(C_{\Ex\Ex}+(\gamma{}g)^2 C_{\In\In}\right)
    \,.
  \end{aligned}
\end{eqnarray}
\par
As an alternative type of feedback manipulation, we assume that all
correlations are equal, irrespective of the neuron type. To this end,
we replace all spike correlations by the population average $C =
(N_\Ex^2 C_{\Ex \Ex} + N_\In^2 C_{\In \In} + 2N_\Ex N_\In C_{\Ex \In}
) / (N_\Ex + N_\In)^2 = (N_\Ex C_{\Ex \Ex} + N_\In C_{\In \In}) /
(N_\Ex + N_\In)$. Thus, the covariance matrix reads
\begin{equation}
  \mat{C_{QQ}}_{ij} = \delta_{ij} A + (1 - \delta_{ij}) C.
\end{equation}
The calculation follows the one leading to the expressions
\prettyref{eq:A_avg_term_III} and \prettyref{eq:C_avg_term_III} and
results in
\begin{eqnarray} \label{eq:A_C_FF_hom_ndim}
  \tilde{A} &=& w^2 K (1 + \gamma g^2) A + (w K)^2 (1 - \gamma g)^2 C + \rho^2\\ \nonumber 
  \tilde{C} &=& w^2 K^2 \left(\frac{1}{N_\Ex} + \frac{(\gamma g)^2}{N_\In} \right) A + (w K)^2 (1 - \gamma g)^2 C
  \,.
\end{eqnarray}

\section{Acknowledgments}
\label{sec:thanks}

We thank the three reviewers for their constructive comments. 

\pdfbookmark[1]{References}{ReferencesPage}

\begin{thebibliography}{10}
\providecommand{\url}[1]{\texttt{#1}}
\providecommand{\urlprefix}{URL }
\expandafter\ifx\csname urlstyle\endcsname\relax
  \providecommand{\doi}[1]{doi:\discretionary{}{}{}#1}\else
  \providecommand{\doi}{doi:\discretionary{}{}{}\begingroup
  \urlstyle{rm}\Url}\fi
\providecommand{\bibAnnoteFile}[1]{%
  \IfFileExists{#1}{\begin{quotation}\noindent\textsc{Key:} #1\\
  \textsc{Annotation:}\ \input{#1}\end{quotation}}{}}
\providecommand{\bibAnnote}[2]{%
  \begin{quotation}\noindent\textsc{Key:} #1\\
  \textsc{Annotation:}\ #2\end{quotation}}
\providecommand{\eprint}[2][]{\url{#2}}

\bibitem{Abeles91}
Abeles M (1991) {Corticonics: Neural Circuits of the Cerebral Cortex}.
\newblock Cambridge: Cambridge University Press, 1st edition.
\bibAnnoteFile{Abeles91}

\bibitem{Aviel05_691}
Aviel Y, Horn D, Abeles M (2005) {{M}emory capacity of balanced networks}.
\newblock Neural Comput 17:691--713.
\newblock Comparative Study.
\bibAnnoteFile{Aviel05_691}

\bibitem{Aviel03a}
Aviel Y, Mehring C, Abeles M, Horn D (2003) On embedding synfire chains in a
  balanced network.
\newblock Neural Comput 15:1321--1340.
\bibAnnoteFile{Aviel03a}

\bibitem{Battaglia07_238106}
Battaglia D, Brunel N, Hansel D (2007) Temporal decorrelation of collective
  oscillations in neural networks with local inhibition and long-range
  excitation.
\newblock Phys Rev Lett 99:238106.
\bibAnnoteFile{Battaglia07_238106}

\bibitem{Bi98}
Bi G, Poo M (1998) Synaptic modifications in cultured hippocampal neurons:
  Dependence on spike timing, synaptic strength, and postsynaptic cell type.
\newblock J Neurosci 18:10464--10472.
\bibAnnoteFile{Bi98}

\bibitem{Bienenstock95}
Bienenstock E (1995) A model of neocortex.
\newblock Network: Comput Neural Systems 6:179--224.
\bibAnnoteFile{Bienenstock95}

\bibitem{Blomquist09}
Blomquist P, Devor A, Indahl UG, Ulbert I, Einevoll GT, et~al. (2009)
  Estimation of thalamocortical and intracortical network models from joint
  thalamic single-electrode and cortical laminar-electrode recordings in the
  rat barrel system.
\newblock PLoS Comput Biol 5:e1000328.
\bibAnnoteFile{Blomquist09}

\bibitem{Boucsein09_1006}
Boucsein C, Tetzlaff T, Meier R, Aertsen A, Naundorf B (2009) Dynamical
  response properties of neocortical neuron ensembles: multiplicative versus
  additive noise.
\newblock J Neurosci 29:{1006--1010}.
\bibAnnoteFile{Boucsein09_1006}

\bibitem{Brunel00_183}
Brunel N (2000) Dynamics of sparsely connected networks of excitatory and
  inhibitory spiking neurons.
\newblock J Comput Neurosci 8:183--208.
\bibAnnoteFile{Brunel00_183}

\bibitem{Brunel01_2186}
Brunel N, Chance FS, Fourcaud N, Abbott LF (2001) Effects of synaptic noise and
  filtering on the frequency response of spiking neurons.
\newblock Phys Rev Lett 86:2186--2189.
\bibAnnoteFile{Brunel01_2186}

\bibitem{Brunel99}
Brunel N, Hakim V (1999) Fast global oscillations in networks of
  integrate-and-fire neurons with low firing rates.
\newblock Neural Comput 11:1621--1671.
\bibAnnoteFile{Brunel99}

\bibitem{Buice09_377}
Buice MA, Cowan JD, Chow CC (2009) Systematic fluctuation expansion for neural
  network activity equations.
\newblock Neural Comput 22:377--426.
\bibAnnoteFile{Buice09_377}

\bibitem{Burkitt07_533}
Burkitt AN, Gilson M, van Hemmen J (2007) Spike-timing-dependent plasticity for
  neurons with recurrent connections.
\newblock Biol Cybern 96:533--546.
\bibAnnoteFile{Burkitt07_533}

\bibitem{Chizhov08_011910}
Chizhov AV, Graham LJ (2008) Efficient evaluation of neuron populations
  receiving colored-noise current based on a refractory density method.
\newblock Phys Rev E 77:011910.
\bibAnnoteFile{Chizhov08_011910}

\bibitem{Cox62}
Cox DR (1962) Renewal Theory.
\newblock London: Methuen.
\bibAnnoteFile{Cox62}

\bibitem{DeLaRocha07_802}
De~la Rocha J, Doiron B, Shea-Brown E, Kresimir J, Reyes A (2007) Correlation
  between neural spike trains increases with firing rate.
\newblock Nature 448:802--807.
\bibAnnoteFile{DeLaRocha07_802}

\bibitem{Diesmann99}
Diesmann M, Gewaltig MO, Aertsen A (1999) Stable propagation of synchronous
  spiking in cortical neural networks.
\newblock Nature 402:529--533.
\bibAnnoteFile{Diesmann99}

\bibitem{Ecker10}
Ecker AS, Berens P, Keliris GA, Bethge M, Logothetis NK (2010) Decorrelated
  neuronal firing in cortical microcircuits.
\newblock Science 327:584--587.
\bibAnnoteFile{Ecker10}

\bibitem{Fourcaud02}
Fourcaud N, Brunel N (2002) Dynamics of the firing probability of noisy
  integrate-and-fire neurons.
\newblock Neural Comput 14:2057--2110.
\bibAnnoteFile{Fourcaud02}

\bibitem{Fourcaud03_11640}
Fourcaud-Trocm{\'e} N, Hansel D, van Vreeswijk C, Brunel (2003) How spike
  generation mechanisms determine the neuronal response to fluctuating inputs.
\newblock J Neurosci 23:11628--11640.
\bibAnnoteFile{Fourcaud03_11640}

\bibitem{Gentet10_422}
Gentet L, Avermann M, Matyas F, Staiger JF, Petersen CC (2010) Membrane
  potential dynamics of {GABA}ergic neurons in the barrel cortex of behaving
  mice.
\newblock Neuron 65:422--435.
\bibAnnoteFile{Gentet10_422}

\bibitem{Gerstner00}
Gerstner W (2000) Population dynamics of spiking neurons: fast transients,
  asynchronous states, and locking.
\newblock Neural Comput 12:43--89.
\bibAnnoteFile{Gerstner00}

\bibitem{Gilson09_1}
Gilson M, Burkitt AN, Grayden DB, Thomas DA, van Hemmen JL (2009) Emergence of
  network structure due to spike-timing-dependent plasticity in recurrent
  neuronal networks. {I. I}nput selectivity - strengthening correlated input
  pathways.
\newblock Biol Cybern 101:81--102.
\bibAnnoteFile{Gilson09_1}

\bibitem{Ginzburg94}
Ginzburg I, Sompolinsky H (1994) Theory of correlations in stochastic neural
  networks.
\newblock Phys Rev E 50:3171--3191.
\bibAnnoteFile{Ginzburg94}

\bibitem{Harris11_509}
Harris KD, Thiele A (2011) Cortical state and attention 12:509--523.
\newblock Doi:10.1038/nrn3084.
\bibAnnoteFile{Harris11_509}

\bibitem{Hebb49}
Hebb DO (1949) The organization of behavior: A neuropsychological theory.
\newblock New York: John Wiley \& Sons.
\bibAnnoteFile{Hebb49}

\bibitem{Helias10}
Helias M, Deger M, Diesmann M, Rotter S (2010) Equilibrium and response
  properties of the integrate-and-fire neuron in discrete time.
\newblock Front Comput Neurosci 3:doi:10.3389/neuro.10.029.2009.
\bibAnnoteFile{Helias10}

\bibitem{Helias10_1000929}
Helias M, Deger M, Rotter S, Diesmann M (2010) Instantaneous non-linear
  processing by pulse-coupled threshold units.
\newblock PLoS Comput Biol 6:e1000929.
\newblock Doi:10.1371/journal.pcbi.1000929.
\bibAnnoteFile{Helias10_1000929}

\bibitem{Helias08_7}
Helias M, Rotter S, Gewaltig M, Diesmann M (2008) Structural plasticity
  controlled by calcium based correlation detection.
\newblock Front Comput Neurosci 2:doi:10.3389/neuro.10.007.2008.
\bibAnnoteFile{Helias08_7}

\bibitem{Hertz10_427}
Hertz J (2010) Cross-correlations in high-conductance states of a model
  cortical network.
\newblock Neural Comput 22:427--447.
\bibAnnoteFile{Hertz10_427}

\bibitem{Jacobsen07_1330}
Jacobsen M, Jensen AT (2007) Exit times for a class of piecewise exponential
  markov processes with two-sided jumps.
\newblock Stoch Proc Appl 117:1330--1356.
\bibAnnoteFile{Jacobsen07_1330}

\bibitem{Jahnke08}
Jahnke S, Memmesheimer R, Timme M (2008) Stable irregular dynamics in complex
  neural networks.
\newblock Phys Rev Lett 100:048102.
\bibAnnoteFile{Jahnke08}

\bibitem{Jahnke09}
Jahnke S, Memmesheimer RM, Timme M (2009) How chaotic is the balanced state?
\newblock Frontiers in Computational Neuroscience 3:1--15.
\bibAnnoteFile{Jahnke09}

\bibitem{Knight72a}
Knight BW (1972) Dynamics of encoding in a population of neurons.
\newblock J Gen Physiol 59:734--766.
\bibAnnoteFile{Knight72a}

\bibitem{Knight72b}
Knight BW (1972) The relationship between the firing rate of a single neuron
  and the level of activity in a population of neurons.
\newblock J Gen Physiol 59:767--778.
\bibAnnoteFile{Knight72b}

\bibitem{Koendgen08_2086}
K{\"o}ndgen H, Geisler C, Fusi S, Wang XJ, L{\"u}scher HR, et~al. (2008) The
  dynamical response properties of neocortical neurons to temporally modulated
  noisy inputs in vitro.
\newblock Cereb Cortex 18:2086--2097.
\bibAnnoteFile{Koendgen08_2086}

\bibitem{Kriener08_2185}
Kriener B, Tetzlaff T, Aertsen A, Diesmann M, Rotter S (2008) Correlations and
  population dynamics in cortical networks.
\newblock Neural Comput 20:2185--2226.
\bibAnnoteFile{Kriener08_2185}

\bibitem{Legenstein07_323}
Legenstein R, Maass W (2007) Edge of chaos and prediction of computational
  performance for neural circuit models.
\newblock Neural Networks 20:323--334.
\bibAnnoteFile{Legenstein07_323}

\bibitem{Lindner05_061919}
Lindner B, Doiron B, Longtin A (2005) Theory of oscillatory firing induced by
  spatially correlated noise and delayed inhibitory feedback.
\newblock Phys Rev E 72:061919.
\bibAnnoteFile{Lindner05_061919}

\bibitem{Lindner01_2934}
Lindner B, Schimansky-Geier L (2001) Transmission of noise coded versus
  additive signals through a neuronal ensemble.
\newblock Phys Rev Lett 86:2934--2937.
\bibAnnoteFile{Lindner01_2934}

\bibitem{Loebel09}
Loebel A, Silberberg G, Helbig D, Markram H, Tsodyks M, et~al. (2009)
  Multiquantal release underlies the distribution of synaptic efficacies in the
  neocortex.
\newblock Front Comput Neurosci 3.
\bibAnnoteFile{Loebel09}

\bibitem{Mehring01a}
Mehring C, Hehl U, Kubo M, Diesmann M, Aertsen A (2002) Activity dynamics and
  propagation of synchronous spiking in locally connected random networks
  (submitted).
\newblock Submitted.
\bibAnnoteFile{Mehring01a}

\bibitem{Meyer02}
Meyer C, van Vreeswijk C (2002) Temporal correlations in stochastic networks of
  spiking neurons.
\newblock Neural Comput 14:369--404.
\bibAnnoteFile{Meyer02}

\bibitem{Monteforte10}
Monteforte M, Wolf F (2010) Dynamical entropy production in spiking neuron
  networks in the balanced state.
\newblock arXiv:cond-mat :arXiv:1003.4410v1.
\bibAnnoteFile{Monteforte10}

\bibitem{Morenobote06_028101}
Moreno-Bote R, Parga N (2006) Auto- and crosscorrelograms for the spike
  response of leaky integrate-and-fire neurons with slow synapses.
\newblock Phys Rev Lett 96:028101.
\bibAnnoteFile{Morenobote06_028101}

\bibitem{MorenoBote08}
Moreno-Bote R, Renart A, Parga N (2008) Theory of input spike auto- and
  cross-correlations and their effect on the response of spiking neurons.
\newblock Neural Comput 20:1651--1705.
\bibAnnoteFile{MorenoBote08}

\bibitem{Murphy09_635}
Murphy BK, Miller KD (2009) Balanced amplification: A new mechanism of
  selective amplification of neural activity patterns.
\newblock Neuron 61:635--648.
\bibAnnoteFile{Murphy09_635}

\bibitem{Naundorf05_297}
Naundorf B, Geisel T, Wolf F (2005) Action potential onset dynamics and the
  response speed of neuronal populations.
\newblock J Comput Neurosci 18:297--309.
\bibAnnoteFile{Naundorf05_297}

\bibitem{Nawrot08_374}
Nawrot MP, Boucsein C, Rodriguez~Molina V, Riehle A, Aertsen A, et~al. (2008)
  Measurement of variability dynamics in cortical spike trains.
\newblock J Neurosci Methods 169:374--390.
\bibAnnoteFile{Nawrot08_374}

\bibitem{Nordlie10}
Nordlie E, Tetzlaff T, Einevoll GT (2010) Rate dynamics of leaky
  integrate-and-fire neurons with strong synapses.
\newblock Front Comput Neurosci 4:149.
\bibAnnoteFile{Nordlie10}

\bibitem{Oppenheim96}
Oppenheim A, Wilsky A (1996) Systems and signals.
\newblock Prentice Hall.
\bibAnnoteFile{Oppenheim96}

\bibitem{Pernice11_e1002059}
Pernice V, Staude B, Cardanobile S, Rotter S (2011) How structure determines
  correlations in neuronal networks.
\newblock PLoS Comput Biol 7:e1002059.
\bibAnnoteFile{Pernice11_e1002059}

\bibitem{Pernice12_031916}
Pernice V, Staude B, Cardanobile S, Rotter S (2012) Recurrent interactions in
  spiking networks with arbitrary topology.
\newblock Phys Rev E 85:031916.
\bibAnnoteFile{Pernice12_031916}

\bibitem{Pfister10_22}
Pfister JP, Tass PA (2010) Stdp in oscillatory recurrent networks: theoretical
  conditions for desynchronization and applications to deep brain stimulation.
\newblock Frontiers in Computational Neuroscience 454:doi:
  10.3389/fncom.2010.00022.
\bibAnnoteFile{Pfister10_22}

\bibitem{Pressley09_63}
Pressley J, Troyer TW (2009) Complementary responses to mean and variance
  modulations in the perfect integrate-and-fire model.
\newblock Biol Cybern 101:63--70.
\bibAnnoteFile{Pressley09_63}

\bibitem{Rajan06}
Rajan K, Abbott L (2006) Eigenvalue spectra of random matrices for neural
  networks.
\newblock Phys Rev Lett 97:188104.
\bibAnnoteFile{Rajan06}

\bibitem{Renart10_587}
Renart A, De~La~Rocha J, Bartho P, Hollender L, Parga N, et~al. (2010) The
  asynchronous state in cortical cicuits.
\newblock Science 327:587--590.
\bibAnnoteFile{Renart10_587}

\bibitem{Richardson10_178102}
Richardson MJE, Swarbrick R (2010) Firing-rate response of a neuron receiving
  excitatory and inhibitory synaptic shot noise.
\newblock Phys Rev Lett 105:178102.
\bibAnnoteFile{Richardson10_178102}

\bibitem{Risken96}
Risken H (1996) The Fokker-Planck Equation.
\newblock Springer Verlag Berlin Heidelberg.
\bibAnnoteFile{Risken96}

\bibitem{Rosenbaum10_00116}
Rosenbaum R, Josic K (2011) Mechanisms that modulate the transfer of spiking
  correlations.
\newblock Neural Comput :doi:10.1162/NECO\_a\_00116.
\bibAnnoteFile{Rosenbaum10_00116}

\bibitem{Rotter99a}
Rotter S, Diesmann M (1999) Exact digital simulation of time-invariant linear
  systems with applications to neuronal modeling.
\newblock Biol Cybern 81:381--402.
\bibAnnoteFile{Rotter99a}

\bibitem{Salinas01}
Salinas E, Sejnowski TJ (2001) Correlated neuronal activity and the flow of
  neural information.
\newblock Nat Rev Neurosci 2:539--550.
\bibAnnoteFile{Salinas01}

\bibitem{Shadlen98}
Shadlen MN, Newsome WT (1998) The variable discharge of cortical neurons:
  Implications for connectivity, computation, and information coding.
\newblock J Neurosci 18:3870--3896.
\bibAnnoteFile{Shadlen98}

\bibitem{Shadlen01_1916}
Shadlen MN, Newsome WT (2001) Neural basis of a perceptual decision in the
  parietal cortex (area {LIP}) of the rhesus monkey.
\newblock J Neurophysiol 86:1916--1936.
\bibAnnoteFile{Shadlen01_1916}

\bibitem{Siegert51}
Siegert AJ (1951) On the first passage time probability problem.
\newblock Phys Rev 81:617--623.
\bibAnnoteFile{Siegert51}

\bibitem{sirovich00_2009}
Sirovich L, Omurtag A, Knight BW (2000) Dynamics of neuronal populations: The
  equilibrium solution.
\newblock SIAM J Appl Math 60:2009--2028.
\bibAnnoteFile{sirovich00_2009}

\bibitem{Softky93}
Softky WR, Koch C (1993) The highly irregular firing of cortical cells is
  inconsistent with temporal integration of random {EPSP}s.
\newblock J Neurosci 13:334--350.
\bibAnnoteFile{Softky93}

\bibitem{Sommers88}
Sommers H, Crisanti A, Sompolinsky H, Stein Y (1988) Spectrum of large random
  asymmetric matrices.
\newblock Phys Rev Lett 60:1895--1898.
\bibAnnoteFile{Sommers88}

\bibitem{Sompolinsky88_259}
Sompolinsky, Crisanti, Sommers (1988) Chaos in random neural networks.
\newblock Phys Rev Lett 61:259--262.
\bibAnnoteFile{Sompolinsky88_259}

\bibitem{Stroeve01}
Stroeve S, Gielen S (2001) Correlation between uncoupled conductance-based
  integrate-and-fire neurons due to common and synchronous presynaptic firing.
\newblock Neural Comput 13:2005--2029.
\bibAnnoteFile{Stroeve01}

\bibitem{Tetzlaff02}
Tetzlaff T, Buscherm{\"o}hle M, Geisel T, Diesmann M (2003) The spread of rate
  and correlation in stationary cortical networks.
\newblock Neurocomputing 52--54:949--954.
\bibAnnoteFile{Tetzlaff02}

\bibitem{Tetzlaff02_673}
Tetzlaff T, Geisel T, Diesmann M (2002) The ground state of cortical
  feed-forward networks.
\newblock Neurocomputing 44--46:673--678.
\bibAnnoteFile{Tetzlaff02_673}

\bibitem{Tetzlaff04_117}
Tetzlaff T, Morrison A, Geisel T, Diesmann M (2004) Consequences of realistic
  network size on the stability of embedded synfire chains.
\newblock Neurocomputing 58--60:117--121.
\bibAnnoteFile{Tetzlaff04_117}

\bibitem{Tetzlaff08_2133}
Tetzlaff T, Rotter S, Stark E, Abeles M, Aertsen A, et~al. (2008) Dependence of
  neuronal correlations on filter characteristics and marginal spike-train
  statistics.
\newblock Neural Comput 20:{2133--2184}.
\bibAnnoteFile{Tetzlaff08_2133}

\bibitem{Toyoizumi10}
Toyoizumi T, Abbott LF (2010) Beyond the edge: Amplification and temporal
  integration by recurrent networks in the chaotic regime.
\newblock Frontiers in Neuroscience :doi:
  10.3389/conf.fnins.2010.03.00155Conference Abstract: Computational and
  Systems Neuroscience 2010.
\bibAnnoteFile{Toyoizumi10}

\bibitem{Toyoizumi09}
Toyoizumi T, Rad KR, Paninski L (2009) Mean-field approximations for coupled
  populations of generalized linear model spiking neurons with markov
  refractoriness.
\newblock Neural Comput 21:1203--1243.
\bibAnnoteFile{Toyoizumi09}

\bibitem{Tripp07_1830}
Tripp B, Eliasmith C (2007) Neural populations can induce reliable postsynaptic
  currents without observable spike rate changes or precise spike timing.
\newblock Cereb Cortex 17:1830--1840.
\bibAnnoteFile{Tripp07_1830}

\bibitem{Trousdale12_e1002408}
Trousdale J, Hu Y, Shea-Brown E, Josic K (2012) Impact of network structure and
  cellular response on spike time correlations.
\newblock PLoS Comput Biol 8:e1002408.
\bibAnnoteFile{Trousdale12_e1002408}

\bibitem{Troyer02_2741}
Troyer TW, Krukowski AE, Miller KD (2002) Lgn input to simple cells and
  contrast-invariant orientation tuning: An analysis.
\newblock J Neurophysiol 87:2741--2752.
\bibAnnoteFile{Troyer02_2741}

\bibitem{Tuckwell88a}
Tuckwell HC (1988) Introduction to Theoretical Neurobiology, volume~1.
\newblock Cambridge: Cambridge University Press.
\bibAnnoteFile{Tuckwell88a}

\bibitem{Vreeswijk96}
van Vreeswijk C, Sompolinsky H (1996) Chaos in neuronal networks with balanced
  excitatory and inhibitory activity.
\newblock Science 274:1724--1726.
\bibAnnoteFile{Vreeswijk96}

\bibitem{Malsburg81}
{von der Malsburg} C (1981) The correlation theory of brain function.
\newblock Internal report 81-2, Max-Planck-Institute for Biophysical Chemistry,
  G{\"o}ttingen, FRG.
\bibAnnoteFile{Malsburg81}

\bibitem{Zhaoping04_198106}
Zhaoping L, Lewis A, Scarpetta S (2004) Mathematical analysis and simulations
  of the neural circuit for locomotion in lampreys.
\newblock Phys Rev Lett 92:198106.
\bibAnnoteFile{Zhaoping04_198106}

\bibitem{Zillmer06}
Zillmer R, Livi R, Politi A, Torcini A (2006) Desynchronization in diluted
  neural networks.
\newblock Phys Rev E 74:036203.
\bibAnnoteFile{Zillmer06}

\bibitem{Zohary94_140}
Zohary E, Shadlen MN, Newsome WT (1994) Correlated neuronal discharge rate and
  its implications for psychophysical performance.
\newblock Nature 370:140--143.
\bibAnnoteFile{Zohary94_140}

\end{thebibliography}


\end{document}